\documentclass[letterpaper,twocolumn,10pt]{article}
\usepackage{usenix-2020-09}
\usepackage{amsmath}
\usepackage{amssymb}
\usepackage{listings}
\usepackage{balance}
\usepackage{booktabs}
\usepackage[most]{tcolorbox}
\usepackage{xcolor}
\usepackage{xspace}
\usepackage{tikz}
\usepackage{multirow}
\usetikzlibrary{arrows.meta,positioning,shapes.geometric}
\usepackage{cleveref}
\usepackage[inline]{enumitem}
\usepackage{siunitx}
\usepackage{pifont}
\usepackage[abbreviations]{foreign}
\usepackage{placeins}

\usepackage{ifthen}
\newboolean{showcomments}
\setboolean{showcomments}{false}

\ifthenelse{\boolean{showcomments}}
{ \newcommand{\mynote}[3]{
		\fbox{\bfseries\sffamily\scriptsize#1}
		{\small$\blacktriangleright$\textsf{\emph{\color{#3}{#2}}}$\blacktriangleleft$}}
	\newcommand{\zzz}[1]{{\setlength{\fboxsep}{2pt}\fcolorbox{black}{yellow}{\textsf{\emph{#1}}}}\xspace}}
{ \newcommand{\mynote}[3]{}
	\newcommand{\zzz}[1]{}}

\usepackage{acronym}
\acrodef{SMR}{state machine replication}
\acrodef{LLM}{large language model}
\acrodef{PQL}{Paxos Quorum Leases}

\newcommand{\code}[1]{\texttt{#1}}

\newcommand{\OurAgent}{\mbox{\textsc{AI-greement}}\xspace}
\newcommand{\OurProtocol}{\mbox{\textsc{A-greement}}\xspace}
\newcommand{\swiftpaxos}{\textsc{SwiftPaxos}\xspace}
\newcommand{\bodega}{\textsc{Bodega}\xspace}
\newcommand{\paxos}{\textsc{Paxos}\xspace}
\newcommand{\multipaxos}{\textsc{Multi-Paxos}\xspace}
\newcommand{\fastpaxos}{\textsc{FastPaxos}\xspace}
\newcommand{\epaxos}{\textsc{EPaxos}\xspace}
\newcommand{\epaxosstar}{\textsc{EPaxos}*\xspace}
\newcommand{\kcensus}{\textsc{KCensus}\xspace}
\newcommand{\codex}{\textsc{Codex}\xspace}
\newcommand{\curp}{\textsc{CURP}\xspace}

\definecolor{promptpurple}{HTML}{6A1B9A}

\graphicspath{ {figures/} }
\usepackage{tikz}
\usepackage[eulergreek]{sansmath}
\usepackage{pgfplots}
\usepackage{pgfplotstable}
\usepackage{cleveref}
\usepackage{comment}
\crefname{assumption}{assumption}{assumptions}

\pgfplotsset{compat=newest}
\usepgfplotslibrary{external,units,colorbrewer,groupplots,fillbetween,statistics}
\usetikzlibrary{arrows.meta, backgrounds, fit, shapes.geometric, positioning}
\tikzsetexternalprefix{figures/}
\tikzset{external/mode=list and make}
\tikzexternaldisable %
\usetikzlibrary{patterns,shapes.misc}

\definecolor{pal1}{HTML}{A6CEE3}\definecolor{pal1dark}{HTML}{1F78B4}
\definecolor{pal2}{HTML}{B2DF8A}\definecolor{pal2dark}{HTML}{33A02C}
\definecolor{pal3}{HTML}{FDBF6F}\definecolor{pal3dark}{HTML}{FF7F00}
\definecolor{pal4}{HTML}{CAB2D6}\definecolor{pal4dark}{HTML}{6A3D9A}
\definecolor{pal5}{HTML}{E31A1C}\definecolor{pal5dark}{HTML}{E31A1C}
\colorlet{palpattern}{black!50}
\pgfdeclarepatterninherentlycolored{pal north east lines}{\pgfqpoint{-1pt}{-1pt}}{\pgfqpoint{4pt}{4pt}}{\pgfqpoint{3pt}{3pt}}{%
  \pgfsetcolor{palpattern}\pgfsetlinewidth{.4pt}%
  \pgfpathqmoveto{0pt}{0pt}\pgfpathqlineto{3.1pt}{3.1pt}\pgfusepathqstroke}
\pgfdeclarepatterninherentlycolored{pal north west lines}{\pgfqpoint{-1pt}{-1pt}}{\pgfqpoint{4pt}{4pt}}{\pgfqpoint{3pt}{3pt}}{%
  \pgfsetcolor{palpattern}\pgfsetlinewidth{.4pt}%
  \pgfpathqmoveto{0pt}{3pt}\pgfpathqlineto{3.1pt}{-.1pt}\pgfusepathqstroke}
\pgfdeclarepatterninherentlycolored{pal crosshatch}{\pgfqpoint{-1pt}{-1pt}}{\pgfqpoint{4pt}{4pt}}{\pgfqpoint{3pt}{3pt}}{%
  \pgfsetcolor{palpattern}\pgfsetlinewidth{.4pt}%
  \pgfpathqmoveto{0pt}{0pt}\pgfpathqlineto{3.1pt}{3.1pt}%
  \pgfpathqmoveto{0pt}{3pt}\pgfpathqlineto{3.1pt}{-.1pt}\pgfusepathqstroke}
\pgfdeclarepatterninherentlycolored{pal dots}{\pgfqpoint{-1pt}{-1pt}}{\pgfqpoint{1pt}{1pt}}{\pgfqpoint{3pt}{3pt}}{%
  \pgfsetcolor{palpattern}%
  \pgfpathqcircle{.5pt}\pgfusepathqfill}
\tikzset{
  pal pattern/.style={postaction={pattern=pal #1}},
  pal1 bar/.style={fill=pal1, pal pattern=north east lines},
  pal2 bar/.style={fill=pal2, pal pattern=north west lines},
  pal3 bar/.style={fill=pal3, pal pattern=crosshatch},
  pal4 bar/.style={fill=pal4, pal pattern=dots},
  pal5 bar/.style={fill=pal5},
  pal1 line/.style={pal1dark, dashed},
  pal2 line/.style={pal2dark, dashdotted},
  pal3 line/.style={pal3dark, densely dotted},
  pal4 line/.style={pal4dark, dash pattern=on 4pt off 1.5pt on 1pt off 1.5pt on 1pt off 1.5pt},
  pal5 line/.style={pal5dark, solid},
}

\newcommand{\inputplot}[1]{\includegraphics{#1.pdf}}

\newcommand{\newgroupwidth}[2]%
{\expandafter\xdef\csname groupwidth#1\endcsname{#2}}

\newcounter{groupwidth}
\newsavebox{\groupwidthbox}
\makeatletter
{\edef\groupnumber{#1}%
	\stepcounter{groupwidth}%
	\@ifundefined{groupwidth\thegroupwidth}{\pgfmathsetlengthmacro{\mywidth}{\linewidth/\groupnumber}}%
	{\expandafter\let\expandafter\mywidth\csname groupwidth\thegroupwidth\endcsname}%
	\begin{lrbox}{\groupwidthbox}%
		\tikzset{/pgfplots/width={\mywidth}}%
		\ignorespaces}%
	{\end{lrbox}%
	\usebox\groupwidthbox
	\pgfmathsetlengthmacro{\mywidth}{\mywidth + (\linewidth - \wd\groupwidthbox)/\groupnumber}
	\immediate\write\@auxout{\string\newgroupwidth{\thegroupwidth}{\mywidth}}}
\makeatother

\usepackage{amsmath,amssymb,amsfonts,amsthm}
\usepackage{physics}
\usepackage{enumitem}
\usepackage{thm-restate}
\usepackage{bbm}
\theoremstyle{definition}

\theoremstyle{remark}

\allowdisplaybreaks

\providecommand{\PreAccept}{\textsc{PreAccept}}
\providecommand{\Commit}{\textsc{Commit}}

\definecolor{ink}{HTML}{202629}
\definecolor{gray}{HTML}{626B70}
\definecolor{mid}{HTML}{A1A9AD}
\definecolor{rule}{HTML}{CBD0D2}
\definecolor{xcol}{HTML}{783F32}
\definecolor{ycol}{HTML}{225078}
\definecolor{reply}{HTML}{617C89}
\definecolor{unsafe}{HTML}{A33A32}
\definecolor{unsafefill}{HTML}{FBF0EE}
\definecolor{finalcol}{HTML}{496F5B}
\definecolor{finalfill}{HTML}{EEF5F1}

\tikzset{
  xmsg/.style={
    draw=xcol,
    line width=1.25pt,
    -{Latex[length=7.5pt,width=6pt]}
  },
  ymsg/.style={
    draw=ycol,
    line width=1.25pt,
    -{Latex[length=7.5pt,width=6pt]}
  },
  replymsg/.style={
    draw=reply,
    line width=1.05pt,
    dash pattern=on 4.5pt off 3pt,
    -{Latex[length=7.5pt,width=6pt]}
  },
  delaymsg/.style={
    draw=xcol,
    line width=1.05pt,
    dash pattern=on 5pt off 3.5pt,
    -{Latex[length=7.5pt,width=6pt]}
  },
  event/.style={
    rounded corners=2.6pt,
    fill=white,
    line width=.65pt,
    minimum width=100pt,
    minimum height=28pt,
    inner sep=2pt,
    font=\fontsize{11.5}{12.3}\selectfont
  },
  eventx/.style={event,draw=xcol},
  eventy/.style={event,draw=ycol},
  unsafeevent/.style={
    event,
    draw=unsafe,
    fill=unsafefill,
    text=unsafe
  },
  finalevent/.style={
    event,
    draw=finalcol,
    fill=finalfill,
    text=finalcol
  },
  replylabel/.style={
    fill=white,
    rounded corners=1.5pt,
    minimum width=34pt,
    minimum height=14pt,
    inner sep=1pt,
    font=\fontsize{10.5}{11.3}\selectfont,
    text=reply
  }
}

\lstdefinestyle{xpaxos}{
  basicstyle=\ttfamily\scriptsize,
  numbers=left, numberstyle=\tiny, numbersep=6pt,
  columns=fullflexible, keepspaces=true,
  breaklines=true, showstringspaces=false,
  frame=single, framerule=0.3pt, framesep=3pt,
  xleftmargin=18pt, xrightmargin=4pt, framexleftmargin=14pt,
  aboveskip=4pt, belowskip=4pt
}

\begin{document}

\date{} \title{The End of Consensus Research?} 
\author{
  {\normalfont Clément Burgelin \quad Rachid Guerraoui \quad Zicong Hong \quad Yinan Hu}\\[0.3em]
  {\normalfont Anne-Marie Kermarrec \quad Rafael Pires \quad Rasmus Moorits Veski \quad Martijn de Vos}\\[0.2em]
  {\normalfont\small Authors listed in alphabetical order}\\[0.5em]
  École Polytechnique Fédérale de Lausanne (EPFL)\\
  Lausanne, Switzerland
}
\maketitle

\begin{abstract}
Consensus is a fundamental problem in distributed computing and at the core of state machine replication (SMR). Despite decades of active human research, discovering practical low-latency solutions to the problem is still open.
This endeavor is very challenging since it requires combining protocol knowledge, correctness reasoning, implementation, and experimental evaluation.
We report on how AI helped us discover \OurProtocol, a new consensus protocol for geo-replicated state machines that significantly outperforms state-of-the-art solutions.
Averaged across six deployments with 5 or 9 replicas, \OurProtocol reduces the mean latency of a geo-distributed key-value store over the fastest baseline by 12.2\% and 33.3\% on write-heavy and read-heavy workloads, respectively, and matches the fastest baseline on read-only workloads.
\OurProtocol combines lease-authorized local reads, direct client learning for leader-ordered writes, and
a fast fixed-quorum per-key write path.
It selects replica roles and write paths according to network topology and workload, accounting for both write completion and read waiting. 
At the heart of our consensus discovery lies \OurAgent, an agent harness for the ``autonomous'' discovery, implementation, experimentation, and verification of candidate protocols.
\OurAgent integrates (1) a concept-oriented protocol knowledge base that explores consensus mechanisms through published protocols, (2) a modular implementation skeleton that structures protocol development, and (3) a \textsc{Lean}-based verification framework with reusable safety definitions, reference specifications, and proofs.
Our harness supports iterative design exploration, using performance measurements and fault-injection feedback to refine protocols.
\OurAgent is of independent interest, beyond \OurProtocol. Its \textsc{Lean} formalization uncovers several previously unreported vulnerabilities
in existing consensus protocols.
\OurAgent provides guidelines on how domain-specific harnesses can turn general-purpose coding agents into effective tools for (distributed) computing research.
Yet we argue that human consensus experts remain essential before, during, and after the search: agents reshape the researcher's role rather than end consensus research.

\end{abstract}

\section{Introduction}

In \ac{SMR}~\cite{lamport1978time,schneider1990smr}, a universal paradigm to build highly-available computing systems, consensus enables replicas to agree on a common sequence of client commands, maintaining consistent state despite concurrent requests and failures.
Consensus is a fundamental building block of modern distributed systems, underpinning replicated databases~\cite{corbett2012spanner,taft2020cockroachdb,huang2020tidb}, storage systems~\cite{bolosky2011datastore}, cloud coordination services~\cite{burrows2006chubby} and blockchain infrastructures~\cite{buchman2018tendermint,yin2019hotstuff}. 
Consensus has been an active research area for over four decades.
However, designing and implementing consensus protocols that are both correct and practically efficient remains an exceptionally challenging task~\cite{chandra2007paxos,ongaro2014search}.
Protocol decisions involving ordering, quorums, recovery, and execution are tightly coupled: an optimization that shortens the common-case path may complicate recovery or introduce subtle correctness violations~\cite{chandra2007paxos,sutra2020correctness,ryabinin2025democracy}, as we also find for the optimizations of EPaxos Revisited~\cite{tollman2021epaxos} (\autoref{sec:bugs}). Translating a protocol design into a working system further requires preserving its safety properties throughout implementation optimizations~\cite{hawblitzel2015ironfleet,wilcox2015verdi,sharma2023grove}, as well as demonstrating performance improvements across diverse network topologies and workloads.
Advancing the state of the art in consensus therefore requires tightly integrating protocol knowledge, correctness reasoning, implementation and experimental evaluation.

The overall question we address in this paper is whether, just like in other fields of computing, AI could help ``crack'' the consensus problem. More specifically, we aim to come up with a new practically effective solution that would, on the one hand, tolerate, besides failures of the processes, failures of the network (\ie, asynchronous periods) and, on the other hand, expedite decisions in common failure-free executions.
In particular, we ask whether we can develop a consensus protocol that would yield an \ac{SMR} scheme in which, in the common case, (a) read-only commands complete at a nearby replica and (b) commands with updates take one round trip to a majority of replicas. This would essentially be optimal. Importantly, and unlike many recent AI-assisted results in mathematics and theoretical computer science, we seek a protocol that is practically useful and hence, besides algorithmic design and proofs,  consider the implementation, deployment and evaluation as part of the AI-assisted task. 

In short, we show in this paper that the answer is positive.
Yet,  we highlight the crucial role of the human researcher in the development of an adequate AI agent, the prompting of the agent as well as the certification of the agent's output. We furthermore argue that this human researcher needs to be a consensus expert to effectively steer and leverage the capabilities of agents. 

Recent advances in the capabilities of \acp{LLM} have indeed enabled coding agents to autonomously perform increasingly complex software engineering tasks~\cite{jimenez2024swe,yang2024sweagent}.
For example, \textsc{Codex} by OpenAI~\cite{openai2026codex}, \textsc{Claude Code} by Anthropic~\cite{anthropic2026claudecode}, and \textsc{Cursor Agent} by Anysphere~\cite{cursor2026agent} can navigate repositories, modify code, execute tests, and iteratively refine implementations based on execution feedback.
However, these general-purpose capabilities do not necessarily translate into the autonomous discovery of consensus protocols.
Existing AI-driven systems research frameworks optimize self-contained policies such as schedulers and load balancers in simulators~\cite{novikov2025alphaevolve,cheng2025adrs}, but do not seem an adequate fit for consensus, whose logic is spread across many communicating, failure-prone nodes~\cite{cheng2025adrs}.
Our preliminary experiment in \autoref{sec:general_agent_enough} also suggests that a coding agent combined with a frontier \ac{LLM}, starting from a single existing consensus codebase, finds only marginal improvements. 
This gap motivates the need for specific support for consensus research beyond general-purpose coding capabilities.

We present \OurAgent, an agent harness that equips general-purpose coding agents with domain-specific support for autonomous consensus research.
\OurAgent addresses three key challenges in consensus discovery.
First, synthesizing new protocols requires understanding and combining mechanisms scattered across the consensus literature.
To support this, a \emph{concept-oriented protocol knowledge base} organizes consensus mechanisms and grounds them in concrete examples from published protocols.
Second, translating novel designs into executable implementations is challenging because consensus mechanisms often have complex protocol states and message flows.
A modular \emph{implementation skeleton} structures this scheme through well-defined components and interfaces, allowing the agent to focus on protocol-specific logic.
Third, subtle correctness violations can escape conventional testing, which covers only a limited set of message orderings and failure scenarios.
Formal verification closes this gap by proving safety for every execution of a protocol model, but writing such models and proofs from scratch for each candidate is too costly.
A \textsc{Lean}-based \emph{verification framework} therefore provides reusable safety definitions, reference specifications, and proof libraries, so the agent can obtain machine-checked evidence that its design, as modeled, is safe.
These components work together in an iterative discovery process, where performance measurements and fault-injection results guide the agent toward more efficient and robust  designs.

We demonstrate the capabilities of \OurAgent through \OurProtocol, a new consensus protocol that significantly outperforms state-of-the-art solutions.
To the best of our knowledge, \OurProtocol is \emph{the first consensus protocol that is designed and implemented by an \ac{LLM} agent.}
When replicating a key-value store, 
\OurProtocol jointly optimizes read and write latency.
\OurProtocol builds on the quorum-lease local reads of \ac{PQL}~\cite{paxosquorumleases}, generalizes \bodega's leader-side gating of writes~\cite{hu2026bodega} to any replica, and adds a leaderless write path in the style of \swiftpaxos~\cite{ryabinin2024swiftpaxos}.
Its design couples write completion with read availability: for frequently accessed keys, a designated completion owner can continue serving reads while a write is pending; for ordinary keys, clients learn leader-ordered writes directly from acceptor votes; and for infrequently accessed keys, writes can bypass leader ordering through a fixed-majority per-key path.
These paths share rules ensuring that local reads do not miss completed writes.
A planner uses network delays and workload statistics to select replica roles and write paths, accounting for both write latency and the time reads spend waiting behind pending writes.

Averaged across six deployments with 5 or 9 replicas, on write-intensive workloads (50\% writes) \OurProtocol reduces mean latency by 12.2\% relative to \swiftpaxos, the best alternative, at the cost of a 13.8\% higher p99 latency.
On read-dominated workloads (5\% writes), \OurProtocol reduces mean and p99 latency by 33.3\% and 30.1\%, respectively, relative to \bodega, the alternative with the lowest mean latency, with a p99 latency within 3.4\% of the best alternative on that metric, \swiftpaxos.
In the common case, \OurProtocol's reads complete at a single nearby replica, and writes to rarely accessed keys complete in one round trip.
\OurAgent completed the discovery process (including the verification) of \OurProtocol in around 24 hours, consuming 778.45 million tokens at an estimated API-equivalent cost of \$198.69.

To summarize, we make the following contributions:
\begin{itemize}
    \item We design \OurAgent, an agent harness designed for the automated discovery of consensus protocols (\autoref{sec:paxosagent}).
    \OurAgent integrates a consensus knowledge base, a modular consensus skeleton, and \textsc{Lean}-based formalization and verification.
    \item Using \OurAgent, we synthesize \OurProtocol, a new consensus protocol ``autonomously'' designed and implemented by \OurAgent.
    \OurProtocol combines localized reads, complementary write paths, as well as topology- and workload-aware planning (\autoref{sec:xpaxos},\ref{sec:apaxos_implementation},\ref{sec:human_enforced_changes}).
    \item We perform an extensive experimental evaluation of \OurProtocol across replica counts, workloads, and network topologies. Averaged across six deployments with 5 or 9 replicas, \OurProtocol reduces mean latency relative to the best alternative by 12.2\% and 33.3\% on write-heavy and read-heavy workloads, respectively, while matching the best-performing baseline on the read-only workload (\autoref{sec:evaluation}).
    \item Using \OurAgent, we describe four previously unreported safety vulnerabilities in existing consensus protocols, found through agent-driven \textsc{Lean} formalization and verification, demonstrating \OurAgent's ability to uncover subtle protocol flaws (\autoref{sec:bugs}).
    \item We reflect on our results and ask whether this marks the end of human-driven consensus research, based on where our agent succeeded, where it needed human steering, and what remains for researchers to do (\autoref{sec:discussion}). Our findings imply that a human consensus expert is (still) badly needed.  %
    Many of these findings should apply to other (distributed) computing problems, especially when the goal is to discover, formally prove and practically evaluate a solution.
\end{itemize}

\section{Background \& Motivation}
\label{sec:background}
We first provide the background needed to motivate \OurAgent by explaining consensus and state machine replication (\autoref{sec:smr}).
We then describe how \ac{LLM} agents and their harnesses can solve complex problems (\autoref{sec:llm_agents}).
Finally, we show that a frontier general-purpose coding agent, given an existing consensus codebase and repeated experimental feedback, rarely improves on existing designs, and we identify the shortcomings that directly motivate the design of our \OurAgent (\autoref{sec:general_agent_enough}) agent.

\subsection{Consensus \& State Machine Replication}
\label{sec:smr}

\Acf{SMR} implements a fault-tolerant service by maintaining copies of its state at multiple replicas~\cite{schneider1990smr}.
The service is modeled as a deterministic state machine: a command and the current state uniquely determine the next state and the response.
Clients submit commands and receive their results through the \ac{SMR} protocol, which coordinates execution across replicas to present the behavior of a single state machine.

This behavior is formalized by \emph{linearizability}: the responses observed by all clients must be consistent with a single sequential execution of their commands by the state machine, preserving real-time order across clients~\cite{herlihy1990linearizability}.
Our verification framework uses linearizability, with the corrected definition of Sela et al.~\cite{sela2021linearizability}, as a uniform proof target for all candidate \ac{SMR} protocols.
Command ordering can exploit commutativity: replicas may choose different execution orders for commuting commands while agreeing on the order of conflicting commands~\cite{lamport2005generalized,pedone1999generic}.
Two commands commute if exchanging their order from any state leaves both the resulting state and each command's response unchanged; otherwise, they conflict.

To coordinate command execution, \ac{SMR} protocols use \emph{consensus}, in which processes propose values and agree on one of them~\cite{lamport2001paxos}.
For example, consensus can be used to commit commands in an order proposed by a leader, as in \multipaxos~\cite{lamport2001paxos}.
It can also establish agreement on command dependencies from which replicas determine an execution order, as in \epaxos~\cite{moraru2013epaxos,tollman2021epaxos,ryabinin2025democracy}.

An \ac{SMR} protocol must also provide \emph{liveness}: requests from non-faulty clients eventually complete under the protocol's stated assumptions.
Our verification currently targets safety only; we discuss the extension to liveness in \autoref{sec:discussion}.

The performance of \ac{SMR} protocols is commonly measured using four metrics.
\begin{enumerate*}[label=(\roman*)]
\item  Mean latency measures the average time between a client submitting a request and receiving its response, while
\item tail latency (P99) measures the 99th percentile of request completion times.
\item Throughput measures the number of requests completed per unit of time.
\item Recovery time measures how long it takes for request latency or throughput to stabilize after the failure of a leader or replica. 
\end{enumerate*}

\begin{figure}[t]
    \centering
    \includegraphics[width=\linewidth]{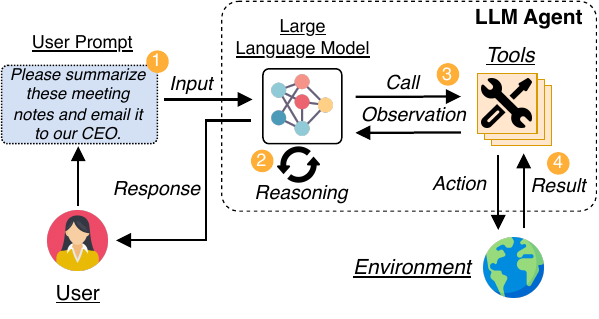}
    \caption{The workflow of an \ac{LLM} agent.}
    \label{fig:llm_agent}
\end{figure}

\subsection{\ac{LLM} Agents}
\label{sec:llm_agents}

\Acfp{LLM} are neural networks trained on large text corpora to predict the next token given the preceding context.
The underlying transformer architecture was originally developed for natural language tasks such as machine translation~\cite{vaswani2017attention}, and early \acp{LLM} were primarily used for tasks such as translation, summarization, and question answering~\cite{brown2020language}.
Instruction tuning and reinforcement learning from human feedback turned these models into general-purpose conversational assistants, which have been popularized by \textsc{ChatGPT}~\cite{openai2022chatgpt}.
As their capabilities improved, \acp{LLM} also became capable at reasoning and more complex tasks such as writing code, extending their use beyond language tasks~\cite{wei2022chain}.

On its own, an \ac{LLM} can only produce text in response to its input prompt.
Agentic AI removes this limitation by placing the model in a loop with \emph{tools} and an \emph{environment}, as shown in \autoref{fig:llm_agent}~\cite{acharya2025agentic}.
At each step, the model decides on an action, such as reading a file, running a command, or querying a search engine; the action is executed in the environment, and its result is returned to the model as an observation that informs the next decision.
This loop lets an \ac{LLM} agent pursue multi-step goals, gather information it was not given up front, and correct its own mistakes based on feedback.
Coding agents, such as \textsc{Codex}~\cite{openai2026codex}, \textsc{Claude Code}~\cite{anthropic2026claudecode}, and \textsc{Cursor Agent}~\cite{cursor2026agent}, apply this paradigm to software development and enable agents to solve complex coding tasks.
Given a task, a coding agent iteratively inspects relevant files, edits code, executes programs, and runs tests, and it revises its approach based on the results until the task is complete.
\ac{LLM} agents are also increasingly used for scientific discovery~\cite{gridach2025agentic}.
Our objective is to use agentic AI and \ac{LLM} agents to synthesize and discover new consensus protocols.

An \emph{agent harness} is the software infrastructure that organizes this interaction between the \ac{LLM}, its tools, and the environment~\cite{ning2026codeagentharness}.
At each step, it supplies the model with instructions and relevant context, dispatches the model's tool calls, and feeds the resulting observations back into subsequent model inputs.
It also manages execution state and stopping conditions across steps.
The model proposes actions and interprets feedback, while the harness makes those actions executable and maintains the interaction loop.
Beyond these general functions, a harness can provide domain-specific knowledge, implementation interfaces, and evaluation tools for a particular class of tasks.
The harness has a large effect on agent performance: changing the harness around a fixed \ac{LLM} can produce a $6\times$ performance gap on the same benchmark~\cite{swe_bench_mobile}.
In this work, we build on this observation by providing a harness specialized for consensus research.

\begin{figure}[t]
    \centering 
    \inputplot{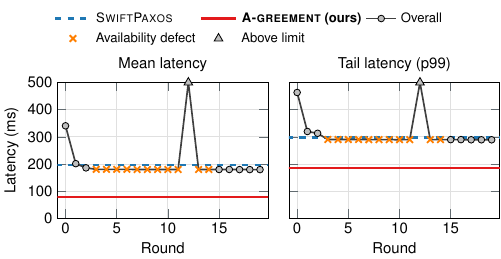}
    \caption{Mean and tail (P99) latency of discovered consensus protocols across 20 search iterations, using the original \codex harness and the \texttt{GPT-6.1 Sol} model (with high reasoning effort).}
    \label{fig:preliminary-evolution}
\end{figure}

\subsection{Limitations of General-Purpose Agents}
\label{sec:general_agent_enough}

We first investigate whether a general-purpose coding agent, given existing consensus implementations and repeated experimental feedback, can discover protocol changes that improve performance.
We choose the \swiftpaxos~\cite{ryabinin2024swiftpaxos} codebase\footnote{See \url{https://github.com/imdea-software/swiftpaxos}.} as our starting point because it provides a high-performance consensus implementation alongside implementations of several baseline protocols.
Using this codebase, we give the \codex harness powered by the \texttt{GPT-6.1 Sol} model with \texttt{High} reasoning effort the prompt in \autoref{app:search-prompt} to conduct an iterative 20-round search for consensus protocols with lower mean and tail (P99) latency.
\autoref{sec:evaluation} provides additional information about our experimental setup.
Unlike \OurAgent, this agent receives no curated protocol knowledge: it may search for the state of the art itself, but the codebase includes neither \bodega nor \ac{PQL}, on which \OurProtocol builds.
This experiment is therefore not a controlled comparison with \OurAgent, but an illustration of the difficulties a general-purpose agent faces.

\autoref{fig:preliminary-evolution} shows the mean (left) and tail (P99) latency (right) across the 20 search iterations, with the latency of \swiftpaxos indicated by a blue dotted line.
For reference, the red line reports the latency of \OurProtocol, discovered with the domain-specific support of our \OurAgent (\autoref{sec:xpaxos}).

We observe an initial latency reduction in the first two rounds, followed by a plateau.
The agent begins with a basic leader-based replication protocol in round 0, in which both reads and writes undergo \paxos-style consensus and execution before clients receive replies.
Our manual inspection of the agent trajectories reveal that the early gains mainly come from applying established optimizations: quorum-validated reads, early write completion, and improved leader placement shorten read and write completion paths.
Later rounds mainly explore secondary engineering optimizations, including batching, message encoding, and caching, which only result in marginal reductions in latency.
Mean latency spikes in round 12 which stems from processing and reply-traffic overhead introduced by periodic snapshot refreshes.
Moreover, we noticed that candidate implementations between rounds 3--14 violate availability despite a surviving majority, due to defects in client registration, recovery transfers, or peer connection handling.
At an offered load of \num{20000} commands per second, the best candidate achieves \SI{179.87}{\ms} mean latency and \SI{289.77}{\ms} average P99, averaged across deployments with 5, 9, and 13 replicas.
These results are only marginally better than the corresponding mean and P99 latencies of \SI{196.16}{\ms} and \SI{297.62}{\ms}, respectively, for \swiftpaxos.

After careful inspection of the traces, we identify three potential problems that limit the ability by the general-purpose agent to discover consensus algorithm with more substantial gains.
The first issue is that the agent relies mainly on conventional optimizations and secondary engineering refinements and appears not to fully understand the principles behind techniques in existing consensus work.
This motivates explicit consensus-design knowledge that explains these techniques' rationale and tradeoffs.
Second, we observed that the agent re-implements the entire protocol each search iteration.
Implementing consensus, however, requires substantial infrastructure for communication, client handling, and recovery, making implementation from scratch complex and error-prone.
The observed defects, including loss of availability despite a surviving majority, motivate a reusable and modular implementation foundation that reduces this burden and lets the agent focus on protocol-specific logic.
Third, experimental feedback measures performance and can expose some correctness violations, but cannot by itself establish protocol safety beyond the tested executions.
This motivates explicit invariants and formal proofs that guarantee safety for all executions permitted by the fault model.

We believe these limitations stem in part from a gap between the general-purpose support provided by the agent's harness and the domain-specific requirements of consensus discovery.
In the following section, we introduce an additional harness layer that builds on the existing software engineering capabilities of state-of-the-art coding agents and augments them with three domain-specific modules for consensus research.

We also conduct a trial with SkyDiscover-Synthesize (\textsc{SkySynth}~\cite{liu2026skysynth}) and use its test-driven agentic workflow to search for \ac{SMR} protocols with lower mean and tail latency.
Despite a requested budget of 20 search rounds, the run remained focused on implementing and repairing its initial candidate.
We terminated the experiment after approximately 11.5 hours, and 486 million input and output tokens across the main agent, workers, and automatic reviews.
Although the workflow produced correctness tests and diagnostic traces, unresolved client-completion failures prevented a complete latency evaluation, and no performance improvement was demonstrated.
This trial further illustrates the difficulty of reaching reliable end-to-end evaluation in the synthesis of consensus protocols, even with a structured agent workflow.

\section{Design of the \OurAgent Harness}
\label{sec:paxosagent}

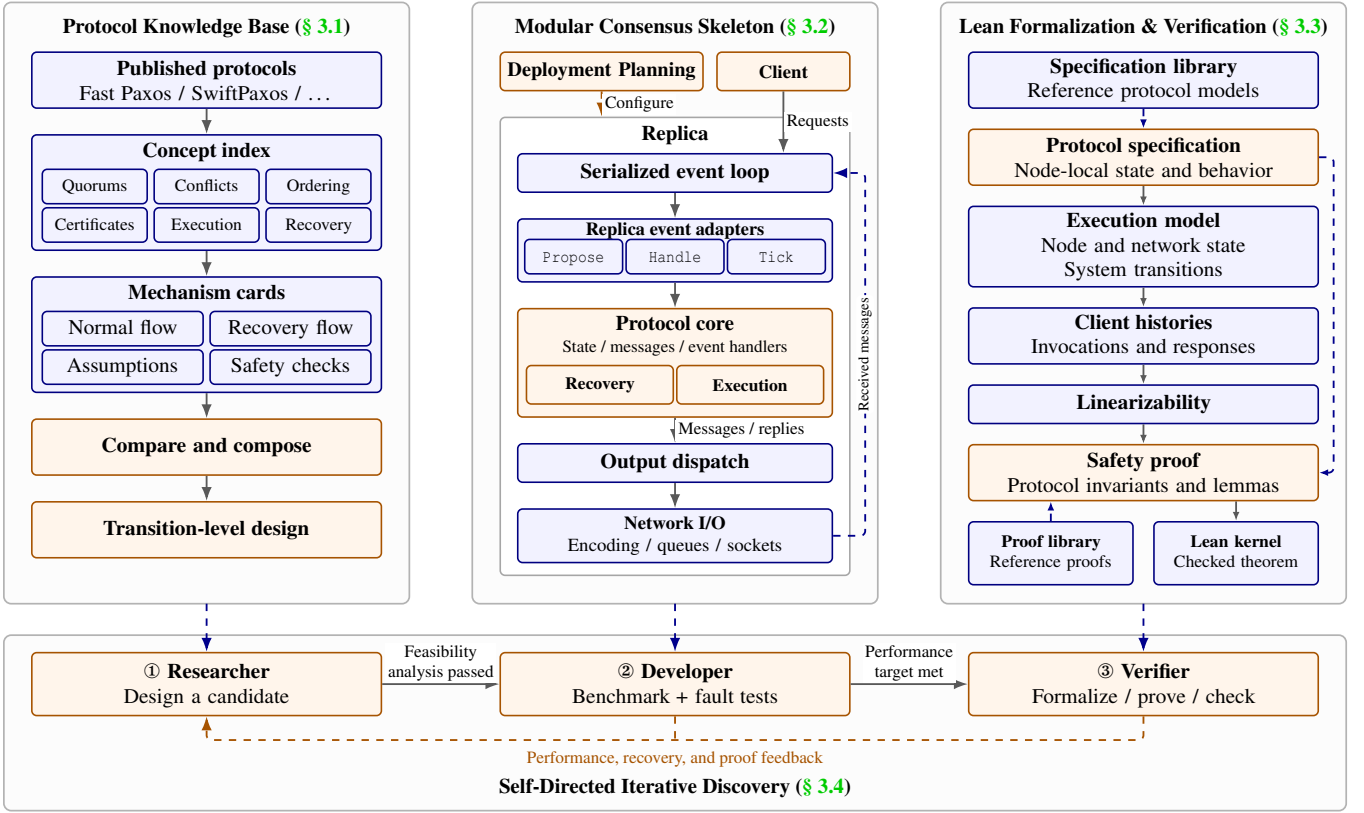
\begin{figure*}[t]
\centering
\resizebox{\textwidth}{!}{%
\begin{tikzpicture}[
  font=\fontsize{8.5}{10}\selectfont,
  hprovided/.style={draw=blue!50!black,fill=blue!5},
  hfilled/.style={draw=orange!65!black,fill=orange!8},
  hbox/.style={rounded corners=2pt,line width=.65pt,
    font=\fontsize{8.5}{10}\selectfont,
    align=center,inner xsep=3pt,inner ysep=3pt,outer sep=.325pt},
  hwide/.style={hbox,minimum width=5cm,
    text width=\dimexpr5cm-6pt\relax},
  hframe/.style={hbox,inner sep=0pt,minimum width=5cm},
  htitle/.style={font=\fontsize{8.5}{10}\selectfont\bfseries,
    text height=7pt,text depth=2pt},
  hrow/.style={text height=6pt,text depth=2pt},
  hcell/.style={hbox,hrow,font=\fontsize{7}{8}\selectfont,
    inner xsep=1pt,minimum width=1.5cm,
    text width=\dimexpr1.5cm-2pt\relax,minimum height=.48cm},
  hthird/.style={hcell,minimum width=1.6cm,
    text width=\dimexpr1.6cm-2pt\relax},
  hhalf/.style={hbox,minimum width=2.35cm,
    text width=\dimexpr2.35cm-6pt\relax,minimum height=1cm},
  hcard/.style={hbox,hrow,minimum width=2.3cm,
    text width=\dimexpr2.3cm-6pt\relax,minimum height=.48cm},
  htwo/.style={text height=7pt,text depth=12pt},
  hthree/.style={text height=7pt,text depth=20pt},
  hflow/.style={-{Latex[length=2mm]},draw=black!65,line width=.75pt},
  hsupport/.style={hflow,dashed,draw=blue!55!black},
  hfeedback/.style={hflow,dashed,draw=orange!65!black},
  hlabel/.style={font=\fontsize{7}{8}\selectfont,
    fill=white,inner sep=2pt,align=center}
]
\node[anchor=west,font=\fontsize{8}{9}\selectfont] at (0,.45)
  {\textcolor{blue!55!black}{$\blacksquare$}\ Expert-designed infrastructure};
\node[anchor=east,font=\fontsize{8}{9}\selectfont] at (19.2,.45)
  {\textcolor{orange!65!black}{$\blacksquare$}\ Agent-designed components};
\foreach \x in {0,6.7,13.4}{
  \draw[rounded corners=3pt,draw=black!30,fill=black!1,line width=.7pt]
    (\x,0) rectangle (\x+5.8,-8.6);
}

\begin{scope}[xshift=2.9cm]
\node[htitle] at (0,-.35) {Protocol Knowledge Base (\autoref{sec:harness-knowledge})};
\node[hwide,hprovided,minimum height=.8cm,anchor=north] (papers) at (0,-.7)
  {\textbf{Published protocols}\\Fast Paxos / SwiftPaxos / \ldots};
\path (papers.south) ++(0,-.36)
  node[hframe,hprovided,minimum height=1.65cm,anchor=north] (concepts) {};
\path (concepts.north) ++(0,-.24) node[htitle] {Concept index};
\path (concepts.north) ++(-1.6,-.75) node[hcell,hprovided] (quorums) {Quorums};
\path (concepts.north) ++(0,-.75) node[hcell,hprovided] (conflicts) {Conflicts};
\path (concepts.north) ++(1.6,-.75) node[hcell,hprovided] (ordering) {Ordering};
\path (concepts.north) ++(-1.6,-1.3) node[hcell,hprovided] (certificates) {Certificates};
\path (concepts.north) ++(0,-1.3) node[hcell,hprovided] (executionconcept) {Execution};
\path (concepts.north) ++(1.6,-1.3) node[hcell,hprovided] (recoveryconcept) {Recovery};
\path (concepts.south) ++(0,-.36)
  node[hframe,hprovided,minimum height=1.65cm,anchor=north] (mechanisms) {};
\path (mechanisms.north) ++(0,-.24) node[htitle] {Mechanism cards};
\path (mechanisms.north) ++(-1.2,-.75) node[hcard,hprovided] (normalcard) {Normal flow};
\path (mechanisms.north) ++(1.2,-.75) node[hcard,hprovided] (recoverycard) {Recovery flow};
\path (mechanisms.north) ++(-1.2,-1.3) node[hcard,hprovided] (assumptions) {Assumptions};
\path (mechanisms.north) ++(1.2,-1.3) node[hcard,hprovided] (safetychecks) {Safety checks};
\path (mechanisms.south) ++(0,-.36)
  node[hwide,hfilled,minimum height=.8cm,anchor=north] (compose)
  {\textbf{Compare and compose}};
\path (compose.south) ++(0,-.36)
  node[hwide,hfilled,minimum height=.8cm,anchor=north] (design)
  {\textbf{Transition-level design}};
\draw[hflow] (papers.south) -- (concepts.north);
\draw[hflow] (concepts.south) -- (mechanisms.north);
\draw[hflow] (mechanisms.south) -- (compose.north);
\draw[hflow] (compose.south) -- (design.north);
\end{scope}

\begin{scope}[xshift=9.6cm]
\node[htitle] at (0,-.35) {Modular Consensus Skeleton (\autoref{sec:harness-skeleton})};
\node[hbox,hfilled,minimum width=2.9cm,
  text width=\dimexpr2.9cm-6pt\relax,minimum height=.55cm,anchor=north,
  font=\fontsize{7.5}{9}\selectfont] (deployment) at (-1.05,-.7)
  {\resizebox{\linewidth}{!}{\textbf{Deployment Planning}}};
\node[hbox,hfilled,minimum width=1.9cm,
  text width=\dimexpr1.9cm-6pt\relax,minimum height=.55cm,anchor=north,
  font=\fontsize{7.5}{9}\selectfont] (client) at (1.55,-.7)
  {\textbf{Client}};

\path (deployment.south) ++(1.05,-.36)
  node[hframe,draw=black!35,fill=white,minimum height=6.55cm,anchor=north]
  (replicaruntime) {};
\path (replicaruntime.north) ++(0,-.24) node[htitle] {Replica};
\path (replicaruntime.north) ++(0,-.52)
  node[hbox,hprovided,minimum width=4.5cm,
  text width=\dimexpr4.5cm-6pt\relax,minimum height=.55cm,anchor=north]
  (loop) {\textbf{Serialized event loop}};
\path (loop.south) ++(0,-.36)
  node[hframe,hprovided,minimum width=4.5cm,minimum height=.9cm,anchor=north]
  (interface) {};
\path (interface.north) ++(0,-.15)
  node[htitle,font=\fontsize{7.5}{9}\selectfont\bfseries] {Replica event adapters};
\path (interface.north) ++(-1.45,-.55)
  node[hcell,hprovided,minimum width=1.4cm,
  text width=\dimexpr1.4cm-2pt\relax] (requests)
  {\texttt{Propose}};
\path (interface.north) ++(0,-.55)
  node[hcell,hprovided,minimum width=1.4cm,
  text width=\dimexpr1.4cm-2pt\relax] (messages)
  {\texttt{Handle}};
\path (interface.north) ++(1.45,-.55)
  node[hcell,hprovided,minimum width=1.4cm,
  text width=\dimexpr1.4cm-2pt\relax] (timers)
  {\texttt{Tick}};

\path (interface.south) ++(0,-.36)
  node[hframe,hfilled,minimum width=4.5cm,minimum height=1.55cm,anchor=north]
  (core) {};
\path (core.north) ++(0,-.23) node[htitle] {Protocol core};
\path (core.north) ++(0,-.58) node[font=\fontsize{7}{8}\selectfont]
  {State / messages / event handlers};
\path (core.north) ++(-1.075,-1.1)
  node[hbox,hfilled,minimum width=2.1cm,
  text width=\dimexpr2.1cm-6pt\relax,minimum height=.55cm,
  font=\fontsize{7}{8}\selectfont] (recovery)
  {\textbf{Recovery}};
\path (core.north) ++(1.075,-1.1)
  node[hbox,hfilled,minimum width=2.1cm,
  text width=\dimexpr2.1cm-6pt\relax,minimum height=.55cm,
  font=\fontsize{7}{8}\selectfont] (execution)
  {\textbf{Execution}};

\path (core.south) ++(0,-.36)
  node[hbox,hprovided,minimum width=4.5cm,
  text width=\dimexpr4.5cm-6pt\relax,minimum height=.55cm,anchor=north]
  (dispatch) {\textbf{Output dispatch}};
\path (dispatch.south) ++(0,-.36)
  node[hbox,hprovided,minimum width=4.5cm,
  text width=\dimexpr4.5cm-6pt\relax,minimum height=.7cm,anchor=north,
  font=\fontsize{7.5}{9}\selectfont] (networkio)
  {\textbf{Network I/O}\\Encoding / queues / sockets};

\draw[hflow,dashed,draw=orange!65!black] (deployment.south)
  -- node[hlabel,right,inner sep=1pt] {Configure}
  (deployment.south |- replicaruntime.north);
\draw[hflow] (client.south)
  -- node[hlabel,right,inner sep=1pt] {Requests}
  (client.south |- loop.north);
\draw[hflow] (loop.south) -- (interface.north);
\draw[hflow] (interface.south) -- (core.north);
\draw[hflow] (core.south) --
  node[hlabel,right,inner sep=1pt] {Messages / replies} (dispatch.north);
\draw[hflow] (dispatch.south) -- (networkio.north);
\draw[hsupport,rounded corners=2pt] (networkio.east) -- (2.72,0 |- networkio.east)
  -- node[hlabel,rotate=90,font=\fontsize{6}{7}\selectfont,inner sep=1pt]
     {Received messages}
  (2.72,0 |- loop.east) -- (loop.east);
\end{scope}

\begin{scope}[xshift=16.3cm]
\tikzset{
  hwide/.append style={inner ysep=2pt},
  hflow/.append style={-{Latex[length=1.3mm,width=1mm]}}
}
\node[htitle] at (0,-.35) {Lean Formalization \& Verification (\autoref{sec:harness-verification})};
\node[hwide,hprovided,minimum height=.8cm,anchor=north] (specifications) at (0,-.7)
  {\textbf{Specification library}\\Reference protocol models};
\path (specifications.south) ++(0,-.28)
  node[hwide,hfilled,minimum height=.8cm,anchor=north] (encoding)
  {\textbf{Protocol specification}\\Node-local state and behavior};
\path (encoding.south) ++(0,-.28)
  node[hwide,hprovided,minimum height=1.15cm,anchor=north] (traces)
  {\textbf{Execution model}\\
   Node and network state\\System transitions};
\path (traces.south) ++(0,-.28)
  node[hwide,hprovided,minimum height=.8cm,anchor=north] (observation)
  {\textbf{Client histories}\\Invocations and responses};
\path (observation.south) ++(0,-.28)
  node[hwide,hprovided,minimum height=.55cm,anchor=north] (linearizability)
  {\textbf{Linearizability}};
\path (linearizability.south) ++(0,-.28)
  node[hwide,hfilled,minimum height=.8cm,anchor=north] (proof)
  {\textbf{Safety proof}\\Protocol invariants and lemmas};
\path (proof.south) ++(-1.325,-.28)
  node[hhalf,htwo,hprovided,minimum height=.9cm,anchor=north,
  font=\fontsize{7}{8}\selectfont] (reference)
  {\textbf{Proof library}\\Reference proofs};
\path (proof.south) ++(1.325,-.28)
  node[hhalf,htwo,hprovided,minimum height=.9cm,anchor=north,
  font=\fontsize{7}{8}\selectfont] (kernel)
  {\textbf{Lean kernel}\\Checked theorem};
\draw[hsupport] (specifications.south) -- (encoding.north);
\draw[hflow] (encoding.south) -- (traces.north);
\draw[hflow] (traces.south) -- (observation.north);
\draw[hflow] (observation.south) -- (linearizability.north);
\draw[hflow] (linearizability.south) -- (proof.north);
\draw[hsupport,rounded corners=2pt] (encoding.east) -- (2.72,0 |- encoding.east)
  -- (2.72,0 |- proof.east) -- (proof.east);
\draw[hsupport] (reference.north) -- (reference.north |- proof.south);
\draw[hflow] (proof.south -| kernel.north) -- (kernel.north);
\end{scope}

\draw[rounded corners=3pt,draw=black!30,fill=black!1,line width=.7pt]
  (0,-9.05) rectangle (19.2,-11.56);
\node[hwide,htwo,hfilled,minimum height=.9cm] (researcher) at (2.9,-9.75)
  {\textbf{\ding{172} Researcher}\\Design a candidate};
\node[hwide,htwo,hfilled,minimum height=.9cm] (developer) at (9.6,-9.75)
  {\textbf{\ding{173} Developer}\\Benchmark + fault tests};
\node[hwide,htwo,hfilled,minimum height=.9cm] (verifier) at (16.3,-9.75)
  {\textbf{\ding{174} Verifier}\\Formalize / prove / check};
\draw[hsupport] (2.9,-8.6) -- (researcher.north);
\draw[hsupport] (9.6,-8.6) -- (developer.north);
\draw[hsupport] (16.3,-8.6) -- (verifier.north);
\draw[hflow] (researcher.east) --
  node[hlabel,above,inner sep=1pt] {Feasibility\\analysis passed} (developer.west);
\draw[hflow] (developer.east) --
  node[hlabel,above,inner sep=1pt] {Performance\\target met} (verifier.west);
\draw[hfeedback,rounded corners=2pt] (verifier.south) -- (16.3,-10.55)
  -- (2.9,-10.55) -- (researcher.south);
\draw[hfeedback,-] (developer.south) -- (9.6,-10.55);

\node[hlabel,text=orange!65!black,fill=black!1] at (9.6,-10.82)
  {Performance, recovery, and proof feedback};
\node[htitle] at (9.6,-11.22)
  {Self-Directed Iterative Discovery (\autoref{sec:harness-discovery})};
\end{tikzpicture}%
}
\caption{\OurAgent harness architecture. Blue components provide reusable knowledge, interfaces, execution semantics, and proof support; orange components contain protocol-specific logic supplied by the agent. In the verification column, protocol specifications and proofs build on a shared execution model and linearizability definition.
}
\label{fig:paxosagent-harness}
\end{figure*}

We design \OurAgent, an agentic harness to support the iterative discovery process of consensus algorithms (\autoref{fig:paxosagent-harness}).
\OurAgent equips general-purpose coding agents with the knowledge, implementation structure, and verification support needed for consensus research.
Given a failure model, performance objective, and search budget, the agent takes researcher, developer, and verifier roles in an iterative process.
We next explain these components and roles in detail.

\subsection{Protocol Knowledge Base}
\label{sec:harness-knowledge}

General-purpose agents can search the web and retrieve relevant papers.
However, access to the literature alone does not provide a systematic way to design new consensus protocols. 
Relevant design ideas are scattered across many papers that differ in their assumptions, terminology, and protocol structures. 
Understanding how a mechanism used by one consensus protocol can be applied to another requires reasoning about its underlying principles, safety conditions, and interactions with other mechanisms.
To address this, we construct a \emph{protocol knowledge base} (the left column in \autoref{fig:paxosagent-harness}) that organizes existing consensus research around reusable designs in Markdown format, instead of supplying individual papers.
This gives the \ac{LLM} agent a structured way to understand and reason about the concepts relevant to consensus research.
The knowledge base is constructed through human-guided, AI-assisted organization and expert review.

\paragraph{Knowledge base construction.}
As a first step, we identified 22 representative papers on foundational and state-of-the-art consensus protocols~\cite{lamport2005generalized,lamport2006fastpaxos,mao2008mencius,moraru2013epaxos,howard2016flexiblepaxos,park2019curp,enes2020atlas,ngo2020copilot,katsarakis2020hermes,uluyol2020pando,ailijiang2020wpaxos,tollman2021epaxos,charapko2021pigpaxos,pan2021rabia,choi2023hydra,ng2023omnipaxos,ryabinin2024swiftpaxos,ryabinin2025democracy,hodsdon2026avicenna,hu2026bodega,tang2026jetpack,burgelin2026kcensus}.
We then define a shared extraction structure covering system assumptions, roles, messages, local state, normal operation, completion conditions, recovery rules, and safety arguments.
Afterwards, we let AI use this structure to produce notes from the published papers.
We manually verify the produced notes against the source papers, correct errors, and add missing explanations. %

\paragraph{Concept-oriented organization.}
The produced notes streamline cross-protocol comparison for the agent.
The \emph{concept index} connects 24 consensus concepts, such as quorums, conflicts, and dependencies, to concrete protocol examples.
The associated \emph{mechanism cards} summarize each example's message flow, assumptions, recovery behavior, and safety conditions.
The shared structure also enables comparisons across 11 design dimensions, including quorum systems, completion rules, recovery mechanisms, and latency.
By organizing mechanisms under common concepts, the knowledge base allows the agent to more easily compare how different protocols address similar design challenges without being constrained by the organization and presentation of individual papers.

\paragraph{Knowledge-guided discovery.}
In the researcher role, \OurAgent uses the concept index during protocol discovery to design a candidate protocol and to identify mechanisms that could cause a performance bottleneck.
It consults the mechanism cards and cross-protocol comparisons to assess their expected benefits, costs, and compatibility conditions.
For example, changing a completion quorum requires examining what evidence remains available during recovery.
These checks guide the \emph{compare and compose} activity, where the agent selects and combines candidate mechanisms.
The resulting \emph{transition-level design} records the candidate's states, transitions, quorum rules, and recovery behavior.

\subsection{Modular Consensus Skeleton}
\label{sec:harness-skeleton}

Many consensus protocols have open-source implementations.
However, their engineering choices differ in code organization, communication primitives, and runtime infrastructure.
Exploring mechanisms across these codebases therefore requires the agent to understand and adapt the surrounding implementation as well as the protocol itself.
We address this with \textsc{ConsensusArena}, a framework written in the Go language in which seven baseline protocols share a common runtime, client framework, and benchmark harness.
We let an AI agent integrate baseline protocols, including classical \paxos~\cite{lamport2001paxos}, \fastpaxos~\cite{lamport2006fastpaxos}, \epaxos~\cite{moraru2013epaxos,tollman2021epaxos,ryabinin2025democracy}, \swiftpaxos~\cite{ryabinin2024swiftpaxos}, \bodega~\cite{hu2026bodega}, \curp~\cite{park2019curp}, and \kcensus~\cite{burgelin2026kcensus}.

On top of this framework, we provide a \emph{modular consensus skeleton} for implementing new candidates (\autoref{fig:paxosagent-harness}, middle).
The \OurAgent (in the developer role) supplies the orange components (deployment planning, the client policy, and the protocol core) while the blue components are shared by all protocols and created by us.
For each candidate, a generator instantiates an independent package with one file per component.
Every agent-supplied component returns an explicit error until the agent replaces it, so a partially implemented candidate cannot produce measurements by silently inheriting a default leader, quorum, commit rule, or routing policy.
Baselines are read-only to the agent, and a finished candidate is registered through a single entry in a log, after which it runs on the same runtime and benchmark as the baselines.
Next, we further describe the components in \autoref{fig:paxosagent-harness} (middle).

\paragraph{Deployment planning.}
Before startup, the planner maps the network topology and workload to an initial leader, quorums, read responders, and one ingress replica per client.
All such policy lives in the candidate's planner rather than in launch scripts, which lets \OurAgent's workload-aware planner (\autoref{sec:xpaxos-planning}) run without changes to the framework.
 
\paragraph{Client.}
Protocols differ in how clients interact with replicas: a \multipaxos client sends each request to the leader, whereas a Fast Paxos client sends it to all replicas.
The client component specifies this behavior: which replicas receive a request, which replies complete it, and whether pending requests are retried after a failure.

\paragraph{Serialized event loop and adapters.}
Each replica processes all events on a single loop, so the protocol core needs
no locks.
The loop drives the core through three adapters: \texttt{Propose} for client requests, \texttt{Handle} for peer messages and local events, and \texttt{Tick} for timers.
 
\paragraph{Protocol core.}
\OurAgent defines the core's state, messages, and handlers, including the quorum checks that determine acceptance and commitment.
Recovery and execution mechanisms are also part of the core: \emph{recovery} collects evidence and selects safe values or orders, and \emph{execution} orders, deduplicates, and applies commands to a shared key-value store.
The skeleton defines accepted, committed, executed, and client-completed as distinct states for a command.
 
\paragraph{Output dispatch and network I/O.}
Each transition returns outgoing messages and client replies, where a reply signals that the protocol has authorized completion.
The dispatcher freezes each message and hands it to a per-connection sender with a bounded, non-blocking queue; received messages return through the event loop.
 
\paragraph{Benchmark and fault tests.}
In the developer role, candidate protocols are evaluated with the \textsc{ConsensusArena} benchmark, which emulates geo-replication on a \textsc{Slurm} cluster using \textsc{Toxiproxy} to emulate network delays and inter-region delays from 35 AWS cloud regions.
The development suite covers 5, 9, and 13 replicas in three topologies under the \textsc{YCSB} workloads.
Candidates that meet the performance target are then crash-tested: replicas are killed until $f=(n-1)/2$ have failed, and throughput must recover to 90\% of its pre-crash level and remain there.
Note that our final evaluation in \autoref{sec:evaluation} uses more advanced topologies and workloads disjoint from this suite to show the generalization.

\subsection{Lean Formalization \& Verification}
\label{sec:harness-verification}

Testing can expose safety violations, but even extensive testing cannot establish safety under all possible event orderings.
Informal correctness arguments must account for interactions among concurrent actions, delayed messages, and failures, making subtle omissions difficult to detect.
For example, \OurAgent found that the dependency-pruning rules of EPaxos Revisited~\cite{tollman2021epaxos} allow replicas to execute conflicting commands in opposite orders, also see \autoref{sec:bugs}.
Proof assistants such as \textsc{Lean}~\cite{demoura2021lean4} and \textsc{Rocq}~\cite{rocqprover} close this gap by mechanically checking that
specified safety properties hold for every execution admitted by a formal model.
However, developing such models and proofs requires substantial human effort, often spanning several person-months~\cite{10.1145/3808293}.

We therefore aim to enable \OurAgent to automate protocol formalization and proof construction with \textsc{Lean} within the discovery loop, using safety properties predefined by experts (\autoref{fig:paxosagent-harness}, right).
The process proceeds in two stages.
First, the agent encodes the candidate protocol as a formal specification in \textsc{Lean}.
Second, it proves that this specification satisfies the predefined linearizability property \autoref{sec:background}, using \textsc{Lean}'s feedback on remaining proof obligations and errors to refine its proof iteratively.
We integrate the reference \textsc{Lean} specifications and proofs or disproofs of linearizability for \textsc{Multi-Paxos}~\cite{lamport2001paxos}, \fastpaxos~\cite{lamport2006fastpaxos}, \epaxosstar~\cite{ryabinin2025democracy} (the corrected version of \epaxos\cite{moraru2013epaxos,tollman2021epaxos}), \swiftpaxos~\cite{ryabinin2024swiftpaxos}, and \bodega~\cite{hu2026bodega}.

The framework lets the agent specify a protocol much as it would in pseudocode, by defining a node's local state and its handlers for messages and timeouts.
A shared transition model uses these handlers to describe how node and network state change as messages are delivered and timeouts occur. Client invocations and responses form the observable history of each execution and provide the basis for the framework's linearizability specification. We then establish protocol safety by proving that every execution starting from the initial state and following these transition rules produces a linearizable history.
Although researchers must still manually check that the specification faithfully captures the intended protocol and its assumptions, the shared foundation reduces this review burden: once the common definitions have been checked, researchers can focus on the protocol-specific state, initialization, and transition rules.

For proving, the agent develops protocol-specific invariants and uses reference proofs and reusable lemmas to establish the predefined safety properties.
\textsc{Lean} provides feedback on unsuccessful proof attempts, and its kernel checks the completed proof against the stated theorem.
An unresolved proof obligation may also prompt \OurAgent to investigate whether the required property can be violated, although proof failure alone does not always result in a bug.
If this investigation yields a counterexample, the agent checks whether it arises from an inaccurate encoding or a flaw in the intended protocol, with human review where needed.
A modeling error requires correcting the specification; a protocol flaw calls for revising the protocol and updating its specification before attempting the proof again.
When \OurAgent establishes a protocol vulnerability, it records the counterexample, the violated property, and any resulting repair, preserving the finding as feedback for subsequent discovery iterations.

We found that reference specifications and proofs are valuable beyond the protocols they verify: they provide modeling patterns and proof strategies that the agent can reuse. Expanding this library helped the agent build more faithful models and complete proofs faster.
More implementation detail, however, is not always helpful. The agent sometimes carries incidental engineering choices into the formal model, adding proof obligations without improving its fidelity to the protocol. We therefore ground formalization in the paper's pseudocode and use open-source implementations only to resolve details that the paper leaves unspecified.

\subsection{\OurAgent for Consensus Discovery}
\label{sec:harness-discovery}

We implement \OurAgent on top of the \codex coding harness.
The search begins with user-defined objectives and a search budget specified as the number of discovery rounds.
The user may target a single metric, such as latency or maximum achievable throughput, or specify tradeoffs among multiple metrics, for example, requiring improvements in both mean and tail latency.

In each discovery round, the agent first acts as a researcher, selecting a parent protocol and forming a hypothesis about the bottleneck limiting the target metrics.
It consults the consensus-specific knowledge base (\autoref{sec:harness-knowledge}) to compare candidate mechanisms, assess their safety risks, and estimate their expected benefits and additional overhead.
The resulting design document specifies messages, local state, quorum and completion rules, execution, fallback, and recovery.

To test each candidate design, the agent then acts as a developer, implementing the candidate design within the skeleton (\autoref{sec:harness-skeleton}) and comparing it with other candidates under the same benchmark.
Instrumentation helps relate the results to the original hypothesis, for example by separating latency spent collecting votes, waiting for execution, and resolving conflicts.
Candidates that meet the user-defined performance criteria proceed to fault-injection testing.
Fault injection separately tests recovery after replica failures.
Candidates that meet the performance criteria and pass the experimental checks proceed to formal verification.
The agent then acts as a verifier, constructing a \textsc{Lean} model and attempting to prove the safety target (\autoref{sec:harness-verification}).

The staged process in \OurAgent rejects unpromising candidates before committing effort to their implementation, experimental evaluation, and formal verification.
Feedback from each stage guides subsequent revision or rejection, concentrating verification effort on promising candidates.
To conserve resources, our current implementation defers formal verification until the end of the search and proves only the final version.

After every discovery round, the agent records the candidate's parent, mechanism changes, expected and observed bottlenecks, measurements, recovery results, and proof status.
It summarizes both successful and unsuccessful attempts into reusable lessons about which mechanisms improved performance, which failed to deliver the expected gains, and which correctness questions remain unresolved.
These accumulated lessons guide the agent's choice of subsequent search strategy, including whether to refine a candidate, revise a bottleneck hypothesis, revisit an earlier parent, or explore a different mechanism family.
For example, when successive changes leave the same bottleneck, the agent reassesses the remaining improvement potential and decides whether to change direction.
By carrying these lessons forward, the agent builds on successful mechanisms and avoids repeating unsuccessful approaches without new evidence.

\section{\OurProtocol Protocol Design}
\label{sec:xpaxos}

\begin{figure*}[t]
    \centering
    \inputplot{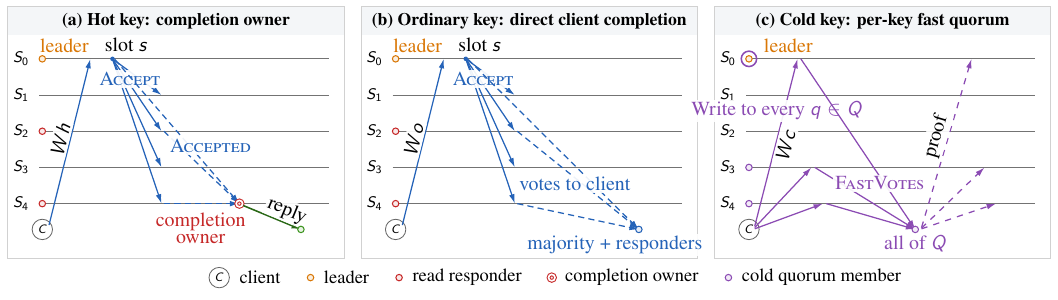}
    \caption{Message propagation in \OurProtocol on the hot (left), ordinary (middle) and cold write (right) paths.}
    \label{fig:message_propagation}
\end{figure*}

Our agent synthesized \OurProtocol, a consensus algorithm that jointly reduces read and write latency.
The design starts from a simple observation: the best way to complete a write depends on how that choice affects subsequent reads.
A frequently accessed key benefits from a nearby replica that can serve reads and control write completion.
A less frequently accessed key may benefit more from sending acceptance evidence directly to the client.
For sufficiently rarely accessed keys, writes can also avoid leader ordering altogether.
\OurProtocol therefore \emph{combines three write paths with local reads}, which allow a replica holding a read lease to answer reads without contacting other replicas.

Because \OurAgent draws on prior consensus designs through its knowledge base (\autoref{sec:harness-knowledge}), many of \OurProtocol's mechanisms are borrowed from existing algorithms; \autoref{sec:xpaxos-delta} points out where each one comes from.
At its core, \OurProtocol is closest to \acf{PQL}~\cite{paxosquorumleases} and \bodega~\cite{hu2026bodega}: a \multipaxos log in which a stable leader orders writes, combined with lease-based local reads at selected replicas.
A write completes only once a majority and every read responder for its key have accepted it, so a responder holding a lease can answer reads from its local state.
\OurProtocol makes two changes to this core.
First, writes need not complete at the leader: depending on the key, they can complete at a responder near the readers or directly at the client.
Second, writes to rarely accessed keys bypass the log entirely and follow a fast path in the style of \swiftpaxos.

Before deployment, \OurProtocol's planner takes as input a workload profile: a sample of key accesses, obtained from historical traces or a profiling phase before the protocol starts serving requests, together with measured network delays between clients and replicas and the expected write fraction.
From this profile, the planner selects the write path for each key and places the replica roles.
The resulting plan is fixed during execution.
A profile that does not match the actual workload can reduce performance but never affects safety: keys absent from the profile are ordered by the shared log, and every path enforces the same completion and read-visibility conditions.
This section explains the high-level design of \OurProtocol; \autoref{app:xpaxos-specification} contains the detailed description, and \autoref{tab:protocol-latency} compares its latency and quorum requirements with prior protocols.

\subsection{\OurProtocol in Simple Terms}
\label{sec:xpaxos-delta}

\OurProtocol divides the key space into three: frequently accessed (hot) keys, regularly accessed (ordinary) keys and infrequent (cold) keys.
Before execution, the planner chooses this partition from the workload profile and assigns the associated replica roles (leader, responder, quorum, etc.) using measured network delays.
All three classes share one read mechanism; they differ in how writes complete and how the replicas serving reads learn about them.
\autoref{fig:message_propagation} shows the write path of each class.
Before describing these paths, we explain how replicas serve local reads, since each write path must ensure that these reads do not miss completed writes.
For each mechanism, we give the main idea, how it works, and where it comes from.

\paragraph{Common ground: a leader-ordered log and local reads.}
Hot and ordinary keys retain a \multipaxos-style shared log ordered by a stable leader~\cite{lamport2001paxos}.
For all three classes, the planner chooses \emph{read responders}: replicas that answer reads without contacting other replicas.
These local reads follow \ac{PQL}~\cite{paxosquorumleases}: replicas grant each other read leases covering all keys at once, a responder may serve local reads while it holds leases from a majority, and read responders are chosen from access statistics.
As in \bodega's roster leases, \OurProtocol binds the lease assignment, that is, the leader and the read responders, to the ballot~\cite{hu2026bodega}.
A lease establishes who may answer a read, not which value is safe to return: a responder must also know that its local value includes every write that could have completed before the read began.
Following \ac{PQL}, a write therefore completes only once a majority and every read responder for its key have accepted it, and a responder that has accepted a write holds reads of that key until it can safely answer.
A local read thus costs a single round trip to one replica, but every responder must learn about each write to the keys it serves.
The three write paths differ in how they achieve this.
To our knowledge, prior protocols that vary the write path by key access pattern~\cite{fonseca2025woc} do not support local reads; \OurProtocol instead chooses each class's write path for how it keeps read responders up to date.

\paragraph{Hot keys: control completion near the reader.}
Hot keys receive the most reads, so their write path aims to keep responders serving reads while writes are pending.
As in \bodega, the leader orders writes in the shared log and broadcasts \textsc{Accept} to the acceptors.
In \bodega, acceptors send \textsc{Accepted} only to the leader, so no other replica can learn that a write committed before the leader does.
The leader therefore implicitly gates every write and can keep serving the committed value while writes are pending~\cite{chandra2007paxos}, whereas the other responders must hold reads of a key once they have accepted a write to it.
\OurProtocol generalizes this gate to a \emph{completion owner}: a responder the planner selects for a range of hot keys, typically close to their readers. A client sends each write for these keys directly to the leader and names the completion owner as the replica that will reply.
Acceptors send \textsc{Accepted} to the leader, the other responders and the completion owner.
Since the completion owner is itself a responder, no write to its keys can complete without its \textsc{Accepted}.
It therefore withholds that message and keeps serving the old value until it has \textsc{Accepted} from a majority and every other responder.
Only then does it switch to the new value and send its own \textsc{Accepted}; it replies to the client once it has executed the write.
Because votes reach several learners rather than only the leader, the gate is no longer implied by the message pattern; withholding the vote enforces it explicitly.
The planner assigns completion owners only where it estimates a latency gain. Other hot keys keep \bodega's arrangement: the leader is their implicit completion owner, and gates the writes it receives itself; writes that clients send through another replica are not gated, and that replica replies once it has executed them.

\paragraph{Ordinary keys: deliver the evidence directly.}
Ordinary keys receive fewer reads, so holding back a write to keep one responder serving the old value gains little; their write path instead minimizes write latency.
The leader orders writes in the same log, but acceptors send their \textsc{Accepted} votes directly to the client, which completes the write once it holds matching votes from a majority that includes every read responder.
This moves \ac{PQL}'s learner into the client: in \ac{PQL}, acceptors also send their votes to the replica near the client, which commits and replies~\cite{paxosquorumleases,castro2002pbft}.
There is no completion-owner shortcut on this path: once a responder accepts a write, it holds later reads of the key until it has itself gathered acceptances from a majority and every responder.

\paragraph{Cold keys: avoid the shared log.}
Writes to infrequently accessed keys rarely conflict, so they can skip leader ordering altogether.
The client sends such a write directly to a fixed majority of replicas, the \emph{cold quorum}.
Instead of a shared log, each cold key has its own version sequence.
Each member votes for the next version of the key and notifies the client directly; without conflicts, the write completes once the client holds votes from every member.
This uncontended case follows the fast path of \swiftpaxos, but omits dependency ordering and the second-vote fallback~\cite{ryabinin2024swiftpaxos}.
Because every member of the cold quorum votes on every write, all of them learn about it, so they also serve as the key's read responders without adding a participant to the fast path.
A member holds reads of the key from its vote until it learns which value was chosen.
Upon a conflict, the leader checks whether either write could already have completed and, if so, keeps it first.
It then proposes the next write in a new per-key round, which every cold quorum member must accept before completion.
If higher-ballot recovery is needed while the cold path is active, the protocol recovers its potentially completed writes and transfers their state into the shared log.
This transfer is permanent; subsequent operations use the log.

\subsection{Model \& Notation}

\OurProtocol replicates a key--value store across a fixed set $R$ of $n=2f+1$ replicas; $m=f+1$ denotes a majority. Replicas may crash but do not behave arbitrarily or restart. Surviving replicas retain accepted commands and duplicate-detection state in memory. The protocol tolerates delayed or repeated messages. Progress requires partial synchrony: eventually, message delays among available replicas stay below the failure-detection and lease timeouts, and a majority remains available; a fast path may additionally depend on particular replicas.

Clients issue read and write operations with unique request identifiers. A write returns an acknowledgment. Operations on distinct keys commute. The target consistency guarantee is linearizability.

The ordinary log operates in \emph{ballots}. Ballot $b$ identifies a leader $L_b$ and the read responders $S_b(k)$ for each key $k$; we call this assignment the \emph{roster}. The planner chooses the roster of the initial ballot. A higher ballot may change the leader and removes failed role holders (\autoref{sec:xpaxos-recovery}), but otherwise keeps the planned roles. A responder needs a valid lease and sufficiently up-to-date state before answering locally. Leases assume bounded clock-rate drift, but not synchronized clocks. When the cold path is enabled, $Q_b$ denotes its fixed majority quorum and its set of read responders.

We distinguish \emph{acceptance}, when a replica records a proposal and votes for it, from \emph{execution}, when it applies the chosen command to its state. A \emph{certificate} is a set of matching votes sufficient to establish that a write has been chosen.

\subsection{Preparation and Role Placement}
\label{sec:xpaxos-planning}

The planner runs once, before deployment, on the workload profile introduced above; it does not run during execution. It first divides keys into hot, ordinary, and cold classes. A key is hot if the profile records more accesses to it than to the average observed key. Among the remaining observed keys, the least frequently accessed ones may be assigned to the cold path. The planner considers up to seven candidate cold cutoffs: the hot boundary itself, and the boundaries that split the remaining keys' cumulative accesses into seven equal parts. All other keys use the shared log. By default, observed keys among them take the ordinary path, while unobserved keys, and hot keys outside every completion-owner range, have the leader as their implicit completion owner; the planner may instead enable direct client certificates for all log-ordered keys, which removes completion owners altogether. Using the sampled read and write counts, measured network delays, and configured workload rate and write fraction, the planner chooses this cold cutoff together with the leader, read responders, completion owners, direct client completion policy, and cold quorum. These choices remain fixed during execution; \autoref{app:xpaxos-planning} gives the exact rules.

The planner balances read and write latency: placing a responder closer to readers can make it more costly for writes to reach that responder. The cold path avoids leader ordering but requires every member of its quorum and incurs recovery work when writes conflict. The planner can therefore leave some non-hot keys on the ordinary log or disable the cold path entirely. Its cost model and search procedure are specified in \autoref{app:xpaxos-planning}; they guide placement rather than guarantee a latency improvement for every workload.

\subsection{Normal Operation}
\label{sec:xpaxos-normal}
\label{sec:xpaxos-output}

This subsection states the precise conditions behind the paths of \autoref{sec:xpaxos-delta}.
Before serving requests, replicas install the planned roles under a common ballot: the leader, the read responders of each key range, the completion owners, and, when enabled, the cold quorum.
Each client also receives from the plan its read responder and its \emph{write ingress}, the replica that replies to its writes; for a key in a completion-owner range, the ingress is that owner.
The client sends each write both to its ingress and directly to the leader, which proposes whichever copy arrives first.
Read responders obtain leases from a majority and catch up before answering locally, and the cold quorum installs the initial per-key state before accepting proposals.

\paragraph{Leader-ordered writes.}
The leader assigns each proposed entry a log slot, and replicas retain the complete entry before voting.
For an entry $E$, let $V(E)$ be the set of matching voters and $S_b(E)$ the union of read responders for the keys modified by $E$.
Every completion of a leader-ordered write, whether at a completion owner, at the client, or at the ingress, requires
\begin{equation}
  |V(E)|\ge m
  \quad\land\quad
  S_b(E)\subseteq V(E).
  \label{eq:xpaxos-ordinary}
\end{equation}
The majority fixes the log value; the coverage ensures that every authorized reader has recorded the write before it completes.
Replicas execute chosen entries in log order and use request identifiers to avoid applying retries twice.

\paragraph{Completion gates.}
A completion owner gates exactly the writes for which it is the ingress: it withholds its vote until it has observed \eqref{eq:xpaxos-ordinary} and advanced its read target to the write, and serves the preceding value meanwhile.
The leader plays the same role for the writes it receives itself on log-ordered keys that use neither a completion owner nor direct certificates.
Batches are split so that each log entry has at most one gate.
If a write that the completion owner does not gate is pending on the key, it holds reads like any other responder.
Without a gate, a responder holds reads of a key from the moment it accepts a write to it until it learns that the write is chosen.

\paragraph{Reading ahead of execution.}
A responder can sometimes return a chosen value while execution is still waiting on earlier log entries. It needs matching votes from a majority and every required read responder for the write, together with evidence that a majority has accepted a contiguous log sequence through its entry. The responder can then check whether the write is a new operation or a retry of an earlier request, and return the key's latest chosen value without waiting for unrelated earlier commands to execute. When the shortcut is unavailable, the read waits. The full eligibility and visibility conditions appear in \autoref{app:xpaxos-normal-details}.

\paragraph{Cold writes.}
A member of $Q_b$ votes for version $v$ of a key only if it has applied every version up to $v-1$ and has not voted for another command at $v$ in the current key round.
Matching votes from every member complete the write; the client then distributes the certificate, and replicas apply certified writes in per-key version order.
Consequently, a write that reaches a member before the certificate of the key's preceding write collides there, even if the two writes did not overlap from the clients' point of view.
Such collisions, like concurrent proposals, are resolved by the leader in a higher key round that preserves any write that might already have completed; the fast path resumes once every member of $Q_b$ has installed the resolved state.

\subsection{Failure Recovery}
\label{sec:xpaxos-recovery}

Failure of the leader or a required read responder triggers a higher ballot. The leader for that ballot revokes conflicting read leases or waits for them to expire. It collects accepted log entries from a majority, which promise to reject proposals from earlier ballots, and preserves any entry that could already have been chosen. Replacement read responders obtain leases and catch up before answering reads. Other replica failures need no role change if a majority and all required role holders remain available.

If the cold path is active, higher-ballot recovery also freezes its voting and collects acceptance records from a majority. A write might have completed before its client distributed the certificate, so a missing certificate does not establish that the write was unchosen. Recovery preserves every write that might have completed and binds the recovered cold state to a bridge entry in the ordinary log. Executing this entry installs that state and imports its request identifiers for duplicate detection. The handoff is permanent: subsequent operations use the ordinary log, and write completion and read visibility require execution of the relevant log prefix. A same-key collision instead uses a leader-coordinated round to resolve the next per-key version, after which the cold quorum can resume the fast path.

Because replicas store the full command before voting, recovery can recover the write even if its client has failed. Retried requests retain the same identifier so that they are not applied twice. The complete recovery and bridge-installation procedures appear in \autoref{app:xpaxos-recovery}.

\section{Engineering Optimization in \OurProtocol}
\label{sec:apaxos_implementation}

Turning a consensus design into an efficient implementation requires engineering optimization beyond pseudocode~\cite{chandra2007paxos}.
When acting as a developer to implement each finalized candidate protocol, we allow \OurAgent to perform up to five rounds of performance tuning.
These rounds use experimental feedback to refine implementation choices while preserving the protocol idea.
The following engineering optimizations reduce communication and waiting in \OurProtocol and limit repeated execution and recovery work.

\subsection{Completion-Aware Batching}
\label{sec:implementation-batching}

The completion owner's read shortcut imposes a constraint on batching: each log entry may have at most one owner withholding its vote.
If an entry depended on several such owners, each could wait for a vote that another owner was withholding.
The leader therefore splits a pending batch when it encounters an eligible write with a different completion owner, preserving command order across the resulting entries.

Only writes eligible for the owner shortcut determine these split points.
Writes eligible for direct client certificates never acquire a gate of their own, even when an owner is assigned to their key. They can, however, be batched into the same log entry as a gated write, and then wait for its gate (\autoref{sec:discussion}).
Replicas also check received entries and disable the shortcut if completion owners are incompatible.
This permits batching among compatible writes while retaining the owner's ability to serve the preceding value during a pending write.
Changes in eligible completion owners can reduce batch size.

\subsection{Selective Vote Dissemination}
\label{sec:implementation-votes}

The leader broadcasts ordinary proposals, while acceptors direct their votes to replicas that collect completion or read-visibility evidence.
For fresh current-ballot entries, the recipient set includes the leader, planned learners, all read responders in the installed roster, and the ingresses of batched requests.
Planned learners include write ingresses and completion owners; previously observed request ingresses are also retained as recipients.
These sites collect evidence directly while votes to other replicas can be omitted.

For eligible ordinary writes, acceptors also send votes directly to the client.
The majority and responder-coverage requirements remain unchanged.
Vote traffic decreases when the recipient set excludes some replicas; if the selected roles cover the entire membership, no votes are omitted.
Recovered entries and configurations without a selective learner set retain dissemination to all replicas.

\subsection{Read Visibility Indexes}
\label{sec:implementation-reads}

A read's target depends on both the state of a write and who controls its completion.
A chosen current-ballot write may be visible before prefix execution, whereas an accepted write that could complete elsewhere must already prevent stale reads.
Each replica therefore maintains per-key indexes for the latest accepted write, the latest committed write, and the latest write whose completion is outside the owner's control.
The third index identifies writes that prevent the owner from continuing to serve the preceding value.
Together, these indexes support the owner's read shortcut while fencing writes that may have completed elsewhere.

The indexes locate a read's target log position without searching the log.
Blocked reads are grouped by that position in two ordered queues: one for prefix execution and another for earlier visibility once the acceptance-prefix evidence permits it.
As each prefix advances, the replica revisits the corresponding groups instead of scanning all waiting readers.
This limits repeated work as blocked reads accumulate while preserving both ways of making a read ready.
Every release still checks the lease, responder assignment, and chosen-value visibility conditions in \autoref{sec:xpaxos-normal}.

\subsection{Planned Read Racing}
\label{sec:implementation-racing}

A nearby responder may return later than a more distant one if a concurrent write holds its read.
For selected hot keys, the planner therefore assigns a second responder using network delays and modeled read waiting.
The client sends the same request to both responders without waiting for a timeout, after checking that the second target is available and authorized for the key.
The assignment applies only while the initial ballot and leader remain unchanged.

Each responder independently checks its lease and read-visibility conditions.
The client returns the first valid result and sends a best-effort cancellation to release waiting state on the other path.
This trades an additional read request for an opportunity to avoid waiting behind a write at one responder.
The planner selects this extra path only for ranges where its model predicts a sufficient gain.

\subsection{Direct Submission to the Leader}
\label{sec:implementation-requests}

Placing write completion near a reader may select an ingress different from the leader.
Sending the command only through that ingress would make ordering wait for an additional forwarding hop.
When the selected ingress differs from an available leader, the client sends an ordinary write directly to the leader as well as to the ingress.
The direct copy identifies the selected ingress, allowing the leader to start ordering while execution replies retain their planned return path.

Both copies retain the original request identifier, so a later forwarded copy is recognized as the same operation.
The leader can thus begin ordering without waiting for ingress forwarding, while the ingress keeps its completion role.
The additional client-to-leader transmission trades bandwidth for earlier ordering; its net benefit depends on the network routes.

\subsection{Recovering Per-Key Write Path}
\label{sec:implementation-snapshots}

The per-key write path can complete a cold-key write before its chosen proof reaches every replica.
Recovery must reconstruct these writes from retained records, but full snapshots would repeatedly transfer command contents held by several participants.
The implementation therefore separates the evidence needed to select the recovered state from the command data needed to install it.

Each replica freezes a metadata report containing per-key versions, request identifiers, command digests, and acceptance evidence, while retaining full commands locally.
The coordinator collects complete reports from a majority and selects the state to preserve according to \autoref{sec:xpaxos-recovery}.
The reports also identify potential command holders, allowing selection without attaching full command contents to every report.

The coordinator distributes a \emph{manifest} listing recovered versions and request identities, with digest references to the required commands.
Recipients check these references against their local command stores and fetch only missing contents from replicas that retain them.
Each recipient tracks missing commands and votes only after retaining and validating all referenced data.
This reduces repeated payload transfer when reports refer to the same command or recipients already cache it; cache misses still require data fetching before installation.

Bridge installation reuses the command store and missing-command tracking, pausing until the referenced contents are complete.
It also imports recovered request identifiers into ordinary duplicate detection, preserving operation identities as pending requests move to the log.
Command transfer continues over ordinary peer streams after the per-key write path is disabled, so installed replicas can help others finish recovery without reactivating that path.

\section{Human-Enforced Changes in \OurProtocol}
\label{sec:human_enforced_changes}

\OurAgent did not create the final version of \OurProtocol fully autonomously, however. Before accepting \OurProtocol's great performance in our measured benchmarks, we reviewed whether its improvements reflected generalizable design choices or hard-coded benchmark-hacking. It is known that agents may use unexpected methods or cheat to reach their goals~\cite{huggingface2026intrusion, skalse2022gaming}. As our discovery process uses a quantitative benchmark to assess performance, the agent might use questionable methods to optimize for latency, resulting in a non-generalizable algorithm with shortcuts. We observed multiple cases where \OurAgent employed such shortcuts, especially when the agent was aware only of the evaluation it could invoke itself.

We mitigated this by introducing a separate test set and telling the agent about it, while keeping its details hidden. While this discouraged hard-coded execution paths, it did not eliminate questionable optimizations. A separate test set can expose overfitting to particular workloads, but it cannot rule out unsupported claims of improvement or shortcuts that exploit assumptions shared by the development and test environments.

Automated checks can enforce explicit requirements, but those requirements may still fail to capture the intended behavior, even in formally verified systems~\cite{goldweber2024ironspec}. Deciding which requirements capture a meaningful consensus experiment therefore still requires domain knowledge. We reviewed the generated implementation and intervened where a latency improvement depended on assumptions we did not consider acceptable. 

For example, our benchmark sampled network jitter independently for each message. The agent exploited this by sending duplicate requests simultaneously over two TCP connections to the same destination and using whichever arrived first. This reduced measured latency by giving each logical request two chances to receive a smaller delay, an advantage that is unlikely to materialize on a real network, where messages sent together to the same destination are rarely reordered. Moreover, due to borrowing ideas from \bodega, \OurProtocol inherited one of its flaws, detailed in \autoref{sec:bugs}. We removed both the inherited flaw and benchmark-hacking before the final evaluation.

\section{Evaluation}
\label{sec:evaluation}

We implement and evaluate the efficiency of \OurProtocol.
Our experiments answer the following questions:

\begin{itemize}
    \item How does failure-free latency of \OurProtocol compare with that of state-of-the-art protocols across workloads, deployment sizes and topologies (\autoref{sec:exp_normal_case})?
    \item Can \OurProtocol keep making progress and recover when replicas fail, despite the role assignments that reduce its normal-case latency (\autoref{sec:evaluation-recovery})?
    \item What does it cost, in terms of tokens and time, for \OurAgent to discover \OurProtocol (\autoref{sec:evaluation-discovery-cost})?
\end{itemize}

\subsection{Implementation} 

We implement \OurAgent as a repository-based harness and run it using the goal mode of \codex powered by \texttt{GPT-6.1 Sol High}, with the prompt in \autoref{app:search-prompt}.
A top-level \texttt{AGENTS.md} encodes the researcher--developer--verifier workflow, consistency requirements across designs, implementations, and proofs, and the performance and recovery criteria for accepting candidates.
Four project-specific skills provide operational guidance for knowledge retrieval, prototype development and benchmarking, \textsc{Lean} formalization, and \textsc{Lean} proof construction.
The skills connect these roles to the three components in \autoref{sec:paxosagent}.

Our implementation is organized into three repositories:
\textsc{ConsensusWiki}, \textsc{ConsensusArena}, and
\textsc{ConsensusVerifier}. \textsc{ConsensusWiki} provides the knowledge base described in \autoref{sec:harness-knowledge}, comprising 97 Markdown documents and approximately 8,600 lines. 
\textsc{ConsensusArena} provides the implementation framework described in \autoref{sec:harness-skeleton}.
It contains approximately 6,800 lines of shared Go runtime code, 19,100 lines across the baseline implementations, and 8,300 lines for \OurProtocol.
\textsc{ConsensusVerifier} provides the verification framework described in \autoref{sec:harness-verification}.
It comprises approximately 135K lines of \textsc{Lean} code, covering shared definitions and formal specifications and proofs for \multipaxos, \fastpaxos, \swiftpaxos, \epaxosstar, \bodega and \OurProtocol.
Of these, approximately 98K lines support the final model and proofs of \OurProtocol.
For \bodega, we prove its linearizability when the early-accept optimization is disabled; when it is enabled, we exhibit a counterexample to linearizability (\autoref{sec:bugs}).

To compare the baselines under common deployment conditions without hardcoded assumptions about topology or key popularity, we adapt their planners to use supplied topology inputs and, where needed, independent pre-measurement workload samples, following the core configuration principles of the respective papers and source code.
We also address implementation issues to support the broader range of workloads and deployment configurations exercised by our benchmarks.

Specifically, for \bodega, whose original evaluation scripts manually select the leader, we reuse our \multipaxos planner to choose the initial leader that minimizes the mean client–leader round-trip latency plus the estimated time to collect a majority of acknowledgments.
When an independent pre-measurement sample contains only reads, we configure every replica as a responder for all keys, enabling local reads throughout the deployment.
In addition, we also disable its early-accept notification optimization.

\subsection{Experimental Setup}

We evaluate \OurProtocol against \swiftpaxos~\cite{ryabinin2024swiftpaxos}, \bodega~\cite{hu2026bodega}, \multipaxos~\cite{lamport2001paxos}, and \kcensus~\cite{burgelin2026kcensus}. We consider four deployment sizes with 5, 9, 13, and 17 replicas, each containing an equal number of clients. At each size, we generate one uniform, one clustered, and one colocated network topology, for a total of twelve configurations. Uniform topologies place all participants independently, clustered topologies group them around three centers, and colocated topologies place each client at the same location as one replica. Network delay, asymmetry, jitter, bandwidth, client placement, leader placement, workload arrival rate, value size, key-space size, and key-access skew vary across these configurations.

\paragraph{Workload.}
We generate key--value workloads with read/write mixes corresponding to \textsc{YCSB} workloads A, B, and C~\cite{cooper2010ycsb}.
Reads and writes correspond to \textsc{Get} and \textsc{Put} operations, respectively. Workload A contains 50\% writes, workload B contains 5\% writes, and workload C is read-only. The per-client arrival rate is sampled independently for each topology between 500 and 1,800 requests per second. Commands carry values of 256, 1,000, or 2,048 bytes and operate over either 250,000 or one million keys. For each topology, we sample a Zipf skew parameter $\theta$ from $[0.65,1.15]$ and hold it fixed throughout the experiment. 
Each request independently selects a popularity rank $r$ with probability $p_r \propto (r+1)^{-\theta}$; ranks are mapped to key identifiers through a fixed random permutation shared by all clients.
Across the twelve generated topologies, the sampled values of $\theta$ range from $0.651$ to $1.120$.
Clients submit requests according to an open-loop Poisson process, meaning that request arrivals are scheduled independently of response completion. After a five-second warm-up, clients generate requests at the configured arrival rates during a ten-second measurement window.
After the measurement window ends, we wait for all requests submitted during that window to complete and include their end-to-end latencies.

\paragraph{Hardware.}
We run all experiments on EPFL's SCITAS Jed CPU cluster using \textsc{Slurm}~\cite{jette2023architecture}\footnote{See \url{https://www.epfl.ch/research/facilities/scitas/jed/}.}.
Experiments use 2--5 compute nodes, depending on the deployment size.
Each node contains two 36-core Intel Xeon Platinum 8360Y processors running at 2.40\,GHz, for a total of 72 physical cores, and 492\,GB of memory.
Replicas, clients, and the coordinating master execute as separate \textsc{Slurm} tasks, each requesting eight CPUs.
Communication uses TCP over the cluster's Ethernet interconnect.
\textsc{ConsensusArena} uses Toxiproxy to emulate the generated network conditions.
For each topology, we select a bandwidth limit of 200\,Mb/s, 1\,Gb/s, or 4\,Gb/s for each TCP connection and direction.
We also select a jitter parameter $J \in \{0,3,10,20\}$\,ms for each topology.
Toxiproxy introduces random variations of approximately $\pm J$ around each nonzero configured directional delay.

\paragraph{Metrics.}
Our primary metrics are mean and p99 end-to-end latency. For each measured read or write, latency is measured from client submission until the client receives sufficient protocol-specific evidence to declare the operation complete. Mean latency averages these measurements, while p99 latency is the value below which 99\% of measured operations complete and therefore captures tail performance.

\subsection{Normal Case Performance}
\label{sec:exp_normal_case}

\begin{figure}[t]
    \centering
    \inputplot{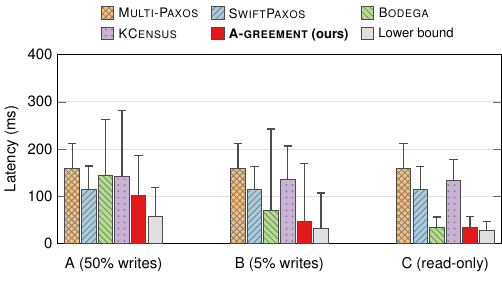}
    \caption{Failure-free latency averaged with equal weight over the six deployments with 5 or 9 replicas. Bars report mean latency, while upper whiskers report p99 latency. Lower bound is the idealized communication baseline.}
    \label{fig:evaluation-latency}
\end{figure}

\autoref{fig:evaluation-latency} reports mean and p99 latency for each workload, averaged across the six deployments with 5 or 9 replicas.
We exclude larger deployments from this average because, at 13 and 17 replicas, saturation of individual baselines dominates it: for example, \swiftpaxos's mean latency reaches 516.8\,ms with 17 replicas in the uniform topology.
Fine-grained results for every algorithm, topology, and workload, including all twelve configurations, are reported in \autoref{app:quantitative_results}, together with the corresponding operation-level read and write latency breakdowns.

On workload A (50\% writes), \OurProtocol reduces mean latency by 12.2\% relative to \swiftpaxos, the alternative with the lowest mean and p99 latency, while its p99 latency is 13.8\% higher than that of \swiftpaxos. 
Notably, the mean read and write latencies of \OurProtocol are 56.5\,ms and 145.4\,ms, respectively, compared with 115.0\,ms and 115.2\,ms for \swiftpaxos. 
Thus, \OurProtocol's overall advantage over \swiftpaxos comes from its local-read path, at the cost of slower writes. 
\bodega also provides fast reads, with a mean read latency of 77.2\,ms, but incurs a substantially higher mean write latency of 212.6\,ms. 
These results show that \OurProtocol combines fast local reads with a write path that avoids \bodega's substantial write-latency penalty.

On workload B (5\% writes), it reduces mean and p99 latency by 33.3\% and 30.1\%, respectively, relative to \bodega, the alternative with the lowest mean latency, while its p99 latency is 3.4\% higher than that of \swiftpaxos, the alternative with the lowest p99 latency. As in workload A, the main weakness of \bodega is its write path, for which it spent on average 207.3\,ms compared to \OurProtocol's 150.5\,ms.

On the read-only workload C, \OurProtocol achieves mean and p99 latencies of 34.0\,ms and 56.3\,ms, respectively, which is comparable to \bodega, because both protocols use lease-based local reads.

The lower bound in \autoref{fig:evaluation-latency} is an analytical communication reference computed from the same network topologies.
For each client, it assigns reads the minimum direct round-trip latency to any replica and writes the time to collect replies from the fastest majority, allowing shortest paths through replicas.
It assumes ideal read freshness and write ordering, and excludes processing, queueing, serialization, bandwidth limits, and jitter.

\subsection{Recovery under Failure}
\label{sec:evaluation-recovery}

\begin{figure}[t]
    \centering
    \inputplot{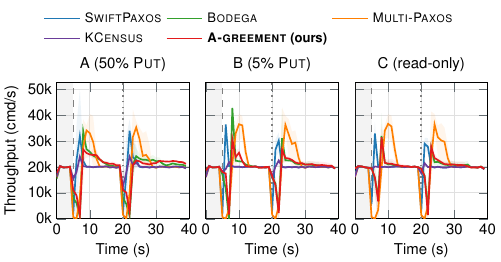}
    \caption{Recovery throughput with five replicas under workloads A, B, and C (left to right). Lines show mean completed requests per second across three topologies; bands show their minimum--maximum range.}
    \label{fig:evaluation-recovery}
\end{figure}

We evaluate recovery with five replicas under workloads A, B, and C across three network topologies.
The offered load is 20,000 requests per second.
Each run lasts 40 seconds, including an initial five-second warm-up.
At 5 seconds, we crash the current leader for protocols with a leader, or a replica for protocols without one.
At 20 seconds, we crash one additional replica, prioritizing the latest reported live leader when available and otherwise selecting the surviving replica.
The two crashes leave a majority of three replicas.
Failed replicas remain stopped for the rest of the run.
\autoref{fig:evaluation-recovery} reports completed requests per second throughout this sequence.

Across all nine runs, \OurProtocol experiences a temporary throughput drop after each failure and recovers to at least 90\% of its initial failure-free throughput within approximately three seconds.
It maintains this level until the next crash or the end of the run.
After the second crash, its throughput settles near 20,000 requests per second through the end of the observation window, as do the baselines.
The temporary peaks above this rate reflect the completion of requests accumulated during the interruption; they do not indicate a higher sustainable throughput.
The curves show that \OurProtocol retains progress with two failed replicas, although it does not consistently recover faster than the other protocols.

\begin{figure*}[t]
    \centering
    \resizebox{\textwidth}{!}{\inputplot{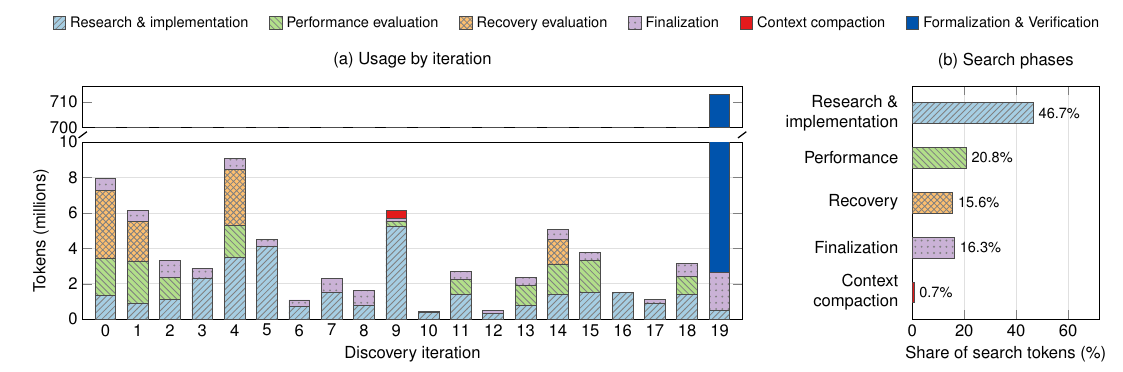}}
    \caption{Stage-attributed token usage, including input and output tokens. (a) Per-iteration usage in millions of tokens, with a broken axis omitting the range from 10 to 700 million. Formalization and verification adds 710.31 million tokens at iteration 19. (b) Phase shares within the search, excluding formalization and verification.
    }
    \label{fig:discovery-token-usage}
\end{figure*}

These failures test whether the role assignments used to reduce normal-case latency prevent progress when replicas become unavailable.
As described in \autoref{sec:xpaxos-recovery}, recovery replaces failed role holders, preserves potentially chosen writes, and reestablishes read authority.
If the per-key write path is active, recovery transfers its retained state into the shared log.
The recovered system can therefore continue processing requests using the surviving majority without requiring the original fast-path participants to return.

\subsection{Discovery Cost}
\label{sec:evaluation-discovery-cost}

The complete discovery process, including 20 search rounds and formalization and verification of the final protocol, took approximately 24 hours in total.
The search rounds consumed 68.14 million tokens across 333 model responses, including 67.83 million input tokens and 0.308 million output tokens. 
Thus, cumulative token usage largely reflects repeated processing of the accumulated context, rather than newly generated content.
Across the complete process, model reasoning accounted for approximately 23 hours.
The remaining elapsed time includes output generation, tool execution, explicit waiting, context compaction, and unattributed runtime intervals.
Formalization and verification of the final protocol consumed an additional 710.31 million tokens, bringing total usage to approximately 778.45 million. 
For the stage breakdown in \autoref{fig:discovery-token-usage}, we attribute this verification usage to iteration 19.

\autoref{fig:discovery-token-usage} attributes token usage to execution stages.
Research and implementation includes literature and code inspection, candidate selection, feasibility analysis, parent-protocol diagnostic experiments, protocol design, coding, compilation, and initial candidate benchmark submission.
Performance evaluation covers subsequent monitoring, troubleshooting, and metric analysis; recovery evaluation covers crash-recovery monitoring and analysis.
Finalization includes evidence preservation and reporting.

Formalization and verification accounts for 91.25\% of the combined token usage.
Within the original search, excluding this additional stage as in \autoref{fig:discovery-token-usage}(b), research and implementation accounts for 46.67\%.
Performance and recovery evaluation account for 20.78\% and 15.60\%, respectively.
Finalization accounts for 16.28\%, and context compaction for 0.67\%.
Ten iterations produced benchmarked candidate prototypes, while ten were rejected before candidate implementation.
Early rejection avoided candidate evaluation, but was not necessarily inexpensive: feasibility assessment could itself require instrumenting and benchmarking a parent protocol.

Individual iterations illustrate these costs.
Iteration~3 used a parent-protocol diagnostic to establish that the proposed execution-lane change had no observed prefix-blocking delay to remove.
Iteration~5 similarly measured an optimistic latency-saving bound below the promotion threshold and rejected the candidate before implementation.
Together, these two iterations consumed 7.36 million tokens.
Iteration~9 investigated acknowledgment queueing and implemented cumulative prefix certificates, but reduced acknowledgment traffic did not translate into improved client latency.
By contrast, iteration~14 combined the rejected iteration~13's read racing mechanism with per-key completion rules, passed the performance gate and all 27 recovery cases, and became the final promoted version.
These results show that discovery cost includes feasibility diagnostics, unsuccessful experiments, and reusable findings from rejected candidates, in addition to implementation of the final design.

\section{Finding Vulnerabilities in Existing Consensus Protocols}
\label{sec:bugs}

While constructing the reference \textsc{Lean} specifications and proofs in \autoref{sec:harness-verification}, \OurAgent uncovered four vulnerabilities in existing consensus protocols that, to the best of our knowledge, were previously unreported.
Two affect the optimizations introduced by EPaxos Revisited~\cite{tollman2021epaxos}: they allow replicas to execute conflicting commands in opposite orders.
Two concern Bodega~\cite{hu2026bodega}: its optional early-accept optimization can expose inconsistent values across read paths, while its client-side timeout retries can execute one client write twice.
\autoref{app:protocol-counterexamples} details the counterexamples and analysis.

\paragraph{Pruning with non-final metadata.}
Dependency pruning uses sequence numbers and dependencies to skip commands that appear to execute after the current command.
However, preaccepted metadata can change before commitment.
A delayed proposal can therefore make a dependency appear safe to prune, allowing one replica to execute commands in an order opposite to another replica's.
The counterexample in \autoref{app:tentative-pruning} shows why pruning requires committed or otherwise immutable metadata.

\paragraph{Pruning that hides unresolved dependencies.}
Committed metadata alone is insufficient: skipping a committed node can also hide commands reachable through its dependency branch.
One replica can consequently execute a command before an unresolved conflicting dependency, while another preserves the opposite order.
\autoref{app:pruning-boundary} presents the counterexample.

\paragraph{Inconsistent visibility across read paths.}
\bodega~\cite{hu2026bodega}'s early-accept notification allows a responder to return a write after collecting majority accept notifications, while a leader reads only committed state.
As mentioned in \autoref{app:bodega-early-accept}, consider \textsc{Write}$(k,7)$ for an initially absent key.
All replicas accept the write, but their acceptance replies to the leader are delayed, so the leader has not committed it or replied to the writing client.
A responder nevertheless receives enough accept notifications to return $7$ to a read.
After that read completes, a second read reaches the leader, which returns absence from its unchanged committed state.
Returning $7$ and then absence in two successive reads violates linearizability.

\paragraph{Duplicate writes after timeout retries.}
The client-side timeout retry path in \bodega~\cite{hu2026bodega}'s implementation assigns a new request identifier to each attempt, so the server can execute the original request and its retry as separate writes.
As mentioned in \autoref{app:bodega-write-retry}, consider a write that sets $x=1$ but whose reply is delayed.
Another client reads $1$ and then completes a write of $2$.
After a timeout, the original client retries with a new identifier; the retry sets $x=1$ again, and a later read returns $1$.
If both attempts represent one client write, the first read requires that write to precede the write of $2$, so the final read must return $2$.
This violates linearizability even though replicas agree on the order of the individual requests.

\section{Related Work}

\paragraph{State-machine replication.} \paxos supplies the ballot-and-quorum safety baseline for replicated agreement~\cite{lamport2001paxos}.
\fastpaxos then lets proposers send values directly to acceptors in prepared fast rounds, enabling two-message-delay learning without collisions at the cost of stronger quorum intersection and recovery rules~\cite{lamport2006fastpaxos}.
Later systems remove different parts of the consensus critical path.
\textsc{Mencius} spreads log slots across rotating coordinators~\cite{mao2008mencius}. \epaxos and \textsc{Atlas} use per-command coordination and dependency ordering, and \textsc{Atlas} additionally sizes its fast quorum around a chosen concurrent site-failure budget~\cite{moraru2013epaxos,enes2020atlas}.
\textsc{Rabia} uses randomized weak multi-valued consensus to avoid a distinguished leader and separate fail-over protocol~\cite{pan2021rabia}.
Later analysis exposed \epaxos's tail-latency and recovery subtleties~\cite{tollman2021epaxos}, while \swiftpaxos reduces the cost of dependency disagreement through leader-including quorums and double voting~\cite{ryabinin2024swiftpaxos}.

Other designs change the system boundary or optimize a particular workload.
\textsc{NOPaxos} and \textsc{Hydra} move common-case ordering into the network~\cite{li2016nopaxos,choi2023hydra}.
\textsc{CURP} uses witnesses for unordered durability when operations commute~\cite{park2019curp}, and \textsc{Jetpack} retrofits a parallel one-RTT path onto compatible consensus protocols~\cite{tang2026jetpack}.
\textsc{Copilot} tolerates one slow replica by redundantly processing commands at two pilots~\cite{ngo2020copilot}.
\bodega combines roster leases with responder-covering write quorums for local linearizable reads during interfering writes~\cite{hu2026bodega}.

\paragraph{AI-driven algorithm and systems discovery.}
\textsc{AlphaEvolve} combines \ac{LLM}-driven evolutionary search with automated evaluators to discover improved algorithms~\cite{novikov2025alphaevolve}.
\textsc{ADRS} studies this approach in systems research, using \textsc{OpenEvolve}~\cite{sharma2025openevolve}, an open-source framework inspired by \textsc{AlphaEvolve}, for tasks such as spot-instance scheduling, mixture-of-experts load balancing, and transaction scheduling~\cite{cheng2025adrs}.
\textsc{Glia} instead uses a multi-agent \ac{LLM} workflow, in which a researcher agent proposes, implements, and evaluates designs in a simulator under a supervisor agent, to design request-routing, scheduling, and autoscaling policies for \ac{LLM}-serving clusters~\cite{hamadanian2026glia}.
These frameworks optimize isolated policies in which every candidate is valid by construction or checked exactly, and a simulator scores it in seconds.
\textsc{ADRS} identifies consensus protocols, whose logic spans state management, communication, and failure handling, as a poor fit for this approach. 
\OurAgent targets this gap, where three challenges arise together: (C1) faster candidates can break safety, (C2) tests cannot establish safety, and (C3) each candidate takes minutes to hours to evaluate. \autoref{tab:ai-discovery} and \autoref{tab:ai-challenges} in \autoref{app:related-comparison} compare these approaches and the challenges they face.

\paragraph{Agent-synthesized verified systems.} The \textsc{SkyDiscover} framework~\cite{liu2026skydiscover}, from the authors of \textsc{ADRS}, extends this line from evolving isolated programs to synthesizing complete systems. Its proof-driven mode, \textsc{IDS}~\cite{agarwal2026ids}, is the work closest to \OurAgent: agents jointly synthesize an implementation and its \textsc{Rocq} proof, use incomplete proofs as a search oracle, and use benchmarks to steer toward efficient designs, producing verified distributed key-value stores for session and causal consistency specifications.
\textsc{IDS}, however, starts from a fixed, human-written specification, so performance only ranks candidates that already meet it (sidestepping C1), and its design freedom lies in data layout and proof structure rather than in new protocol mechanisms.
Its targets involve neither fault-tolerant agreement nor recovery.
\OurAgent instead designs new consensus protocols under a performance objective. Conversely, \textsc{IDS} proves the extracted implementation, whereas our \textsc{Lean} proofs cover a protocol model whose correspondence to the prototype is additionally checked by a human consensus expert. %

\paragraph{AI for consensus.} Prior AI work on consensus finds bugs in existing protocols, generates small protocols, or assists human protocol designers. \textsc{Agora}'s agents inspect \textsc{Raft}, \epaxos, \textsc{HotStuff}, and \textsc{BullShark} codebases, synthesize attacks, and generate executable tests.
\textsc{Agora} reports 15 new defects classified as protocol-level logic bugs because the observed executions violate protocol properties~\cite{liu2026agora}.
\textsc{GGMS} learns protocols without an \ac{LLM}, combining Monte Carlo tree search, a transformer action encoder, and a model checker~\cite{hui2026ggms}.
It targets synchronous consensus and atomic commit with crash failures, and verifies candidates by exhaustive model checking for at most four processes.
For consensus, it rediscovers \textsc{FloodSet}~\cite{lynch1996distributed}.
By dropping the performance objective and restricting the search to four processes, it sidesteps C1 and keeps evaluation cheap (C3), but its guarantees do not extend beyond the checked instance sizes. An \ac{LLM} also substantially assisted the design and proofs of a communication-efficient Byzantine broadcast protocol~\cite{efron2026bb}.
There, however, humans drive an interactive collaboration toward an asymptotic bound established by pen-and-paper proof; no candidate is implemented or measured, so C3 does not arise and safety checking (C2) rests on the authors.
\OurAgent instead faces all three challenges: it designs protocols under a latency objective and evaluates them on deployments of up to 17 replicas.

\section{Conclusion: The End of Consensus Research?}
\label{sec:discussion}

Our answer is no; this is just a new beginning.
We believe this work marks a turning point in how consensus research is conducted. The low latency achieved might still be decreased in specific situations we did not focus on.  More generally, our consensus solution could be extended to other failure models and settings. 
Our results show that AI agents can take on substantial parts of the research process, including retrieving protocol knowledge, designing and implementing candidates, as well as constructing formal arguments and counterexamples.
Human expertise, however, remains essential for defining performance objectives and safety requirements, designing evaluation environments, and reviewing agent behavior and research outcomes.
A crucial question therefore is: \emph{how do our results reshape the role of consensus researchers?}

\subsection{Where Is Human Judgment Needed?}

From our experiences, we found that human judgment was needed before, during, and after the agentic search.

\paragraph{Before the search.}
Human expertise is required before the quest starts.
Researchers need to develop the agent harness by checking the protocol knowledge base, defining the implementation interfaces, and choosing the fault model, correctness properties, performance objectives, and evaluation conditions (\autoref{sec:paxosagent}).
Although an agent drafted the knowledge-base notes, we verified them against the source papers and corrected errors.
This preparation determines which designs the agent can express and what its feedback measures.

\paragraph{During the search.}
Inspecting an agentic run while it is in progress allows researchers to steer the agent early and avoids restarting a long run that wastes many tokens.
Giving the agent an explicit budget, such as a fixed amount of time to search for a new candidate, can reduce the need for such manual checks.

\paragraph{After the search.}
Human review remains necessary when interpreting improvements.
As described in \autoref{sec:human_enforced_changes}, the agent duplicated requests over two TCP connections to the same destination to exploit independently sampled network jitter.
The resulting latency reduction reflected an opportunity in the benchmark's delay model, and we removed the mechanism before the final evaluation.
A separate test set helps expose overfitting, but shared assumptions in the development and test environments can still reward such shortcuts.
Researchers must therefore examine both candidate protocols and their evaluation environments to reduce the risk of reward hacking.

Human review can also expose inefficiencies that the agent's own tuning left in place.
In \OurProtocol's batching (\autoref{sec:implementation-batching}), the leader cuts a batch into log entries in arrival order.
This couples some writes to the slower completion of a hot key and fragments batches when completion owners alternate.
Grouping each batch by completion owner would remove both inefficiencies and is safe, but the agent chose not to do it.
Moreover, upon a ballot change \OurProtocol does not restore the cold quorum (\autoref{sec:xpaxos-recovery}), moving all write-related actions to the common log, essentially sacrificing a part of its optimization for no reason.
We did not fix either of these inefficiencies, so that the evaluated protocol stays as close as possible to the agent's design, with only correctness-critical fixes.

Human review is also necessary for formal verification.
\textsc{Lean} checks proofs against the encoded specification; researchers must check that this specification faithfully captures the protocol and its implementation.
In our experience, formalizing a protocol from pseudocode presents substantially different challenges from formalizing it from implementation code.
The discovery loop could generate an implementation first and then formalize its behavior, or develop both the implementation and the formal specification from shared pseudocode.
Which approach produces more faithful specifications with less human intervention remains an open question.

\subsection{The Changing Role of Researchers}

As agents take on more research tasks, we believe that human researchers retain three central responsibilities.
First, they must define meaningful problems and translate practical requirements into explicit research objectives.
This includes deciding which workloads and deployment conditions matter, what safety and availability guarantees are required, and how to prioritize latency, throughput, and resource costs.
These decisions give the agent a concrete target and determine which tradeoffs are acceptable.
Second, they must test and review agent outputs, which may still contain hallucinations, implementation errors, or unsupported conclusions.
Generated code, explanations, and experimental reports require independent scrutiny before they can support a research claim.
Researchers must assess whether the evidence justifies the claimed improvement, investigate failures and unexpected behavior, and determine how broadly the conclusions apply.
Third, they remain important for proposing new ideas and opening directions beyond the current search.
The search reported here primarily combined and adapted known mechanisms, showing how agents can explore combinations that are costly to investigate manually.
Human researchers can contribute new abstractions, question assumptions inherited from existing protocols, and formulate hypotheses that expand the design space.
Agents can then help develop these ideas into protocols, construct implementations and proofs, and evaluate their consequences, allowing researchers to pursue more alternatives and refine the ideas using concrete evidence.

\clearpage
\begingroup
\small
\bibliographystyle{plainurl} 
\bibliography{ref}
\endgroup

\clearpage
\appendix
\section{Detailed \OurProtocol Specification} \label{app:xpaxos-specification}

This appendix specifies the three write paths introduced in \autoref{sec:xpaxos}: completion at a designated replica, direct client learning from the ordinary log, and direct client learning from the cold per-key history. The first two paths share log state and recovery. We specify their different completion and read-visibility rules explicitly. Message names below describe logical actions; transport framing, compression, and retransmission do not change the evidence required by a handler.

\subsection{State, Identity, and Authority}
\label{app:xpaxos-state}

\paragraph{Requests and retained state.} A request $c$ contains an identifier $\operatorname{id}(c)$, its operation, key, and original payload. Retries preserve this identity and command body. An ordinary request also retains its designated reply replica. This identity is needed to determine whether the completion owner controls the request's reply. The cold fast path completes at the client and has no completion-owner role. Participants are non-Byzantine. Vote sets count distinct replica identities; hashes bind content and are not signatures. We assume that different retained bodies do not collide under the content hash $H$.

\paragraph{Ordinary state.} Ballot $b$ installs leader $L_b$ and responder sets $S_b(k)$. Slot $s$ contains an ordered batch $E=(b,s,C,\mathit{fresh})$; $C$ includes the original requests and their logical reply destinations. A fresh entry was newly proposed in the current ballot; a recovered entry is not fresh. Replicas retain accepted entries, votes, completion identifiers, and the highest contiguous executed slot \texttt{executedPrefix}. The completion-owner assignment, when applicable, is denoted $O_b(k)$; a delegated owner must remain an authorized responder. Otherwise, the leader supplies the default owner used by the read shortcut. The ordinary executed-reply route still follows the request's retained reply destination.

For a batch, define its required reader coverage as
\[
 S_b(E)=\bigcup_{c\in C:\,\operatorname{op}(c)=\textsc{Write}}
                   S_b(\operatorname{key}(c)).
\]
An ordinary completion certificate contains matching acceptance evidence $V(E)$ satisfying $|V(E)|\ge m$ and $S_b(E)\subseteq V(E)$. A Paxos majority fixes the log value; responder coverage is the additional condition used here before completion or exposure of a not-yet-executed value. Below, an ordinary entry marked \emph{covered-chosen} satisfies both conditions. A trusted current-ballot commit notification can convey this outcome, but it does not supply a missing batch body or authorize reading a different ballot's entry.

For a fresh current-ballot entry $E$, a \emph{gate candidate} is a non-certificate write whose designated reply replica is its completion owner. If the candidates in $E$ all name the same owner $p$, define the completion gate $g(E)=p$; if there is no candidate or candidates name different owners, let $g(E)=\bot$. Normal batching splits entries when candidate ownership changes, and an unsplit conflicting entry conservatively has no gate. A local acceptance counts in $V(E)$ immediately, but $g(E)$ does not publish its \textsc{Accepted} or any client vote for the entry until it has itself marked $E$ covered-chosen. Thus the gate advances its read target before its vote can allow another learner to complete the entry.

\paragraph{Cold state.} Cold control state has generation $G=(b,e)$, where $e$ is an epoch within ballot $b$. Epochs distinguish freeze and resume barriers, including those used by scans. Each key also has a key-round number $a$. Rounds $(b,e,a)$ are ordered lexicographically; version $v$ identifies the position in the key's history, independently of its round. A cold vote binds
\[
 (b,e,a,k,v,\operatorname{id}(c),H(c),
   \operatorname{replyTo}(c),\operatorname{timestamp}(c),Q_b).
\]
Here $H(c)$ hashes the original operation, key, and payload; the remaining fields bind its request and routing identity. The fast quorum $Q_b$ has size $m$ and remains fixed across key rounds and scan epochs within an installed ballot. A new ballot may select a different quorum for recovery. For each key, a replica retains its promise, chosen prefix, accepted tail, chosen evidence, pending original requests, duplicate identifiers, and readiness state. A bare payload or routing hint supplies none of this authority.

\paragraph{Lease and installation preconditions.} Local reads require an installed roster and a majority of unexpired lease grants whose catch-up thresholds the responder has executed. Self-grants count only when the local threshold is met. A leader must additionally have completed ordinary preparation. Grantor and holder deadlines include a conservative clock-drift margin; delayed renewal replies cannot restore a revoked ballot. Before replacing conflicting read authority, grantors revoke the old grants and receive acknowledgments, or wait until the grants expire. Failure suspicion starts this process but cannot replace it. Cold readers also require the installed generation to equal the promised generation and the key to be ready.

\paragraph{Completion policy.} Let $\eta$ be the measured hot boundary and $\tau\ge\eta$ the selected cold boundary, defined in \autoref{app:xpaxos-planning}. Under the reference policy, fresh ordinary writes to observed ranks below $\eta$ complete through their reply destination after execution; ranks at or above $\eta$ are eligible for direct certificates unless routed to the cold path. The alternative policy enables direct certificates for all ordinary keys, including hot and unobserved keys. Unobserved keys never enter the cold path. Recovered entries are ineligible for early ordinary completion, and bridge acceptance disables new early votes. Thus heat classification, completion-owner placement, and certificate eligibility are related planning decisions, not interchangeable definitions.

\subsection{Normal-Operation Details}
\label{app:xpaxos-normal-details}

\paragraph{Ordinary read visibility.} Each \textsc{Accepted} vote reports \texttt{acceptedPrefix}, the highest slot through which the sender has accepted a contiguous current-ballot sequence. The $m$th largest report, counting missing reports as zero, is \texttt{majorityFixedPrefix}. It fixes the preceding entry identities and the first occurrences of duplicate request identifiers; it does not establish responder coverage by itself.

The target of a read depends on whether its responder gates completion. At replica $p$, let $A_p(k)$ be the greatest accepted slot touching $k$, $C_p(k)$ the greatest covered-chosen slot touching $k$, and $D_p(k)$ the greatest accepted slot touching $k$ for which the write is not a gate candidate for $p$ or $g(E)\ne p$. Missing positions are zero. Direct-certificate writes, writes with another reply destination, recovered entries, different-ballot entries, and entries without a unique gate therefore contribute to $D_p(k)$. The read target is
\[
 t_p(k)=
 \begin{cases}
   \max\{C_p(k),D_p(k)\}, & p=O_b(k),\\
   A_p(k), & \text{otherwise}.
 \end{cases}
\]
The gate can therefore bypass its gated pending entries while its vote remains unpublished: no other learner can yet establish responder coverage. It cannot bypass an accepted entry without this gate. After collecting majority and coverage evidence, it marks the entry covered-chosen and advances $C_p(k)$ before releasing its vote; subsequent reads target the chosen write.

After checking read authority, a responder returns applied state if $t_p(k)\le\texttt{executedPrefix}$. Otherwise, it can return the last write to $k$ in that slot only when the entry is covered-chosen, fresh, in the current ballot, and within \texttt{majorityFixedPrefix}. This reads a chosen value without executing an arbitrary log suffix. If any required condition is missing, the read waits. \texttt{strictPrefix} disables this shortcut after handoff. No read may bypass an unapplied bridge. Authority and visibility are rechecked when a waiting read is released.

\paragraph{Cold fast-path details.} Each cold key advances through versions $v$ in order. The round $(b,e,a)$ distinguishes voting authority from version order; the leader coordinates preparation and conflict recovery. The client sends its original write in \textsc{ColdPropose} to all of $Q_b$ and sends \textsc{Intent} to notify the coordinator of the pending write. Acceptors require the installed, prepared generation and a fast-enabled key round. Each proposes its locally next version $v=\texttt{keyPrefix}(k)+1$; differing prefixes cannot produce matching votes. They retain the full command, block local reads of the key, and send \textsc{FastVote} directly to the client. Within $(b,e,a,k,v)$, they vote for only one command; retries repeat that vote. Each vote binds the ballot, epoch, key round, key, version, request identifier, command digest, original reply destination, client timestamp, and fixed quorum. Matching \textsc{FastVote}s from all of $Q_b$ form the fast certificate, completing the write. The client asynchronously disseminates this certificate as a \textsc{ChosenProof}. Identical retries of chosen requests may receive \textsc{ColdDone}, binding the original identifier, key, and digest. Replicas execute in per-key order, fetching missing proofs or bodies while retaining pending commands.

Each round permits exactly one fast quorum, reducing Fast Paxos's recovery-intersection requirement~\cite{lamport2006fastpaxos} to
\[
  P\cap Q_b\ne\varnothing
  \qquad\text{for every majority }P.
\]
Only the designated quorum can vote in a fast round. A member's failure blocks this path and triggers roster recovery.

\paragraph{Collision fallback and cold reads.} Competing writes trigger a higher key round. The coordinator collects complete snapshots from a majority and preserves chosen prefixes and tails that might already have been chosen. It then enters a classic round, proposing one safe value for replica acceptance. The value requires matching votes from every member of $Q_b$. Losing inputs remain pending for later versions. Fast resumption requires a higher prepare, a safe empty remaining tail, and catch-up and \textsc{Ready} replies from all of $Q_b$ confirming installation of the prepared key round.

Cold reads require caught-up members of the installed $Q_b$ and leases from the parent roster. The predicate \texttt{keyReady(k)} records installation and catch-up of the chosen key prefix in the prepared round. An unresolved accepted tail blocks reads until a valid chosen prefix is installed. Completed fast writes fence all cold readers, even before proof delivery; neither routing hints nor payloads restore reads.

\begin{figure}[t]
\begin{lstlisting}[style=xpaxos]
function submit(c):
  retain pending[id(c)] = exact original c
  if c.op = READ:
    send c to selected read responder
  else if coldEnabled and coldKey(c.key)
     and c.op = WRITE and not migrated:
    send ColdPropose(c) to every q in Q_b
    send Intent(c) to current coordinator
  else:
    submit Request(c) for leader ordering
      with its designated reply replica

when ClientVote(b,s,id,hE,hc,S) arrives from p:
  pre: p in R, pending[id] is eligible WRITE
       and hc = H(pending[id]), b>0, s>0,
       nonempty S subset R
  cert := (b,s,id,hE,hc,S)
  voters[cert] := voters[cert] union {p}
  if |voters[cert]| >= m
     and S subset voters[cert]:
    completeOnce(id, acknowledgement)

when FastVote(b,e,a,k,v,id,h,origin,timestamp,Q_b)
     arrives from p:
  pre: p in Q_b, exact pending[id] is WRITE,
       k = pending[id].key, h = H(pending[id]),
       origin matches pending[id].replyReplica,
       b>0, v>0, |Q_b|=m, Q_b subset R
  x := (b,e,a,k,v,id,h,origin,timestamp,Q_b)
  fastVoters[x] := fastVoters[x] union {p}
  if Q_b subset fastVoters[x]:
    disseminate ChosenProof(x,Q_b,pending[id]) to R
      asynchronously
    completeOnce(id, acknowledgement)

when a valid read/executed reply or ColdDone
     with (id,result) arrives:
  pre: reply satisfies its identity/output checks
  completeOnce(id, result)
\end{lstlisting}
\caption{\OurProtocol client. Vote sets count each replica once; $h_E$ binds the ordered batch contents and reply destinations, $h_c=H(c)$, and $S=S_b(E)$. The function \texttt{completeOnce} completes each original request at most once.}
\label{fig:xpaxos-client}
\end{figure}

\begin{figure*}[t]
\begin{minipage}[t]{0.485\textwidth}
\begin{lstlisting}[style=xpaxos]
when Request(c) arrives at prepared leader L_b:
  pre: active installed roster and ordinary route
  append exact c to next ordered batch E
  perform local acceptance handler for E
  send Accept(E) to the other replicas

when Accept(E) arrives from installed L_b:
  pre: matching ballot and immutable slot value
  retain full E before acknowledging
  update pending-key indexes and read fences
  record self in V[b,s]
  update local contiguous acceptedPrefix
  if g(E) != self:
    publishVotesOnce(E)
  tryChoose(E)

when Accepted(b,s,prefix) arrives from replica p:
  pre: p in R, matching installed ballot/slot
  V[b,s] := V[b,s] union {p}
  retain p's highest reported accepted prefix
  majorityFixedPrefix := mth largest report
  tryChoose(E)

function tryChoose(E):
  if full local E exists and |V[b,s]| >= m
     and S_b(E) subset V[b,s]
     and not coveredChosen(E):
    mark E covered-chosen; advance read targets
    execute contiguous covered-chosen prefix once
    deliver executed replies from completion replicas
    if g(E) = self:
      publishVotesOnce(E)
    if self = L_b:
      send Commit(b,s) to the other replicas
    reconsider held reads

when Commit(b,s) arrives from installed L_b:
  if local E at s has ballot b:
    mark E covered-chosen; advance read targets
    execute contiguous covered-chosen prefix once
    reconsider held reads

function publishVotesOnce(E):
  send Accepted(b,s,acceptedPrefix)
     to replicas learning the entry
  if fresh(E) and not migrated
     and no Bridge entry is present:
    for each certificate-eligible WRITE c in E:
      send ClientVote(b,s,id(c),batchDigest(E),
                      H(c),S_b(E)) to client(c)

when READ(k) arrives at ordinary responder p:
  pre: p in S_b(k), valid majority leases,
       caught up through grantor thresholds
  t := target t_p(k) defined above
  if t <= executedPrefix:
    reply from applied state
  else if coveredChosen(t) and fresh(t)
          and ballot(t)=b
          and t <= majorityFixedPrefix
          and not strictPrefix
          and no unapplied Bridge at or before t:
    reply from chosen key value
  else:
    hold request until safe or forward it
\end{lstlisting}
\end{minipage}\hfill
\begin{minipage}[t]{0.485\textwidth}
\begin{lstlisting}[style=xpaxos]
when ColdPropose(c) arrives at q in Q_b:
  pre: installed generation (b,e), fast key round a,
       matching generation hint, known predecessor,
       no freeze or handoff
  if exact id/body is already chosen:
    send ColdDone to client(c)
  else if next version v is unvoted in (b,e,a):
    retain full original c
    invalidate key k; accept (b,e,a,k,v,c,Q_b)
    send FastVote(b,e,a,k,v,id(c),H(c),
                  origin(c),timestamp(c),Q_b)
      to client(c)
  else if the same c was accepted in (b,e,a,v):
    resend its unchanged FastVote
  else if a different c competes:
    retain c; notify coordinator of collision

when a valid ChosenProof(x,Q_b,c) arrives:
  pre: all Q_b votes match x
  retain hash-checked body
  if migrated or a Bridge is present: return
  retain proof and chosen identifier
  if chosen predecessor of x is installed:
    apply next per-key version once
    recompute key readiness; release safe reads
  else:
    buffer proof and fetch missing predecessor

when READ(k) arrives at cold member q:
  pre: q in installed Q_b, matching installed/promised generation,
       stable lease, fully caught-up key,
       no freeze or handoff
  if keyReady(k) and no unresolved accepted tail:
    reply from chosen per-key state
  else:
    hold or forward to authorized Q_b member

when collision recovery prepares key round a':
  promise a' across remaining versions
  collect m complete key snapshots
  preserve chosen prefix; select safe tail
    by the highest (b,e,a) and intersection rule
  propose one classic value at each next version
  choose only with matching votes from all Q_b
  apply in per-key order; replay losing inputs
  before fast resumption: higher prepare,
    safe empty tail, catch-up and all-Q_b Ready
\end{lstlisting}
\end{minipage}
\caption{Ordinary (left) and cold (right) handlers. Votes require full bodies; reads need leases, catch-up, and visibility checks.}
\label{fig:xpaxos-normal}
\end{figure*}

\subsection{Leader and Cold-Plane Recovery}
\label{app:xpaxos-recovery}

\paragraph{Leadership change and ordinary recovery.} Failure suspicion triggers a higher-ballot leader or replacement of an unavailable required read responder; suspicion alone cannot select values or authorize reads. Replicas suspend old read authority, revoke grants or await expiration, and retain accepted evidence. The new leader runs Paxos phase one: it sends \textsc{Prepare}, collects complete promises from $m$ replicas, and adopts the highest-ballot accepted entry at each remaining log slot. Only unreported gaps receive internal no-ops. It re-accepts recovered entries under the new roster with $\mathit{fresh}$ cleared. New read responders require majority grants and catch-up through reported thresholds.

\paragraph{Recovering possible cold choices.} The coordinator sends \textsc{ColdPrepare} for a frozen generation $G'=(b',e')$ with $b'>b$ to freeze the cold plane: replicas stop old cold-path votes and promise $G'$ over all admitted cold keys and their future versions. A majority $P$ returns \textsc{PromiseCatalog} replies containing complete immutable catalogs of chosen prefixes, duplicate identifiers, accepted tails, historical ballots, key rounds, quorums, and references to command bodies.

For an open version, recovery first selects the highest reported accepted round $(b_0,e_0,a)$, ordered lexicographically. Classic rounds retain their unique accepted value. Fast rounds using $Q_{b_0}$ retain $x$ if
\[
  \forall p\in P\cap Q_{b_0}:\quad
  p\text{ reports acceptance of }x\text{ at }(b_0,e_0,a).
\]
This condition identifies a \emph{possible} old fast choice, not a new certificate. Because $|P|=|Q_{b_0}|=m$ and $|R|=2f+1$, the intersection is nonempty. Any value previously accepted by all of $Q_{b_0}$ must appear at every member of this intersection, provided each promise is complete. A missing acceptance in a complete snapshot rules out that candidate; a missing snapshot or body does not. Recovery uses the identities of intersecting replicas, not just a count of matching reports. If no value could have been fast-chosen, the coordinator deterministically selects a retained original input and leaves others pending. It records recovered prefixes, selected tails, and body references in a manifest that remains immutable for generation $G'$. Missing bodies must be fetched from original holders; their absence cannot justify discarding a possible choice.

\paragraph{Recovered state and its certificate.} The frozen manifest $M=(G',Q_{b'},\ldots)$ records recovered prefixes, chosen identifiers, and selected classic tails. Together, these form the foundation state to be transferred into the ordinary log. The coordinator broadcasts $M$ and $H(M)$ in \textsc{BulkAccept}. Each \textsc{BulkVote} requires a matching promise, local retention of every referenced body, installation of valid chosen-prefix evidence, and acceptance of every selected tail. Matching voters $V_M$ must satisfy $|V_M|\ge m$ and $Q_{b'}\subseteq V_M$; $Q_{b'}$ is selected from complete promise respondents. \textsc{BulkCommit} certifies the tails for per-key application. Replicas acknowledge application with \textsc{BulkApplied}. All of $Q_{b'}$ and the coordinator must finish application; \textsc{Ready} replies from all of $Q_{b'}$ confirm installation before the log transition is proposed. Bulk recovery shares evidence transfer across keys.

\paragraph{One-way handoff.} Handoff permanently moves cold keys to the ordinary log. To bind this transition to the recovered foundation, the coordinator proposes a bridge entry
\[
  F=\textsc{Bridge}(M,H(M),V_M).
\]
The bridge has an internal identifier and no client response. Ordinary Paxos phase one and choice preserve its exact metadata. Accepting it votes for the foundation entry without re-accepting referenced user writes. Bulk voters retain the bound bodies, with a survivor after at most $f$ crashes.

A replica cannot execute past $F$ until it validates the foundation and obtains all hash-checked bodies. Executing $F$ installs recovered cold state, imports chosen identifiers into ordinary deduplication, and atomically marks the replica migrated. Installation occurs once, preventing replay from overwriting later writes with an old foundation. Local bridge acceptance, mode hints, and body receipt cannot authorize migration.

After migration, all keys use the ordinary log in strict-prefix mode: reads wait for their relevant slot to execute, and write replies require actual execution. Direct acceptance certificates and cold admission stay disabled. Stale cold requests retain their identities when rerouted to the log. Stale control and proof messages cannot alter migrated state. Historical bodies support reconstruction but grant no authority.

\paragraph{Further failures.} Later leaders run ordinary phase one, preserve selected bridges, and resume prefix replay and body fetching without starting a separate cold history. A locally accepted but unselected bridge awaits phase-one resolution; acceptance alone cannot establish handoff. After migration, failures use ordinary-log and roster recovery.

\begin{figure*}[t]
\begin{minipage}[t]{0.485\textwidth}
\begin{lstlisting}[style=xpaxos]
function recover(b') at new leader:
  choose b' above observed ballots
  fence old leases and install new roster
  keep reads disabled until grants/catch-up
    and required preparation complete
  send Prepare(b') to R

when complete promises arrive from majority P:
  for each ordinary suffix slot s:
    adopt highest-ballot accepted entry in P
    use internal no-op only for unreported gap
    mark adopted entry non-fresh
  re-accept suffix with ordinary quorum rules
  if migrated or suffix has a selected Bridge:
    continue its prefix replay and body fetch
  else if cold plane is enabled:
    start frozen recovery at generation G'=(b',e')

when ColdPrepare(G') arrives:
  promise G' over every key and future version
  disable old cold read and fast authority
  retain old bodies, votes, choices, and IDs
  send complete immutable PromiseCatalog

when complete PromiseCatalogs arrive from majority P:
  Q_b' := majority selected from actual responders
  keep highest valid chosen prefixes and IDs
  for each open tail:
    (b0,e0,a) := highest reported accepted round
    if that round is classic: retain its value
    else if every p in P intersect Q_b0 reports
            the same x accepted at (b0,e0,a): retain x
    else: deterministically select retained input
  fetch and verify every referenced body
  freeze one immutable manifest M=(G',Q_b',...)
  send BulkAccept(M,H(M)) to R

when complete BulkAccept(M,h) arrives:
  pre: current promise/leader, same M, h=H(M),
       not migrated and no Bridge present
  fetch every missing referenced full body
  only after all bodies are locally retained:
    install valid old chosen-prefix/ID evidence
    invalidate and accept each selected tail
    send BulkVote(G',Q_b',h)
\end{lstlisting}
\end{minipage}\hfill
\begin{minipage}[t]{0.485\textwidth}
\begin{lstlisting}[style=xpaxos]
when matching BulkVotes arrive from V_M:
  pre: |V_M| >= m and Q_b' subset V_M
  send BulkCommit(M,H(M),V_M) to R

when matching BulkCommit arrives:
  pre: locally retained/accepted complete M,
       matching certificate: |V_M|>=m, Q_b' subset V_M,
       current promise, no migration or Bridge
  attach choice evidence to selected tails
  apply recovered tails in per-key order
  send BulkApplied(G',Q_b',H(M))

when all Q_b' and coordinator have applied M:
  complete all-Q_b' installation/Ready barrier
  pre: frozen recovery
  propose Bridge(M,H(M),V_M) as separate log entry

when prefix execution reaches Bridge F:
  pre: ordinary chosen entry in prefix order
  verify exact manifest, hash, and |V_M| >= m
    with Q_b' subset V_M
  fetch missing hash-bound bodies from holders
  if any required body is missing: block prefix
  if foundation has not already been installed:
    install recovered cold key state and tails
    import chosen original IDs into log dedup
    atomically set migrated and strictPrefix
  disable cold control and early client votes
  retain historical bodies for other replicas
  replay pending originals through ordinary log

when migrated READ(k) arrives:
  require ordinary lease/catch-up authority
  wait until relevant slot <= executedPrefix
  reply from applied state

when migrated WRITE completes:
  require actual ordinary prefix execution
  emit only the identity-bound executed result
\end{lstlisting}
\end{minipage}
\caption{One-way handoff. Votes require full bodies and acceptance; bridge acceptance does not authorize reads.}
\label{fig:xpaxos-recovery}
\end{figure*}

\subsection{Client and Replica Failure Details}
\label{app:xpaxos-other-failures}

\paragraph{Client failures.} Both ordering mechanisms retain full commands at replicas before voting. If a client fails before disseminating a proof, phase one preserves possibly chosen values from acceptance records; proof absence cannot rule out a choice.

\paragraph{Replica failures.} Ordinary writes can continue after nonmandatory replica failures if a majority and all required responders survive. Losing an ordinary read responder or a member of $Q_b$ requires a higher roster ballot, even with a live leader. Old leases are fenced and replacement responders catch up before activation. The cold plane uses the recovery and handoff in \autoref{app:xpaxos-recovery}; its fast quorum stays fixed throughout ballot $b$, including changes of key round.

\paragraph{Payload and response recovery.} Replicas fetch hash-bound bodies and predecessor proofs from original holders. Transfers are idempotent and confer no vote or read authority. \textsc{ColdDone} requires the original command's chosen evidence and, after migration, ordinary prefix execution. Retries retain original identifiers and complete at most once. Eligible writes may complete through certificates; executed replies require prefix execution at the planned completion replica.

\subsection{Completion and Read-Execution Details}
\label{app:xpaxos-output}

\paragraph{Early completion.} Ordinary certificates shorten evidence delivery to clients; the leader still orders commands. The cold path also removes this ordering hop for enabled, uncontended per-key instances. Both certificate paths require recoverable acceptance certificates before write completion. Write acknowledgements need no speculative execution of unchosen commands. Read results require valid read visibility.

\paragraph{Read execution.} Normal-mode reads beyond the applied prefix use chosen key values under the freshness, fixed-prefix, and lease conditions in \autoref{app:xpaxos-normal-details}, without speculative suffix execution. Broader certificate eligibility can hold more reads behind accepted writes that may already have completed elsewhere. Read races preserve visibility checks, so placement accounts for write-evidence delivery and read waiting.

\paragraph{The handoff boundary.} Bridge acceptance disables early client votes; ordered application installs the foundation and permanently enables strict-prefix execution. Reads cannot pass an unapplied bridge, and later writes cannot combine cold votes with ordinary acceptances. Later operations follow one recovered foundation and one ordinary executed prefix.

\subsection{Routing and Delivery Optimizations}
\label{app:xpaxos-optimizations}

\paragraph{Ingress and direct admission.} The designated reply replica is implemented as a logical ingress. The planner chooses this ordinary write ingress using the modeled end-to-end write cost, rather than simply choosing the nearest replica. A designated completion owner, when usable, becomes the ingress for its key. The client can also submit the same write directly to the leader while retaining that ingress identity. Duplicate submissions are coalesced. A leader's executed-result notification does not let the ingress acknowledge a write it has not itself executed. These routes change message arrival times, not certificate conditions.

\paragraph{Selective acceptance delivery.} Accepting replicas send ordinary votes to the leader and the learners selected to complete writes; eligible writes additionally generate client votes. Restricting recipients reduces traffic. It does not reduce the number or identity requirements of the votes a recipient must collect. Commit notices and repair messages let other replicas learn and execute the chosen history. Recipient selection is an optimization; delaying publication of a completion gate's own vote until it has locally established coverage is a protocol rule.

\paragraph{Read races.} A client may send the same read to more than one authorized responder and use the first valid reply. Each responder independently checks its lease, catch-up state, and visibility predicate. Later replies do not create another logical completion. Racing requests cannot make an unsafe value visible.

\paragraph{Streams and body transfers.} Protocol messages, including recovery traffic, share one protocol stream per peer pair. Shared control/RTT RPCs are separate from these protocol streams. Fragments, retransmissions, and fetched bodies remain bound to their identifiers and hashes; receiving them alone neither chooses a value nor restores read authority.

\subsection{Detailed Workload-Aware Planning}
\label{app:xpaxos-planning}

\paragraph{Inputs and decisions.} Let $\mathcal{K}$ be the key domain. From an independent profiling sample of $N$ offered requests, the system records
\[
  N(k)=N^R(k)+N^W(k)
\]
accesses to each observed key $k$, where $N^R(k)$ and $N^W(k)$ are its sampled read and write counts. Observed keys are ordered by decreasing $N(k)$, with key identifier used only to break ties. If $k_j$ is the key at popularity rank $j$, its empirical probability is
\[
  p_j=\frac{N(k_j)}{N}.
\]
The runtime retains the inverse mapping $\rho(k_j)=j$. An unobserved key has no rank and always uses the ordinary log. Its direct-certificate eligibility depends on the selected reference or all-key policy.

\paragraph{Measured boundaries.} Let $U=|\{k:N(k)>0\}|$ be the number of observed keys and $\overline N=N/U$ their mean sampled count. The hot boundary is
\[
 \eta=|\{k:N(k)>\overline N\}|.
\]
Thus ranks $0,\ldots,\eta-1$ are hot. Integer implementation compares a key's count with $\lfloor N/U\rfloor$, which is equivalent to the strict comparison above. Equal-frequency keys remain on the same side of the boundary. A uniform sample has $\eta=0$. This is a fixed empirical classification rule, not an optimality claim or an online estimate that changes during execution.

The cold-path search starts with candidate cutoff $\eta$. It computes cumulative request mass over the remaining observed ranks and adds boundaries at the first complete frequency group crossing each of the fractions $1/7,\ldots,6/7$ of that mass. A boundary at the end of all observed keys is omitted, and coincident boundaries are merged. There are at most seven candidates. The selected cutoff $\tau\ge\eta$ may leave a non-hot ordinary region $\eta\le\rho(k)<\tau$. If the cold path is disabled, every key uses the ordinary log. Neither boundary depends on a numeric key identifier. The heat profile is hash-pinned, loaded before planning, and immutable across recovery; original keys, rather than their ranks, remain in commands.

For $N_c$ clients, placement additionally uses directed delays $d_{xy}$, request rates, and write fraction $w$. A plan $\pi$ selects leader $\ell$, read responders $S$ containing $\ell$, client read responders $r_i\in S$, write ingresses $h_i$, completion owners, certificate eligibility, and an optional cold quorum $Q$. A rank cutoff $\tau$ assigns observed keys with $\rho(k)\ge\tau$ to the cold path.

The model groups popularity ranks into cohorts $g$, with probability mass $P_g$ and effective write rate $\lambda_g$:
\[
  P_g=\sum_{j\in g}p_j,\qquad
  \lambda_g=\lambda\frac{\sum_{j\in g}p_j^2}{P_g},
\]
where $\lambda$ is the aggregate configured write rate per millisecond. This grouping approximates the per-key distribution. Ordinary candidate search uses configured client weights; the final combined screen averages over clients uniformly.

\paragraph{Ordinary write costs.} Let $z_v(x)=d_{\ell v}+d_{vx}$ be the time for replica $v$'s acceptance to reach recipient $x$ after leader propagation. If $\operatorname{ord}_m$ denotes the $m$th smallest arrival time, the required evidence reaches $x$ at
\begin{equation}
  \begin{split}
    E_\pi(x)=\max\bigl\{&\operatorname{ord}_m\{z_v(x):v\in R\},\\
                       &\max_{s\in S}z_s(x)\bigr\}.
  \end{split}
  \label{eq:xpaxos-plan-evidence}
\end{equation}
The two terms capture majority acceptance and coverage of every authorized read responder. A nearby responder can lower read delay while raising write-evidence time.

For client $i$, the model accounts for forwarding and direct submission using
\[
  a_i=\min\{d_{ih_i}+d_{h_i\ell},\ d_{i\ell}+1\,\mathrm{ms}\}.
\]
Certificate and ingress-return costs are
\begin{align}
  T_i^C &= a_i+E_\pi(i)+2\,\mathrm{ms},\\
  T_i^H &= \max\{d_{ih_i},a_i+E_\pi(h_i)\}
           +d_{h_i i}.
\end{align}
An eligible command has modeled cost $\min\{T_i^C,T_i^H\}$; other commands use $T_i^H$. A cohort may designate a completion owner, the replica returning its executed replies, in place of the default ingress. Ordinary candidate search also adds $1$\,ms to read-network costs and selected write costs. These are fixed heuristic offsets, not measured processing delays or required runtime timers. In particular, the direct-admission implementation sends immediately; the model's $1$\,ms term is not an enforced delay.

\paragraph{Read waiting.} The interval between acceptance and evidence arrival at responder $r$ is approximated by
\[
  H_r=[E_\pi(r)-d_{\ell r}]_+,
  \qquad [x]_+=\max\{x,0\}.
\]
Let $N_{ir}=d_{ir}+d_{ri}$ and $\lambda^f_{gr}=\lambda_g s_{gr}$, where $s_{gr}$ scales the rate of writes imposing read fences. It is one for certificate-eligible cohorts. Otherwise, it is zero at a designated completion owner, reflecting entries for which that owner gates publication of its vote; it is the estimated remote-write fraction at a leader without such an owner, and one at other responders. With $\mathbf{1}$ denoting an indicator, the exponential approximation gives the read cumulative distribution function (CDF):
\begin{equation}
  F^R_{igr}(t)=\mathbf{1}_{t\ge N_{ir}}
    \exp\!\left(-\lambda^f_{gr}[H_r+N_{ir}-t]_+\right).
  \label{eq:xpaxos-plan-read}
\end{equation}
The extra waiting time $W_{gr}$ has expectation
\[
  \mathbb{E}[W_{gr}]
    =H_r-\frac{1-e^{-\lambda^f_{gr}H_r}}{\lambda^f_{gr}},
\]
with value zero when $\lambda^f_{gr}=0$. This residual-wait expression is used in preliminary mean estimates. Broader certificate eligibility can lower write latency while increasing read waiting. For a planned race between responders $r$ and $s$, the combined screen uses $F_r(t)+F_s(t)-F_{rs}(t)$, where $F_{rs}$ accounts for overlapping fencing intervals caused by the same writes.

\paragraph{Cold-path costs.} A cold write requires every member of $Q$, so its uncontended cost for client $i$ is
\[
  T_i^F=\max_{q\in Q}(d_{iq}+d_{qi})+2\,\mathrm{ms}.
\]
The conflict approximation uses $V_Q=\max_i T_i^F+200$\,ms and $\rho_g=\min\{1,\lambda_g V_Q\}$. Let $A$ be the majority leader--replica RTT, $B$ the largest leader--$Q$ RTT, and $D$ the largest client--replica RTT. The additional recovery-cost proxy is $\Delta_Q=2A+2\max\{A,B\}+B+D+20$\,ms. The cold-write CDF is
\begin{equation}
  F^{W,\mathrm{cold}}_{ig}(t)
    ={}(1-\rho_g)\mathbf{1}_{t\ge T_i^F}
      +\rho_g\mathbf{1}_{t\ge T_i^F+\Delta_Q}.
\end{equation}
Cold reads select a nearby authorized member of $Q$. Their waiting model replaces the fixed interval $H_r$ in \autoref{eq:xpaxos-plan-read} with a write-duration distribution $B_g$: each of the $N_c$ clients contributes durations $T_j^F$ and $T_j^F+\Delta_Q$ with weights $(1-\rho_g)/N_c$ and $\rho_g/N_c$. The exponent becomes $-\lambda_g\mathbb{E}[(B_g+N_{ir}-t)_+]$, capturing both uncontended and recovering writes.

\paragraph{Joint objective and search.} Let $F^R_{ig,\pi}$ and $F^W_{ig,\pi}$ be the modeled CDFs for cohort $g$. The final screen combines cohort masses and uniform client weights:
\begin{equation}
  F_\pi(t)=\sum_g\frac{P_g}{N_c}\sum_i
  \bigl[(1-w)F^R_{ig,\pi}(t)+wF^W_{ig,\pi}(t)\bigr].
  \label{eq:xpaxos-plan-mixture}
\end{equation}
The implemented midpoint sum with $h=0.25$\,ms estimates the truncated mean
\[
  \mathbb{E}\!\left[\min\{T,1000\,\mathrm{ms}\}\right].
\]
Quantiles are defined by the CDF and estimated with $35$ bisection iterations on $[0,1000]$\,ms:
\begin{align}
  \widehat\mu_\pi &= h\sum_{j=0}^{3999}[1-F_\pi((j+\tfrac12)h)],\\
  q_\pi(u) &= \inf\{t:F_\pi(t)\ge u\}.
\end{align}
This mixture evaluates overlapping read/write effects together; individual path savings are not added separately.

For memberships up to $13$ replicas, the ordinary search considers leader/responder combinations, chooses a nearby responder for each client, and compares ingress and completion assignments. An alternative stage selects a best-mean candidate allowing certificates for all ordinary keys. Above $13$ replicas, this all-key exploration is skipped; the inherited planner uses a bounded topology-derived base search and subsequent refinement. These are implementation search restrictions, not limits on the quorum rules. The cold stage ranks size-$m$ cold quorums by a workload-weighted combination of cold-read and uncontended-write cost, retains up to $64$ candidates, and searches the measured cutoff candidates defined above. This stage screens a coarse mean-gain proxy and nearby tail quantiles. Its tail search uses a $10{,}000$\,ms upper bound, distinct from the $1000$\,ms final comparison below. It prefers modeled $0.99$-quantile improvement, breaking ties by mean gain; it can also disable the cold plane.

Candidate $\pi_1$ is compared with reference plan $\pi_0$. The reference plan permits direct ordinary certificates only for observed keys with $\rho(k)\ge\eta$, whereas $\pi_1$ permits them for all ordinary-path writes. Keys absent from the empirical profile remain on the ordinary path and use the conservative reference behavior unless the all-key policy is selected. Define $\Delta\mu=\widehat\mu_{\pi_0}-\widehat\mu_{\pi_1}$ and $\Delta q(u)=q_{\pi_0}(u)-q_{\pi_1}(u)$. When the alternative enables all-key certificates, select it when
\begin{equation}
  \Delta\mu>\varepsilon
  \quad\land\quad
  \min_{u\in\mathcal{U}}\Delta q(u)\ge-\gamma\Delta\mu,
  \label{eq:xpaxos-plan-selection}
\end{equation}
where the current heuristic uses $\varepsilon=0.05$\,ms, $\gamma=5$, and $\mathcal{U}=\{0.9898,0.9900,0.9902\}$. The selected plan is fixed before normal operation, and its roles become active through the lease, promise, and catch-up transitions in \autoref{sec:xpaxos-normal}.

\paragraph{Scope of the model.} The numerical grid, truncation bounds, and fixed offsets above are performance heuristics. They do not grant votes, leases, or read authority, and the selection test does not guarantee an improvement in measured mean or tail latency. The rank aggregation and moment-expansion boundaries in the implementation control numerical approximation independently of hot membership and candidate cold cutoffs. The profile counts both reads and writes; the model uses a configured write fraction to estimate per-key write pressure rather than inferring a separate write fraction for every key. A workload change can therefore make a fixed plan inefficient without changing its safety conditions.

\onecolumn

\section{Prompt Used for Consensus Discovery by the General-purpose Agent}
\label{app:search-prompt}

\begin{tcolorbox}[
  colframe=promptpurple,
  colback=white,
  coltitle=white,
  fonttitle=\bfseries,
  title=Prompt,
  boxrule=1.2pt,
  arc=3pt,
  left=6pt,
  right=6pt,
  top=6pt,
  bottom=6pt
]
Run a 20-round iterative protocol search to discover, implement, and verify a new state-machine replication protocol with lower mean and p99 latency.

\medskip Assume a non-Byzantine crash-failure model with n = 2f + 1 replicas, tolerating any f replica failures.

\medskip
Promote a protocol version only if one latency metric improves by at least 1\% while the other degrades by no more than 0.2\% compared with the current best promoted version. 
\end{tcolorbox}

\section{Qualitative Analysis on Latency}

\begin{table}[htbp]
\centering
\small
\renewcommand{\arraystretch}{1.1}
\setlength{\tabcolsep}{6pt}
\resizebox{\textwidth}{!}{%
\begin{tabular}{@{}lclcccc@{}}
\toprule
\textbf{Protocol} & \textbf{Leader} & \textbf{Path}
  & \textbf{W} & \textbf{Write quorum} & \textbf{R} & \textbf{Read quorum} \\
\midrule
\paxos & Yes & Normal & $4\delta$ & $f+1$ & $4\delta$ & $f+1$ \\
\midrule
\multirow{2}{*}{\fastpaxos} & \multirow{2}{*}{Fallback only}
  & Fast & $\ge 3\delta$ & $>3n/4$ & $\ge 3\delta$ & $>3n/4$ \\
\cmidrule(l){3-7}
  & & Slow & $\ge 4\delta$ & $f+1$ & $\ge 4\delta$ & $f+1$ \\
\midrule
\multirow{2}{*}{\epaxos} & \multirow{2}{*}{Per command}
  & Fast & $\ge 4\delta$ & $f+\lfloor(f+1)/2\rfloor$
  & $\ge 4\delta$ & $f+\lfloor(f+1)/2\rfloor$ \\
\cmidrule(l){3-7}
  & & Slow & $\ge 6\delta$ & $f+1$ & $\ge 6\delta$ & $f+1$ \\
\midrule
\multirow{2}{*}{\textsc{CURP}} & \multirow{2}{*}{Yes}
  & Fast & $\ge 2\delta$ & $f+\lceil f/2\rceil+1$
  & $\ge 2\delta$ & $f+\lceil f/2\rceil+1$ \\
\cmidrule(l){3-7}
  & & Slow & $\ge 4\delta$ & $f+1$ & $\ge 4\delta$ & $f+1$ \\
\midrule
\multirow{2}{*}{\swiftpaxos} & \multirow{2}{*}{Yes}
  & Fast & $2\delta$ & Fixed $f+1$ or any $>3n/4$
  & $2\delta$ & Fixed $f+1$ or any $>3n/4$ \\
\cmidrule(l){3-7}
  & & Slow & $3\delta$ & $f+1$ or corrected fast quorum
  & $3\delta$ & $f+1$ or corrected fast quorum \\
\midrule
\multirow{2}{*}{\kcensus} & \multirow{2}{*}{Fallback only}
  & Fast & $\ge 3\delta$ & $f+1\sim 2f+1$
  & \multirow{2}{*}{$\ge 4\delta$} & \multirow{2}{*}{$f+1$} \\
\cmidrule(l){3-5}
  & & Slow & $\ge 5\delta$ & $f+1$ & & \\
\midrule
\bodega & Yes & Normal & $\ge 4\delta$ & Any $f+1$ $\cup$ all $S$
  & $\ge 2\delta$ & $1$ \\
\midrule
\multirow{4}{*}{\textbf{\OurProtocol (ours)}} & \multirow{4}{*}{Yes}
  & Cold fast & $2\delta$ & \multirow{2}{*}{Fixed $f+1$}
  & \multirow{4}{*}{$\ge 2\delta$} & \multirow{4}{*}{$1$} \\
\cmidrule(l){3-4}
  & & Cold slow & $4\delta$ & & & \\
\cmidrule(l){3-5}
  & & Ordinary cert & $\ge 3\delta$
  & \multirow{2}{*}{Any $f+1$ $\cup$ all $S$} & & \\
\cmidrule(l){3-4}
  & & Ordinary exec & $\ge 4\delta$ & & & \\
\bottomrule
\end{tabular}%
}
\caption{Qualitative comparison of latency and quorum requirements.
$W$ and $R$ denote write and read latency, respectively; $\delta$ is a one-way
message delay, $n=2f+1$, and $S$ is the set of authorized read responders.
``Cert'' and ``exec'' denote certificate-based completion and completion
after execution, respectively.}
\label{tab:protocol-latency}
\end{table}

\clearpage

\section{Quantitative Results Extended} \label{app:quantitative_results}

\begin{table*}[htbp]
  \centering
  \scriptsize
  \setlength{\tabcolsep}{3pt}
  \resizebox{\textwidth}{!}{%
  \begin{tabular}{rrll*{12}{r}}
    \toprule
    Replicas & Clients & Topology & Workload
      & \multicolumn{2}{c}{\swiftpaxos}
      & \multicolumn{2}{c}{\bodega}
      & \multicolumn{2}{c}{\multipaxos}
      & \multicolumn{2}{c}{\kcensus}
      & \multicolumn{2}{c}{\textbf{\OurProtocol (ours)}}
      & \multicolumn{2}{c}{Lower bound} \\
    \cmidrule(lr){5-6}
    \cmidrule(lr){7-8}
    \cmidrule(lr){9-10}
    \cmidrule(lr){11-12}
    \cmidrule(lr){13-14}
    \cmidrule(lr){15-16}
     & & & & mean & p99 & mean & p99 & mean & p99
     & mean & p99 & mean & p99 & mean & p99 \\
    \midrule
    5 & 5 & clustered & A & 80.955 & \textbf{116.199} & 121.120 & 223.832 & 113.834 & 151.313 & 94.267 & 180.063 & \textbf{66.412} & 139.075 & 44.250 & 101.500 \\
    5 & 5 & clustered & B & 80.960 & \textbf{115.654} & 54.306 & 217.560 & 113.693 & 151.195 & 97.384 & 119.678 & \textbf{24.687} & 134.241 & 19.905 & 98.000 \\
    5 & 5 & clustered & C & 80.908 & 115.247 & 19.602 & 28.499 & 113.637 & 151.255 & 97.763 & 119.604 & \textbf{19.598} & \textbf{28.490} & 17.200 & 25.000 \\
    \addlinespace
    5 & 5 & uniform & A & 168.977 & \textbf{244.626} & 233.824 & 352.695 & 214.035 & 264.545 & 230.193 & 403.652 & \textbf{166.515} & 263.005 & 118.700 & 176.500 \\
    5 & 5 & uniform & B & 168.956 & 244.639 & 144.595 & 346.727 & 214.058 & 264.543 & 225.345 & 384.629 & \textbf{122.017} & \textbf{238.409} & 99.710 & 170.500 \\
    5 & 5 & uniform & C & 168.789 & 244.591 & \textbf{99.664} & 173.462 & 213.860 & 264.618 & 220.368 & 314.027 & 99.672 & \textbf{173.441} & 97.600 & 170.500 \\
    \addlinespace
    5 & 5 & colocated & A & 153.695 & \textbf{208.565} & 161.040 & 332.649 & 229.564 & 305.973 & 150.883 & 371.423 & \textbf{117.837} & 263.490 & 52.750 & 136.000 \\
    5 & 5 & colocated & B & 153.724 & 208.115 & 55.159 & 298.163 & 228.324 & 306.047 & 142.547 & \textbf{196.971} & \textbf{13.598} & 236.153 & 5.275 & 120.000 \\
    5 & 5 & colocated & C & 153.808 & 208.484 & \textbf{1.172} & \textbf{1.941} & 228.488 & 304.936 & 142.128 & 172.418 & 1.175 & 1.944 & 0.000 & 0.000 \\
    \addlinespace
    9 & 9 & clustered & A & 79.813 & \textbf{103.833} & 87.442 & 161.186 & 117.727 & 154.410 & 124.762 & 229.154 & \textbf{70.493} & 124.253 & 34.694 & 62.500 \\
    9 & 9 & clustered & B & 79.883 & 104.306 & 46.822 & 132.945 & 117.225 & 153.806 & 121.158 & 155.653 & \textbf{43.330} & \textbf{101.677} & 21.219 & 54.000 \\
    9 & 9 & clustered & C & 79.775 & 103.899 & 39.072 & \textbf{63.176} & 116.788 & 153.267 & 117.940 & 153.957 & \textbf{38.934} & 63.322 & 19.722 & 30.500 \\
    \addlinespace
    9 & 9 & uniform & A & 122.627 & 185.610 & 174.834 & 324.223 & 162.447 & 233.553 & 155.954 & 290.973 & \textbf{105.987} & \textbf{174.604} & 63.000 & 139.000 \\
    9 & 9 & uniform & B & 122.701 & 185.536 & 84.364 & 299.766 & 162.025 & 233.145 & 151.243 & 209.582 & \textbf{55.114} & \textbf{178.455} & 36.250 & 118.500 \\
    9 & 9 & uniform & C & 122.856 & 185.277 & 43.658 & \textbf{68.721} & 162.114 & 233.574 & 148.907 & 200.658 & \textbf{43.601} & 68.897 & 33.278 & 52.500 \\
    \addlinespace
    9 & 9 & colocated & A & 84.417 & \textbf{123.636} & 91.433 & 182.198 & 112.918 & 158.605 & 97.413 & 217.760 & \textbf{78.660} & 153.258 & 29.861 & 92.500 \\
    9 & 9 & colocated & B & 84.388 & \textbf{123.739} & 30.395 & 157.426 & 112.048 & 158.044 & 79.084 & 172.399 & \textbf{18.657} & 126.553 & 2.986 & 78.500 \\
    9 & 9 & colocated & C & 84.265 & 123.641 & \textbf{0.993} & \textbf{1.760} & 111.907 & 157.370 & 74.374 & 108.523 & 0.995 & 1.769 & 0.000 & 0.000 \\
    \addlinespace
    13 & 13 & clustered & A & 150.335 & \textbf{181.058} & 235.239 & 431.051 & 205.402 & 290.612 & 194.503 & 405.152 & \textbf{133.820} & 257.339 & 73.788 & 163.500 \\
    13 & 13 & clustered & B & 150.315 & \textbf{181.092} & 102.102 & 417.248 & 205.353 & 290.481 & 166.134 & 318.501 & \textbf{50.181} & 241.444 & 22.575 & 151.000 \\
    13 & 13 & clustered & C & 150.335 & 181.062 & \textbf{19.146} & \textbf{38.409} & 205.528 & 290.659 & 155.950 & 186.375 & 19.158 & 38.444 & 16.885 & 35.000 \\
    \addlinespace
    13 & 13 & uniform & A & 151.033 & \textbf{200.002} & 189.139 & 345.329 & 204.076 & 273.261 & 188.260 & 350.724 & \textbf{121.836} & 224.449 & 86.865 & 177.500 \\
    13 & 13 & uniform & B & 151.037 & \textbf{199.996} & 92.750 & 307.940 & 201.136 & 259.295 & 165.960 & 281.061 & \textbf{60.988} & 201.785 & 48.563 & 163.500 \\
    13 & 13 & uniform & C & 151.091 & 199.981 & 46.742 & 83.545 & 201.257 & 259.212 & 159.612 & 203.161 & \textbf{46.717} & \textbf{83.492} & 44.308 & 81.000 \\
    \addlinespace
    13 & 13 & colocated & A & 119.665 & \textbf{181.827} & 151.955 & 301.688 & 165.229 & 236.281 & 123.742 & 281.655 & \textbf{103.693} & 236.762 & 38.462 & 107.000 \\
    13 & 13 & colocated & B & 119.781 & 182.130 & 66.961 & 268.403 & 163.679 & 234.861 & 110.213 & \textbf{148.136} & \textbf{9.939} & 199.710 & 3.846 & 99.000 \\
    13 & 13 & colocated & C & 119.769 & 181.937 & 1.113 & 2.041 & 163.658 & 234.713 & 109.325 & 144.376 & \textbf{1.105} & \textbf{2.020} & 0.000 & 0.000 \\
    \addlinespace
    17 & 17 & clustered & A & 59.082 & \textbf{94.893} & 94.162 & 187.876 & 78.638 & 121.907 & 64.853 & 121.383 & \textbf{51.768} & 114.666 & 31.338 & 78.000 \\
    17 & 17 & clustered & B & 59.107 & 94.891 & 38.903 & 171.718 & 78.175 & 121.491 & 63.897 & 98.869 & \textbf{20.723} & \textbf{94.450} & 14.437 & 75.500 \\
    17 & 17 & clustered & C & 59.106 & 94.823 & 15.098 & 24.921 & 78.158 & 121.346 & 64.040 & 99.049 & \textbf{15.064} & \textbf{24.918} & 12.559 & 19.500 \\
    \addlinespace
    17 & 17 & uniform & A & 516.777 & 2213.449 & 259.884 & 444.251 & 250.378 & \textbf{307.976} & 263.976 & 486.963 & \textbf{229.048} & 401.577 & 81.412 & 158.500 \\
    17 & 17 & uniform & B & 245.141 & 777.406 & 110.600 & 285.145 & 242.196 & 298.475 & 232.480 & 382.520 & \textbf{95.506} & \textbf{279.663} & 46.629 & 137.500 \\
    17 & 17 & uniform & C & 200.599 & 331.342 & 67.349 & \textbf{103.689} & 241.667 & 297.822 & 221.816 & 275.872 & \textbf{67.346} & 103.860 & 42.765 & 72.000 \\
    \addlinespace
    17 & 17 & colocated & A & 124.224 & \textbf{185.319} & 172.521 & 363.182 & 169.596 & 266.422 & 119.298 & 245.966 & \textbf{105.594} & 234.027 & 49.706 & 149.000 \\
    17 & 17 & colocated & B & 124.170 & 185.336 & 74.361 & 311.960 & 169.546 & 266.466 & 113.237 & \textbf{173.443} & \textbf{15.891} & 229.748 & 4.971 & 128.000 \\
    17 & 17 & colocated & C & 124.371 & 185.224 & 1.240 & \textbf{2.046} & 169.700 & 265.668 & 113.243 & 173.545 & \textbf{1.232} & 2.047 & 0.000 & 0.000 \\
    \bottomrule
  \end{tabular}%
  }
  \caption{Normal-case latency results in milliseconds (mean and p99). The Lower bound is analytical and excluded from ranking. Within each row, \textbf{bold} marks the best result among protocols, separately for mean and p99 latency.}
  \label{tab:evaluation-full-latency}
\end{table*}

\begin{table*}[!htbp]
  \centering
  \scriptsize
  \setlength{\tabcolsep}{3pt}
  \resizebox{\textwidth}{!}{%
  \begin{tabular}{llrrrrrrrrrrrr}
    \toprule
    Workload & Command & \multicolumn{2}{c}{\swiftpaxos} & \multicolumn{2}{c}{\bodega} & \multicolumn{2}{c}{\multipaxos} & \multicolumn{2}{c}{\kcensus} & \multicolumn{2}{c}{\textbf{\OurProtocol (ours)}} & \multicolumn{2}{c}{Lower bound} \\
    \cmidrule(lr){3-4} \cmidrule(lr){5-6} \cmidrule(lr){7-8} \cmidrule(lr){9-10} \cmidrule(lr){11-12} \cmidrule(lr){13-14}
     & & mean & p99 & mean & p99 & mean & p99 & mean & p99 & mean & p99 & mean & p99 \\
    \midrule
    A (50\% writes) & Read & 114.975 & 164.010 & 77.220 & 176.345 & 158.315 & 211.431 & 138.746 & 210.334 & \textbf{56.544} & \textbf{129.864} & 27.967 & 46.417 \\
     & Write & \textbf{115.185} & \textbf{163.568} & 212.630 & 265.170 & 158.526 & 211.380 & 145.781 & 297.498 & 145.359 & 188.035 & 86.452 & 118.000 \\
     & All & 115.081 & \textbf{163.745} & 144.949 & 262.797 & 158.421 & 211.400 & 142.245 & 282.171 & \textbf{100.984} & 186.281 & 57.209 & 118.000 \\
    \addlinespace
    B (5\% writes) & Read & 115.081 & 163.654 & 61.927 & 164.521 & 157.869 & 211.112 & 136.461 & 200.008 & \textbf{40.669} & \textbf{96.328} & 27.967 & 46.417 \\
     & Write & \textbf{115.481} & \textbf{163.833} & 207.254 & 262.435 & 158.382 & 211.527 & 129.774 & 268.959 & 150.453 & 187.524 & 86.452 & 118.000 \\
     & All & 115.102 & \textbf{163.665} & 69.273 & 242.098 & 157.895 & 211.130 & 136.127 & 206.485 & \textbf{46.234} & 169.248 & 30.891 & 106.583 \\
    \addlinespace
    C (read-only) & Read (= All) & 115.067 & 163.523 & 34.027 & \textbf{56.260} & 157.799 & 210.837 & 133.580 & 178.198 & \textbf{33.996} & 56.310 & 27.967 & 46.417 \\
    \bottomrule
  \end{tabular}%
  }
  \caption{Operation-level normal-case latency in milliseconds, averaged with equal weight over the six deployments with 5 or 9 replicas. The Lower bound is analytical and excluded from ranking. Within each row, \textbf{bold} marks the best result among protocols, separately for mean and p99 latency.}
  \label{tab:evaluation-operation-latency}
\end{table*}

\FloatBarrier

\clearpage

\section{Protocol Counterexamples}
\label{app:protocol-counterexamples}

\tikzset{
  tlbox/.style={rounded corners=2.5pt,draw=reply,fill=white,
    line width=.8pt,inner sep=4pt,align=center,font=\fontsize{15}{17}\selectfont},
  tlbad/.style={tlbox,draw=unsafe,fill=unsafefill,text=unsafe},
  tlfinal/.style={tlbox,draw=finalcol,fill=finalfill},
  tltext/.style={fill=white,inner sep=2pt,align=center,font=\fontsize{14}{16}\selectfont},
  tlmsg/.style={-{Latex[length=7pt,width=5pt]},line width=1.2pt,draw=ycol},
  tlreply/.style={tlmsg,draw=reply,dash pattern=on 5pt off 3pt},
  tldelay/.style={tlmsg,draw=xcol,dash pattern=on 5pt off 3pt},
  tlexec/.style={tlbox,draw=ink,font=\fontsize{16}{18}\selectfont\bfseries}
}
\newcommand{\tllane}[3]{%
  \draw[mid,line width=.8pt] (60,#2)--(#3,#2);
  \node[anchor=east,font=\fontsize{18}{20}\selectfont\bfseries] at (48,#2) {$r_{#1}$};}
\newcommand{\tlphase}[4]{%
  \node[align=center,font=\fontsize{14}{16}\selectfont\bfseries,text=gray] at (#1,27) {#2};
  \draw[rule,dash pattern=on 2pt off 3pt] (#1,49)--(#1,#3);
  \node[tltext,text=gray,font=\fontsize{12}{14}\selectfont] at (#1,#4) {};}
\newcommand{\tltime}[3]{%
  \draw[gray,-{Latex[length=6pt]},line width=.8pt] (60,#1)--(#2,#1);
  \node[anchor=east,text=gray,font=\fontsize{14}{16}\selectfont] at (#2,#3) {time};}

\subsection{Tentative Metadata Authorizes an Unsafe Pruning Decision}
\label{app:tentative-pruning}

\paragraph{Audit question.}
Is the metadata used to justify pruning already final? 
EPaxos Revisited~\cite{tollman2021epaxos} uses sequence numbers and dependencies to prune commands that appear to execute after the current instance.
A tentative \code{PREACCEPTED} tuple can differ from the tuple fixed by \code{COMMIT}, so using it to authorize a cutoff can expose an incorrect order.

\begin{figure}[!htbp]
\centering
\resizebox{\linewidth}{!}{%
\begin{tikzpicture}[x=1pt,y=-.8pt,text=ink]
\path[use as bounding box] (0,0) rectangle (1000,450);
\tlphase{180}{$X$ preaccepted}{405}{425}
\tlphase{475}{$Y$ commits}{405}{425}
\tlphase{680}{Tentative pruning}{405}{425}
\tlphase{900}{Final metadata arrives}{405}{425}
\tllane{0}{95}{990}\tllane{1}{210}{990}\tllane{2}{325}{990}
\draw[tldelay] (80,95) .. controls (65,240) and (95,275) .. (250,285)
  .. controls (420,260) and (550,270) .. (600,325);
\draw[tldelay] (386,79) .. controls (500,42) and (840,135) .. (880,303);
\fill[xcol] (80,95) circle[radius=3pt];
\draw[tlmsg,draw=xcol] (80,95)--(125,210);
\node[tltext,text=xcol] at (125,130) {$\PreAccept(X)$};
\node[tlbox] at (180,210) {$X:(0,\emptyset)$\\PREACCEPTED};
\draw[tlreply] (240,210)--(285,95);
\node[tltext,text=reply] at (270,162) {$\mathit{OK}_X$};
\node[tlfinal] at (335,95) {$X:(0,\emptyset)$\\COMMITTED};
\node[tlexec] at (440,95) {execute $X$};
\fill[ycol] (260,325) circle[radius=3pt];
\draw[tlmsg] (260,325)--(310,210);
\node[tltext,text=ycol] at (282,277) {$\PreAccept(Y)$};
\node[tlbox] at (365,210) {$Y:(1,\{X\})$\\PREACCEPTED};
\draw[tlreply] (425,210)--(460,325);
\node[tltext,text=reply] at (457,260) {$\mathit{OK}_Y$};
\node[tlfinal] at (500,325) {$Y:(1,\{X\})$\\COMMITTED};
\draw[tlmsg] (550,325) .. controls (562,255) and (565,160) .. (580,95);
\node[tltext,text=ycol] at (572,163) {$\Commit(Y)$};
\node[tlexec] at (635,95) {execute $Y$};
\node[tltext,text=xcol] at (425,292) {delayed $\PreAccept(X)$};
\node[tlbad] at (660,325) {$X:(2,\{Y\})$\\PREACCEPTED};
\draw[unsafe,-{Latex[length=5pt]},line width=.8pt] (660,347)--(660,365);
\node[tlbad] at (660,385) {prune $X$};
\draw[unsafe,-{Latex[length=5pt]},line width=.8pt] (700,385)--(730,385);
\node[tlexec,draw=unsafe] at (775,385) {execute $Y$};
\node[tltext,text=xcol] at (770,157) {delayed $\Commit(X)$};
\node[tlfinal] at (900,325) {$X:(0,\emptyset)$\\COMMITTED};
\draw[unsafe,-{Latex[length=5pt]},line width=.8pt] (725,325)--(843,325);
\node[tltext,text=unsafe] at (784,306) {overwritten};
\draw[unsafe,-{Latex[length=5pt]},line width=.8pt] (900,347)--(922,368);
\node[tlexec,draw=unsafe] at (935,385) {execute $X$};
\tltime{420}{990}{444}
\end{tikzpicture}}
\caption{Pruning with tentative metadata reverses conflicting command order:
$r_0$ executes $X$ before $Y$, while $r_2$ prunes $X$ using its tentative
tuple and executes $Y$ before $X$.}
\label{fig:pruning-counterexample-corrected}
\end{figure}
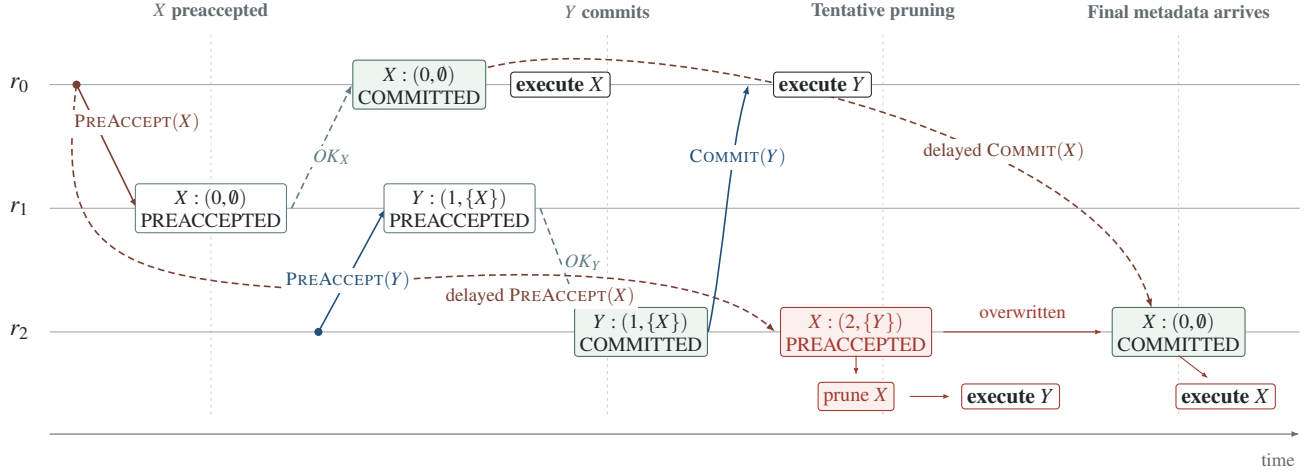

\paragraph{Setup.}
Three replicas $r_0,r_1,r_2$ run EPaxos with dependency pruning enabled.
Conflicting commands $X$ and $Y$ originate at $r_0$ and $r_2$, respectively.
No replica fails; selected \code{PREACCEPT} and \code{COMMIT} messages are delayed.

\paragraph{Counterexample.}
In \autoref{fig:pruning-counterexample-corrected}, command $X$ first commits at $r_0$ with no dependencies. 
Command $Y$ subsequently commits with a dependency on $X$, so $r_0$ executes $X$ before $Y$. At $r_2$, however, a delayed \code{PREACCEPT} for $X$ arrives only after $Y$ has committed. 
The tentative local tuple gives $X$ a higher sequence number and a dependency on $Y$. 
If the pruning rule treats that tentative tuple as final evidence, it removes $X$ from $Y$'s dependency chain and permits $r_2$ to execute $Y$ first. 
The later \code{COMMIT} for $X$ overwrites the tentative tuple, but it arrives too late to undo the order already exposed at $r_2$.
Replica $r_2$ subsequently executes $X$, reversing $r_0$'s order and violating execution consistency.

\subsection{Pruning a Committed Node Can Hide an Earlier Dependency}
\label{app:pruning-boundary}

\paragraph{Audit question.}
Does finality of the pruned node justify discarding its entire dependency
branch in EPaxos Revisited~\cite{tollman2021epaxos}? Unlike \autoref{app:tentative-pruning},
this counterexample already requires the pruned node to be committed.

\begin{figure}[!htbp]
\centering
\resizebox{\linewidth}{!}{%
\begin{tikzpicture}[x=1pt,y=-1pt,text=ink]
\path[use as bounding box] (0,0) rectangle (1000,425);
\tlphase{130}{Committed prefix}{370}{390}
\tlphase{330}{$B$ preaccepted}{370}{390}
\tlphase{555}{$B$ commits}{370}{390}
\tlphase{800}{Different execution orders}{370}{390}
\tllane{0}{95}{990}\tllane{1}{220}{990}\tllane{2}{345}{990}
\draw[tlmsg] (80,345) .. controls (230,310) and (240,160) .. (265,95);
\node[tltext,text=ycol] at (175,285) {$\PreAccept(B)$};
\node[tlfinal] at (130,95) {$C:(1,\{A\})$\\COMMITTED};
\node[tlfinal] at (130,220) {$A:(1,\{B\})$\\COMMITTED};
\node[tlbox] at (325,95) {$B:(2,\{A,C\})$\\PREACCEPTED};
\draw[tlreply] (410,95)--(470,345);
\node[tltext,text=reply] at (450,170) {$\mathit{OK}_B$};
\node[tlfinal] at (535,345) {$B:(2,\{A,C\})$\\COMMITTED};
\draw[tlmsg] (610,345)--(635,220);
\node[tltext,text=ycol] at (590,285) {$\Commit(B)$};
\node[tlbad] at (675,220) {prune $B$\\$C$ is hidden};
\node[tlexec,draw=unsafe] at (760,220) {execute $A$};
\node[tlexec] at (800,95) {execute $C,A,B$};
\node[tltext] at (800,132) {after learning $B$};
\draw[tldelay] (205,95) .. controls (400,40) and (780,40) .. (875,65)
  .. controls (900,75) and (900,165) .. (900,220);
\node[tltext,text=xcol] at (590,65) {delayed $\Commit(C)$};
\node[tlexec,draw=unsafe] at (935,265) {execute $C$};
\draw[unsafe,-{Latex[length=5pt]}] (900,220)--(922,246);
\tltime{390}{990}{415}
\end{tikzpicture}}
\caption{Pruning committed $B$ hides $C$, allowing $r_1$ to execute $A$
before $C$ while $r_0$ executes $C$ before $A$.}
\label{fig:appendix-pruning-boundary}
\end{figure}
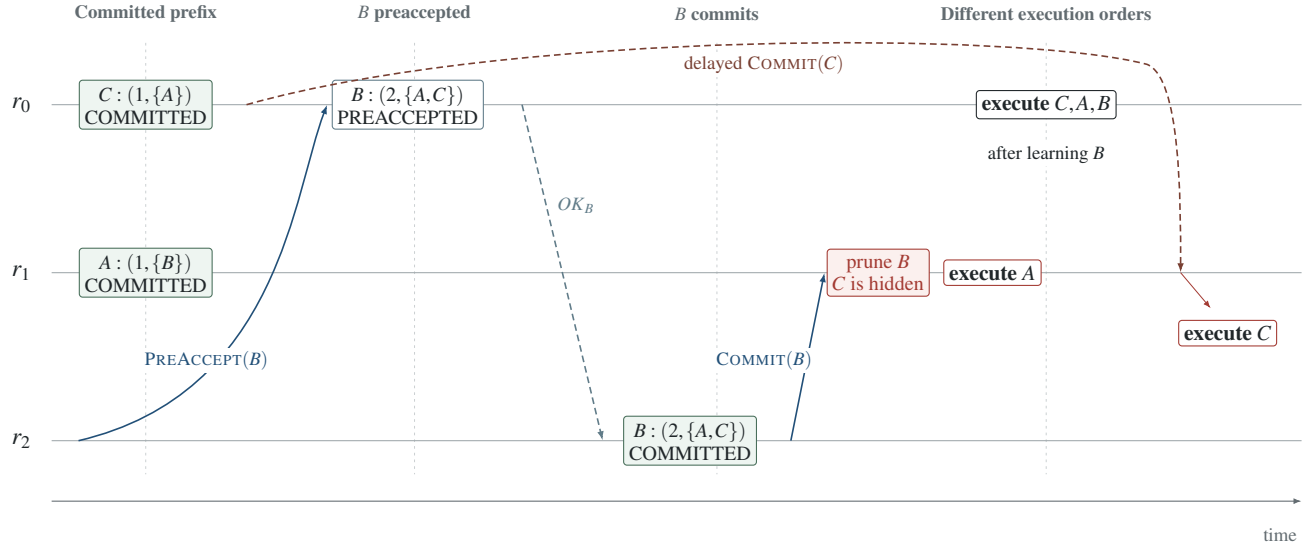

\paragraph{Setup.}
Three replicas originate conflicting commands $C$, $A$, and $B$ at owners
$r_0$, $r_1$, and $r_2$, respectively.
Dependency pruning is enabled; no replica fails and no recovery occurs.

\paragraph{Counterexample.}
Normal-path processing and fast decisions produce the tuples shown in
\autoref{fig:appendix-pruning-boundary}.
The full committed graph is one SCC. Commands $C$ and $A$ tie at sequence 1,
and the replica-ID tie-breaker places $C$ before $A$; $B$ has sequence 2.
Replica $r_0$ selects $B$ as its execution root and executes $C,A,B$.
Replica $r_1$, however, learns the committed $B$ tuple while executing $A$.
The reciprocal dependency and strict sequence inequality allow it to prune
$B$. The cutoff hides $C$, whose final tuple is not yet known locally.
It executes $A$, later learns $C$, and executes $C$, reversing the order of
the conflicting pair $A,C$ and violating execution consistency despite the
pruned node's committed metadata.

\subsection{Early Reads Expose a Value Missing from the Leader's Read View}
\label{app:bodega-early-accept}

One of the authors had identified this issue before the agent but the agent received no hint about it and found it independently.

\paragraph{Audit question.}
Does every read path respect a value already exposed by an early read, even before the leader commits that write?
Bodega's main algorithm~\cite[\S3.2.2 and Fig.~7]{hu2026bodega} permits a non-leader responder to return an accepted value after collecting majority accept notifications (lines~38--39), but directs a stable leader to return the last committed value (line~36).

\begin{figure*}[!htbp]
\centering
\resizebox{\textwidth}{!}{%
\begin{tikzpicture}[x=1pt,y=-.72pt,text=ink,
  bstate/.style={tlbox,minimum width=100pt},
  bread/.style={tlexec,minimum width=88pt},
  bnote/.style={tltext,font=\fontsize{15}{18}\selectfont}]
\path[use as bounding box] (0,0) rectangle (1200,610);
\node[bnote,anchor=west] at (100,18)
  {Three replicas; majority size $2$; responder $A$; $k$ initially absent; valid leases.};
\node[bnote,font=\fontsize{16}{19}\selectfont\bfseries] at (340,58)
  {Accept $w=\textsc{Write}(k,7)$};
\node[bnote,font=\fontsize{16}{19}\selectfont\bfseries] at (730,58)
  {Early read at $A$};
\node[bnote,font=\fontsize{16}{19}\selectfont\bfseries] at (1040,58)
  {Later read at $L$};
\draw[rule,dash pattern=on 2pt off 3pt] (590,80)--(590,547);
\draw[rule,dash pattern=on 2pt off 3pt] (865,80)--(865,547);

\foreach \y/\name in {140/{Leader $L$},255/{Replica $B$},385/{Responder $A$},525/{Reader $R$}}{
  \draw[mid,line width=.8pt] (100,\y)--(1180,\y);
  \node[anchor=east,align=right,font=\fontsize{16}{18}\selectfont\bfseries]
    at (90,\y) {\name};
}

\node[bstate] (lvote) at (215,140) {accept $w$};
\node[bstate] (bvote) at (340,255) {accept $w$};
\node[bstate] (avote) at (340,385) {accept $w$};
\node[bstate,draw=finalcol,fill=finalfill] (anotes) at (535,385)
  {notes $\{L,B\}$\\readable: $7$};
\node[bstate,draw=unsafe,fill=unsafefill,text=unsafe,minimum width=155pt]
  (lread) at (1040,140) {committed state empty\\return absence};
\node[bstate,draw=reply!65,text=gray,minimum width=160pt,
  font=\fontsize{14}{17}\selectfont] at (650,175)
  {$A,B\to L$: accept replies\\remain delayed; no commit};

\node[bnote] at (190,98) {$W$ invokes $w$};
\draw[tlmsg,draw=xcol] (190,112)--(lvote.north);
\draw[tlmsg] (lvote.east)--(bvote.north west);
\draw[tlmsg] (lvote.south)--(avote.north west);
\node[bnote] at (221,285) {Accept};
\draw[tlmsg] (285,140)--(anotes.north);
\node[bnote] at (445,220) {$\mathsf{AcceptNote}_L$};
\draw[tlmsg] (bvote.east)--(anotes.north west);
\node[bnote] at (420,325) {$\mathsf{AcceptNote}_B$};

\node[bread] (first) at (805,525) {$r_1\to 7$};
\node[bread,draw=unsafe,fill=unsafefill,text=unsafe] (second) at (1130,525)
  {$r_2\to\varnothing$};
\draw[tlmsg] (625,525)--(665,385);
\node[bnote] at (616,451) {$\textsc{Read}_1(k)$};
\draw[tlreply] (715,385)--(first.north west);
\node[bnote] at (733,440) {$7$};
\draw[tlmsg] (first.east)--(lread.south west);
\node[bnote] at (916,368) {$\textsc{Read}_2(k)$};
\draw[tlreply,draw=unsafe] (lread.south east)--(second.north);
\node[bnote,text=unsafe] at (1118,330) {absence};

\draw[gray,-{Latex[length=6pt]},line width=.8pt] (100,565)--(1180,565);
\node[bnote,anchor=west,text=gray] at (100,593)
  {$w$ remains pending; $r_1$ completes before $r_2$ starts; each channel preserves FIFO.};
\node[bnote,anchor=east,text=gray] at (1180,593) {event order};
\end{tikzpicture}}
\caption{Counterexample to Bodega's early-accept read optimization.
All replicas accept $w$, but delayed acceptance replies prevent $L$ from committing it.
Accept notifications let $A$ return $7$, while a subsequent read at $L$ returns absence.}
\label{fig:bodega-early-accept-counterexample}
\end{figure*}
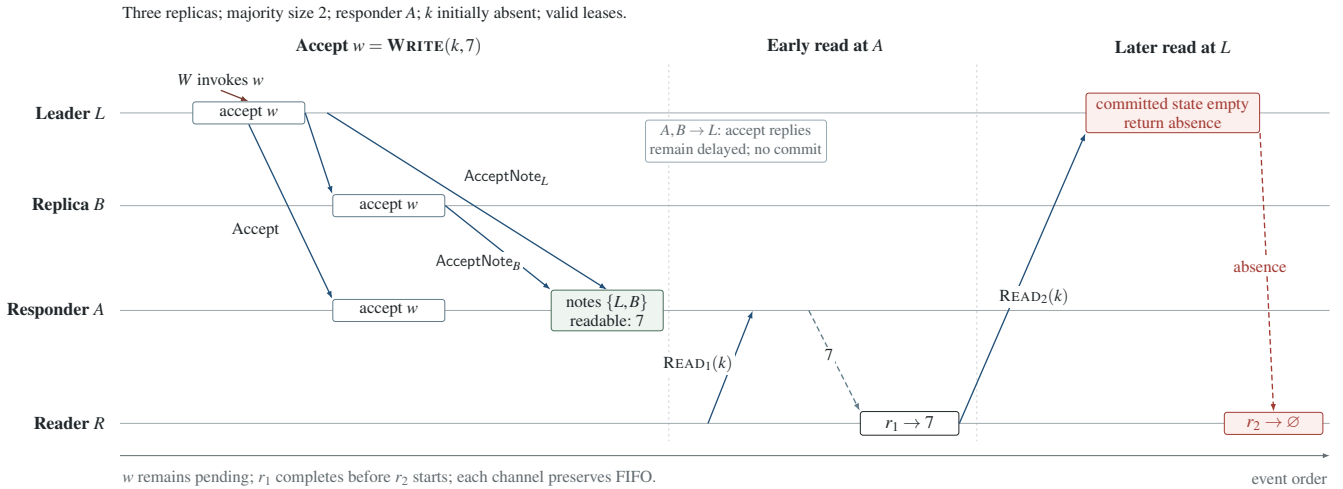

\paragraph{Setup.}
Three replicas $L,A,B$ tolerate one failure; the trace itself has no failures.
The fixed ballot and roster designate $L$ as leader and $A$ as the responder for key $k$, initially absent.
The roster is installed, the leader is prepared, and read-serving replicas hold valid majority-backed leases with catch-up thresholds zero.
Early accept is enabled. No lease expiration or roster change occurs.

\paragraph{Counterexample.}
As shown in \autoref{fig:bodega-early-accept-counterexample}, Writer $W$ invokes $w=\textsc{Write}(k,7)$, which $L$ proposes at log slot zero after preparation.
All three replicas accept $w$.
The accept replies from $A$ and $B$ to $L$ remain delayed, so the leader cannot commit the write.
$A$ collects accept notifications from the majority $\{L,B\}$.
Reader $R$ then invokes $r_1=\textsc{Read}(k)$.
At $A$, the notifications authorize an early reply with $7$ while the write remains uncommitted locally.
After receiving that result, $R$ invokes $r_2=\textsc{Read}(k)$.
When this read reaches $L$, its stable-leader read path uses the committed state, which is still empty, and returns absence.
The second read can reach the leader after the client unhold timeout while its response from $A$ is delayed.
The write remains pending throughout.

\subsection{Duplicate Writes from Client Retries}
\label{app:bodega-write-retry}

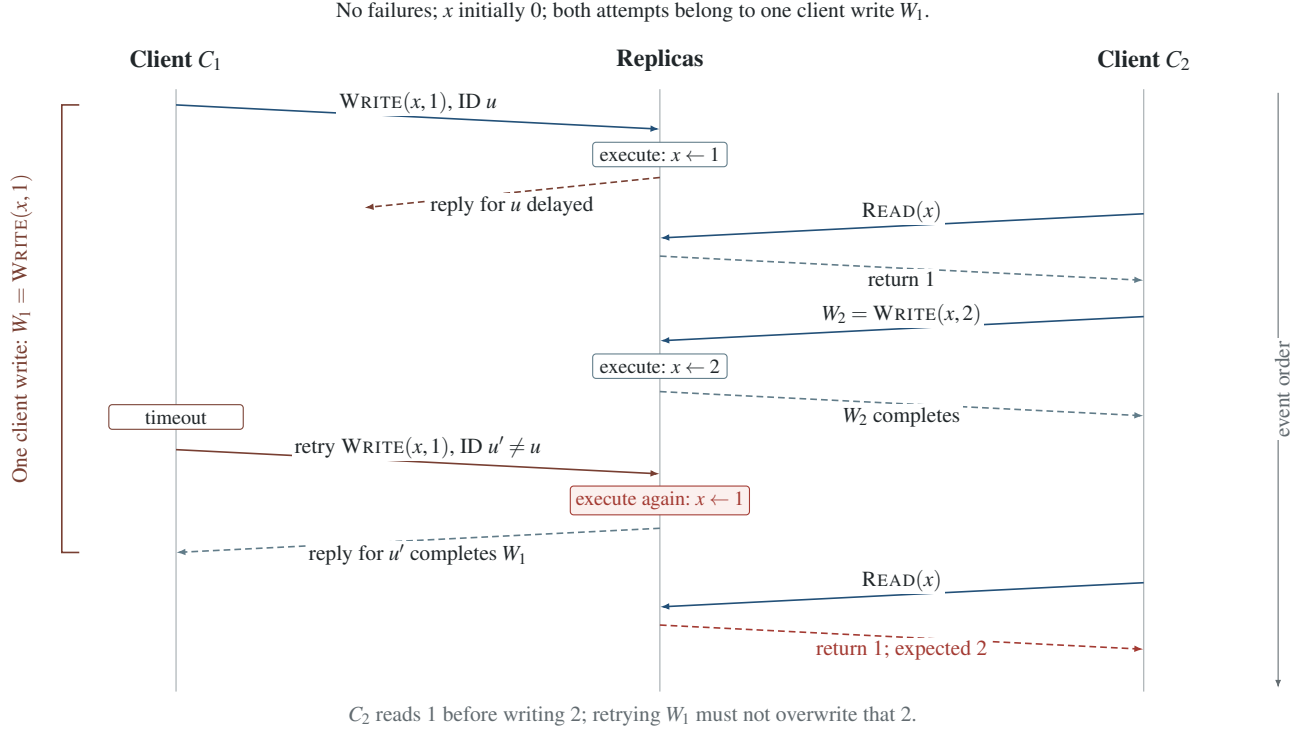
\begin{figure*}[t]
\centering
\resizebox{\textwidth}{!}{%
\begin{tikzpicture}[x=1pt,y=-.9pt,text=ink,
  rlabel/.style={tltext,font=\fontsize{16}{18}\selectfont},
  rstate/.style={tlbox,minimum width=100pt},
  rhead/.style={font=\fontsize{18}{21}\selectfont\bfseries,align=center}]
\path[use as bounding box] (0,0) rectangle (1000,610);
\node[rlabel] at (500,18)
  {No failures; $x$ initially $0$; both attempts belong to one client write $W_1$.};

\foreach \x/\name in {160/{Client $C_1$},520/{Replicas},880/{Client $C_2$}}{
  \node[rhead] at (\x,58) {\name};
  \draw[mid,line width=.8pt] (\x,82)--(\x,580);
}
\draw[gray,-{Latex[length=6pt]},line width=.8pt] (980,85)--(980,577);
\node[rlabel,rotate=90,text=gray] at (983,330) {event order};

\draw[draw=xcol,line width=1.2pt] (88,95)--(75,95)--(75,465)--(88,465);
\node[rlabel,rotate=90,text=xcol] at (45,280)
  {One client write: $W_1=\textsc{Write}(x,1)$};

\draw[tlmsg] (160,95)--(520,115)
  node[rlabel,midway,above] {$\textsc{Write}(x,1)$, ID $u$};
\node[rstate] at (520,136) {execute: $x\gets1$};
\draw[tldelay] (520,155)--(300,180)
  node[rlabel,midway,below] {reply for $u$ delayed};

\draw[tlmsg] (880,185)--(520,205)
  node[rlabel,midway,above] {$\textsc{Read}(x)$};
\draw[tlreply] (520,220)--(880,240)
  node[rlabel,midway,below] {return $1$};

\draw[tlmsg] (880,270)--(520,290)
  node[rlabel,midway,above] {$W_2=\textsc{Write}(x,2)$};
\node[rstate] at (520,311) {execute: $x\gets2$};
\draw[tlreply] (520,332)--(880,352)
  node[rlabel,midway,below] {$W_2$ completes};

\node[rstate,draw=xcol] at (160,353) {timeout};
\draw[tlmsg,draw=xcol] (160,380)--(520,400)
  node[rlabel,midway,above] {retry $\textsc{Write}(x,1)$, ID $u'\ne u$};
\node[rstate,draw=unsafe,fill=unsafefill,text=unsafe] at (520,422)
  {execute again: $x\gets1$};
\draw[tlreply] (520,445)--(160,465)
  node[rlabel,midway,below] {reply for $u'$ completes $W_1$};

\draw[tlmsg] (880,490)--(520,510)
  node[rlabel,midway,above] {$\textsc{Read}(x)$};
\draw[tlreply,draw=unsafe] (520,525)--(880,545)
  node[rlabel,midway,below,text=unsafe] {return $1$; expected $2$};
\node[rlabel,text=gray] at (500,600)
  {$C_2$ reads $1$ before writing $2$; retrying $W_1$ must not overwrite that $2$.};
\end{tikzpicture}}
\caption{A Bodega client retry can repeat a write that has already taken effect.
The first attempt sets $x=1$, but its reply is delayed.
After $C_2$ reads $1$ and completes a write of $2$, a retry with a new identifier overwrites $x$ with $1$.
If both attempts represent one client write, the final read must return $2$.}
\label{fig:bodega-write-retry-counterexample}
\end{figure*}

\paragraph{Audit question.}
Can retrying a timed-out write cause the same client operation to take effect twice?
In Bodega~\cite{hu2026bodega}, the client's timeout retry path assigns a new request identifier to each write attempt.
The server treats the original request and its retry as separate writes, so both can execute.

\paragraph{Setup.}
Two clients access a key $x$, initially $0$, through a leader in a three-replica deployment.
Client $C_1$ invokes $W_1=\textsc{Write}(x,1)$ with timeout retries enabled.
The original request and its retries represent this single write and should take effect only once.
The reply to the first attempt is delayed beyond the client timeout; all reads return committed values.

\paragraph{Counterexample.}
As shown in \autoref{fig:bodega-write-retry-counterexample}, the first attempt of $W_1$, with request identifier $u$, sets $x=1$, but its reply is delayed.
Client $C_2$ reads $x$ and obtains $1$, then completes $W_2=\textsc{Write}(x,2)$.
After timing out, $C_1$ resubmits \textsc{Write}$(x,1)$ with a new identifier $u'\ne u$.
The server executes the retry, overwriting $2$ with $1$, and its reply completes $C_1$'s write.
A subsequent read by $C_2$ returns $1$.
The first read requires $W_1$ to take effect before $W_2$.
If both attempts belong to the same write $W_1$, there is no additional write after $W_2$, so the final read must return $2$.
Returning $1$ therefore violates linearizability for these client operations.

\clearpage

\section{Comparison with AI-Driven Discovery Approaches}
\label{app:related-comparison}

\autoref{tab:ai-discovery} compares \OurAgent with prior AI-based work that generates algorithms or protocols automatically.
AI-driven discovery has already produced new algorithms, but mostly for single-node code or for policies that a simulator can score.
For distributed systems, prior AI work either synthesizes verified implementations of a fixed specification, finds bugs in existing protocols, learns small protocols checked only for a bounded number of processes, without a performance objective, or assists humans in designing protocols proved by hand.

\begin{table*}[htbp]
\centering
\small
\renewcommand{\arraystretch}{1.15}
\setlength{\tabcolsep}{4pt}
\resizebox{\textwidth}{!}{%
\begin{tabular}{@{}lllll>{\raggedright\arraybackslash}p{6cm}@{}}
\toprule
\textbf{Work} & \textbf{Search} & \textbf{Target} & \textbf{Correctness check} & \textbf{Objective} & \textbf{New algorithm or protocol?} \\
\midrule
\multicolumn{6}{@{}l}{\emph{Algorithm and policy discovery}} \\
AlphaDev~\cite{mankowitz2023alphadev} & Deep RL & Sorting, hashing (assembly) & Tests & Latency & Yes: sorting routines merged into LLVM libc++ \\
FunSearch~\cite{romeraparedes2024funsearch} & LLM + evolution & Math, bin packing & Evaluator & Score & Yes: new cap-set constructions and packing heuristics \\
AlphaEvolve~\cite{novikov2025alphaevolve} & LLM + evolution & Math, kernels, scheduling & Evaluator & Score & Yes: new algorithms (e.g., matrix multiplication) and data-center heuristics \\
ADRS~\cite{cheng2025adrs} & LLM + evolution & Systems policies & Simulator, tests & Cost, latency & Yes: single-node policies (scheduling, load balancing), consensus deemed a poor fit \\
Glia~\cite{hamadanian2026glia} & Multi-agent LLM & LLM-serving routing, scaling & Simulator & Completion time & Yes: request-routing algorithm \\
SkyDiscover~\cite{liu2026skydiscover,liu2026evox} & Adaptive LLM evolution & 200+ tasks (math, systems) & Evaluator & Score & Yes: improved policies (e.g., cross-cloud transfer, MoE placement), no distributed protocol \\
\midrule
\multicolumn{6}{@{}l}{\emph{Protocol synthesis, verification, and testing}} \\
IDS~\cite{agarwal2026ids} & LLM agents & Verified key-value stores & Rocq proof of code & Throughput & No new mechanism: data layout and proofs for a fixed specification \\
GGMS~\cite{hui2026ggms} & MCTS + model checker (no LLM) & Sync.\ consensus, atomic commit ($n\le4$, crash) & Bounded model checking & None & Consensus: rediscovers FloodSet~\cite{lynch1996distributed}, one atomic-commit variant claimed new \\
Agora~\cite{liu2026agora} & LLM agents & Raft, EPaxos, HotStuff, BullShark & Tests & --- & No: finds 15 bugs in existing implementations \\
Efron et al.~\cite{efron2026bb} & Human-led, LLM-assisted & Byzantine broadcast (sync., $t<n$) & Pen-and-paper proof & Communication & Yes: communication-efficient Byzantine broadcast protocol \\
\midrule
\textbf{\OurAgent (ours)} & LLM agents & WAN consensus (crash) & Fault injection, Lean & Read/write latency & Yes: \OurProtocol \\
\bottomrule
\end{tabular}%
}
\caption{AI-driven discovery of algorithms and protocols. Before \OurAgent, no automated system had produced a new fault-tolerant consensus protocol that improves measured performance.}
\label{tab:ai-discovery}
\end{table*}

To compare discovery procedures, we adopt the nomenclature of SkyDiscover~\cite{liu2026skydiscover}, which decomposes AI-driven discovery into four components, and use it to describe all systems in \autoref{tab:ai-procedure}, including \OurAgent.
A \emph{context builder} assembles the prompt from the problem and from earlier solutions, a \emph{solution generator} produces candidates, an \emph{evaluator} scores them (by simulation, benchmarking, testing, or proof checking), and a \emph{solution selector} decides which earlier solutions guide the next iteration.
\OurAgent follows the same loop.
It differs in three ways: its context is a curated consensus knowledge base rather than prior programs alone, its evaluator stages benchmarks, fault injection, and Lean proofs so that cheaper checks reject candidates first, and its selector is the agent itself, which chooses a parent from recorded lessons rather than from a population by a fixed rule.

The selectors of FunSearch, AlphaEvolve, ADRS, and SkyDiscover come from evolutionary computation and operate on a population of candidates.
In the \emph{island model}, the population is split into subpopulations (islands) that evolve independently: each new candidate is built from parents of a single island, and islands occasionally exchange their best candidates or are reseeded from stronger ones, which keeps the search from converging prematurely on one line of solutions.
\emph{MAP-Elites} preserves diversity differently, by keeping the best candidate for each cell of a grid of solution features, such as program length or behavior.
None of the other systems maintains such a population: AlphaDev and GGMS select by tree search, IDS by a coordinator, and Glia and \OurAgent by the agent's own judgment.

\begin{table*}[htbp]
\centering
\footnotesize
\renewcommand{\arraystretch}{1.15}
\setlength{\tabcolsep}{4pt}
\begin{tabular}{@{}>{\raggedright\arraybackslash}p{1.9cm}*{4}{>{\raggedright\arraybackslash}p{3.65cm}}@{}}
\toprule
\textbf{Work} & \textbf{Context builder} & \textbf{Solution generator} & \textbf{Evaluator} & \textbf{Solution selector} \\
\midrule
AlphaDev~\cite{mankowitz2023alphadev} & Current assembly program and its CPU state & Policy network adding one instruction at a time & Correctness tests and measured latency & Monte Carlo tree search \\
FunSearch~\cite{romeraparedes2024funsearch} & Prompt with the best programs sampled from one island & Pretrained code LLM & Runs the program on problem instances and scores it & Island-based evolutionary database \\
AlphaEvolve~\cite{novikov2025alphaevolve} & Prompt with prior programs, scores, and human-written context, with prompts that can themselves evolve & Ensemble of Gemini models emitting code diffs & User-supplied evaluation function, applied as a cascade of tests & Program database combining MAP-Elites and islands \\
ADRS~\cite{cheng2025adrs} (OpenEvolve) & Fixed prompt template with problem statement and prior solutions & Ensemble of LLMs editing code & Simulator replaying workload traces & MAP-Elites and islands \\
Glia~\cite{hamadanian2026glia} & Task description and shell access to the codebase & Researcher LLM agent, guided by a Supervisor agent & Simulator (request completion time) & Researcher's own analysis, then best of independent runs \\
SkyDiscover~\cite{liu2026skydiscover,liu2026evox} & Problem context with feedback and reflections from prior solutions & LLM calls with optional tool use & User-supplied scoring function & Pluggable (Top-K, MAP-Elites, Pareto), adaptive (AdaEvolve), or evolved (EvoX) \\
IDS~\cite{agarwal2026ids} & Fixed specification, current strategy, proof state, and failed strategies & Deductive agents writing code and Rocq proofs, and an inductive agent proposing new strategies & Rocq checker on partial proofs, then benchmarks & Coordinator reverts on stagnation and keeps the fastest verified candidate \\
GGMS~\cite{hui2026ggms} & Partial protocol as a game state & Tree search with a transformer action encoder (no LLM) & Bounded model checker & Tree-search values and global depth-first backtracking \\
\midrule
\textbf{\OurAgent (ours)} & Protocol knowledge base (concept index, mechanism cards), parent protocol, and lessons from earlier attempts & Coding agent writing a design document, then a Go implementation in the consensus skeleton & Staged: emulated geo-replicated benchmark, crash fault injection, then Lean proof of safety & Agent chooses the parent and direction from recorded lessons and user-defined latency targets \\
\bottomrule
\end{tabular}
\caption{Discovery procedures, described in the nomenclature of SkyDiscover~\cite{liu2026skydiscover}.}
\label{tab:ai-procedure}
\end{table*}

\DeclareRobustCommand{\faced}{\tikz[baseline=-0.6ex]\fill (0,0) circle (0.6ex);}
\DeclareRobustCommand{\notfaced}{\tikz[baseline=-0.6ex]\draw (0,0) circle (0.6ex);}
\DeclareRobustCommand{\sidestepped}{\tikz[baseline=-0.6ex]{\draw (0,0) circle (0.6ex); \fill (0,-0.6ex) arc (-90:90:0.6ex) -- cycle;}}

Three challenges set consensus apart from the problems that AI-driven discovery has addressed so far.
\textbf{(C1) Performance conflicts with correctness:} the cheapest way to reduce latency is to shrink a quorum or skip a round, which breaks agreement, so a performance score alone steers the search toward incorrect protocols.
\textbf{(C2) Tests cannot establish correctness:} safety violations arise in rare interleavings of messages, failures, and recovery that benchmarks seldom reach, so safety requires a proof, and a proof is only as good as its specification.
\textbf{(C3) Evaluation is expensive:} each candidate requires an implementation, a geo-distributed benchmark, fault injection, and a proof attempt, which takes minutes to hours rather than seconds, so thousands of candidates cannot be evaluated.
\autoref{tab:ai-challenges} shows which of these challenges prior work faces.
In classical algorithms and in distributed-systems policies, every candidate is either valid by construction or checked exactly and cheaply, so none of the three arises.
The few AI-driven approaches to distributed protocols sidestep them by dropping the performance objective, fixing the specification, or restricting the search to three or four processes.

\begin{table*}[htbp]
\centering
\small
\renewcommand{\arraystretch}{1.15}
\setlength{\tabcolsep}{5pt}
\begin{tabular}{@{}llccc>{\raggedright\arraybackslash}p{7.3cm}@{}}
\toprule
\textbf{Work} & \textbf{Problem} & \textbf{C1} & \textbf{C2} & \textbf{C3} & \textbf{Why the challenges do not arise, or how they are sidestepped} \\
\midrule
\multicolumn{6}{@{}l}{\emph{Classical algorithms}} \\
AlphaDev~\cite{mankowitz2023alphadev} & Sorting, hashing & \notfaced & \notfaced & \notfaced & Output is checked by tests on small deterministic inputs. Latency is measured in microseconds \\
FunSearch~\cite{romeraparedes2024funsearch} & Cap sets, bin packing & \notfaced & \notfaced & \notfaced & Every packing is valid and every construction is checked exactly \\
AlphaEvolve~\cite{novikov2025alphaevolve} & Matrix multiplication, scheduling & \notfaced & \notfaced & \notfaced & Decompositions are checked exactly, and every schedule is valid \\
\midrule
\multicolumn{6}{@{}l}{\emph{Distributed-systems policies}} \\
ADRS~\cite{cheng2025adrs} & Spot scheduling, load balancing & \notfaced & \notfaced & \notfaced & Optimizations rarely alter system semantics. Candidates are scored in a simulator \\
Glia~\cite{hamadanian2026glia} & Request routing & \notfaced & \notfaced & \notfaced & Every route is valid, and candidates are scored in a simulator \\
SkyDiscover~\cite{liu2026skydiscover,liu2026evox} & Cloud transfer, GPU placement & \notfaced & \notfaced & \notfaced & Every placement is valid, and candidates are scored by an evaluator \\
\midrule
\multicolumn{6}{@{}l}{\emph{Distributed protocols}} \\
Hui et al.~\cite{hui2023zero} & Atomic commit ($n=3$) & \sidestepped & \sidestepped & \sidestepped & No performance objective. Exhaustive model checking of three processes \\
GGMS~\cite{hui2026ggms} & Consensus, atomic commit ($n\le4$) & \sidestepped & \sidestepped & \sidestepped & No performance objective. Bounded model checking of at most four processes \\
IDS~\cite{agarwal2026ids} & Verified key-value stores & \sidestepped & \faced & \faced & Fixed human-written specification. Performance only ranks verified candidates \\
\textbf{\OurAgent (ours)} & Fault-tolerant WAN consensus & \faced & \faced & \faced & Staged evaluation (benchmark, fault injection, Lean proof) rejects candidates early. Humans define the safety properties \\
\bottomrule
\end{tabular}
\caption{Challenges of consensus discovery (C1: performance conflicts with correctness, C2: tests cannot establish correctness, C3: evaluation is expensive) faced by AI-driven discovery approaches. \faced~faced, \sidestepped~sidestepped by restricting the problem, \notfaced~does not arise.}
\label{tab:ai-challenges}
\end{table*}

\end{document}